\documentclass{article}

\usepackage{arxiv}

\usepackage[utf8]{inputenc} 
\usepackage[T1]{fontenc}    
\usepackage{url}            
\usepackage{booktabs}       
\usepackage{nicefrac}       
\usepackage{microtype} 
\usepackage{lipsum}
\usepackage{graphicx}
\RequirePackage[OT1]{fontenc}
\RequirePackage{amsmath, amsfonts, amssymb}
\RequirePackage[numbers]{natbib}
\RequirePackage[colorlinks,citecolor=blue,urlcolor=blue]{hyperref}
\usepackage{sectsty, secdot}
\usepackage[normalem]{ulem}
\usepackage{wrapfig}
\usepackage{subcaption}
\sectionfont{\fontsize{12}{14pt plus.8pt minus .6pt}\selectfont}

\subsectionfont{\fontsize{12}{14pt plus.8pt minus .6pt}\selectfont}

\usepackage{amsmath, amsthm, amsfonts}
\usepackage{amssymb,mathrsfs}
\usepackage{float, multirow, url}
\usepackage{array, hhline}
\usepackage{caption}
\usepackage{sidecap}

\def\be{\begin{equation}}
\def\ee{\end{equation}}
\newtheorem{theorem}{Theorem}
\newtheorem{lemma}{Lemma}
\newtheorem{corollary}{Corollary}

\theoremstyle{definition}

\newtheorem*{lemma*}{Lemma}
\newtheorem{example}{Example}

\def\l{{\vert }}

\def\bOmg{\boldsymbol{\Omega}}
\def\bgam{\boldsymbol{\Gamma}}
\def\bsig{\boldsymbol{\Sigma}}
\def\bX{\mathbf{X}}

\def\bW{\mathbf{W}}
\def\bZ{\mathbf{Z}}
\def\bS{\mathbf{S}}
\def\bA{\mathbf{A}}
\def\bB{\mathbf{B}}

\def\bI{\mathbf{I}}
\def\bM{\mathbf{M}}
\def\bm{\mathbf{m}}
\def\bN{\mathbf{N}}
\def\bH{\mathbf{H}}
\def\ba{\mathbf{a}}
\def\bb{\mathbf{b}}

\def\by{\mathbf{y}}
\def\bV{\mathcal{V}}
\def\bE{\mathbf{e}}
\def\bbE{\mathcal{E}}
\def\bU{\mathbf{U}}
\def\bu{\mathbf{u}}
\def\bR{\mathbf{R}}

\def\bw{\mathbf{w}}

\def\bxi{\boldsymbol{\xi}}
\def\bt{\boldsymbol{t}}
\def\bu{\boldsymbol{u}}
\def\bmu{\boldsymbol{\mu}}

\def\bsig{\boldsymbol{\Sigma}}

\def\bsig{\boldsymbol{\Sigma}}
\def\bSn{\boldsymbol{S}_n}

\def\bVn{\boldsymbol{V}_n}

\def\bS{\boldsymbol{S}}

\def\x1bi{\bX_{1\bB_i}}

\def\bgam{\boldsymbol{\Gamma}}
\def\bmu{\boldsymbol{\mu}}
\newcommand{\E}{\mathrm{E}}
\def\T{{ \mathrm{\scriptscriptstyle T} }}

\def\T{{ \mathrm{\scriptscriptstyle T} }}

\title{Testing the Graph of a Gaussian Graphical Model}

\author{
  Thien-Minh Le\\
  Department of Biostatistics\\
 Harvard University\\
 Massachusetts, U.S.A.\\
  \texttt{thle@hsph.harvard.edu} \\
   \And
 Ping-Shou Zhong\\
  Department of Mathematics, Statistics and Computer Science\\ University of Illinois at Chicago\\
  Illinois,  U.S.A.\\
  \texttt{pszhong@uic.edu} \\
  \And 
   Chenlei Leng \\
Department of Statistics\\
University of Warwick\\ 
Coventry, UK \\
\texttt{C.Leng@warwick.ac.uk}
}

\begin{document}
\maketitle

\begin{abstract}
The Gaussian graphical model is routinely employed to model the joint distribution of multiple random variables. 
The graph it induces is not only useful for describing the relationship between random variables but also critical for 
improving statistical estimation precision. In high-dimensional data analysis, despite an abundant literature on estimating this graph structure, tests for the adequacy of its specification at a global level is severely underdeveloped. To make progress, this paper proposes a novel goodness-of-fit test that is computationally easy and theoretically tractable. Under the null hypothesis, it is shown that asymptotic
distribution of the proposed test statistic follows a Gumbel distribution. Interestingly the location parameter of this limiting Gumbel distribution depends  on the dependence structure under the null. We further develop a novel consistency-empowered test statistic when the true structure is nested in the postulated structure, by amplifying the noise incurred in estimation. Extensive simulation illustrates that the proposed test procedure
has the right size under the null, and is powerful under the alternative. As an application, we apply the test to the analysis of a COVID-19 data set, demonstrating that our test can serve as a valuable tool in choosing a graph structure to
improve estimation efficiency.
\end{abstract}

\keywords{Dependence; Goodness of fit test; Gaussian graphical model; Gumbel distribution; High-dimensional data.}

\section{Introduction}
The Gaussian graphical model is commonly used for describing the joint distribution of multiple random variables (Lauritzen, 1996). 
The graph structure induced by this model not only delineates the conditional dependence between these variables, but also is critical  for improving estimation precision. In estimating regression parameters in generalized estimating equations (GEE) for example, Zhou $\&$ Song (2016) found that incorporating a suitable dependence structure can  
improve estimation efficiency, sometimes substantially. In another example, Li $\&$ Li (2008) showed that 
a correctly specified dependence structure is also useful to improve estimation efficiency in regularized 
estimation and variable selection in linear regression models. On the other hand however, 
a mis-specified dependence structure affects efficiency negatively (Zhou $\&$ Song, 2016). 
Therefore, specifying an appropriate graph is critical for efficiently estimating a parameter of interest.

In practice, the underlying graph structure of a given data set is rarely known and is often estimated or assumed a priori. Regardless how this graph becomes available, a natural question is whether it is adequate to describe the data from a statistical perspective. This paper aims to develop a novel goodness-of-fit test to address this issue, in the context of high-dimensional data in which dimensionality can exceed the sample size.

There is abundant literature focusing on estimating the underlying graph in the Gaussian graphical model. For fixed-dimensional data, Edwards (2000) studied this problem by using a model selection approach that employs stepwise likelihood ratio tests, while Drton $\&$ Perlman (2004) developed a multiple testing procedure using partial correlations. For high-dimensional data, a popular approach is to employ a penalized likelihood approach, with a penalty explicitly formulated to encourage the sparsity of the resulting precision matrix that induces the underlying dependence structure. On this, we refer to Yuan $\&$ Lin (2007),  Friedman et al. (2007), Cai et al. (2011), Liu $\&$ Wang (2017), and Eftekhari et al. (2021),  among many others. On testing the graphical structure itself, there exist methods for testing elements of the graphical structure. For example, Liu (2013) proposed a bias-corrected estimator of the precision matrix and applied it to test individual components of the precision matrix. 
Similar tests for individual components in a 
precision matrix are also discussed in Jankov\'a and Geer (2017), Ren et al. (2015) and Ning $\&$ Liu (2017). {There are also some existing global tests for precision matrices. For example, Xia et al. (2015) and Cheng et al. (2017). However, these is no existing general specification test for precision matrices, and the existing methods can not be directly applied to global specification tests considered in this paper for testing the entire graph structure. }  

{\color{black}
Our work is also related to a growing body of literature on testing specific covariance structures for high dimensional data. For example, Chen et al. (2010) considered testing sphericity and identity structures, Qiu $\&$ Chen (2012) and Wang et al. (2022) developed tests for bandedness structures, Zhong et al. (2017) developed tests for some parameterized covariance structures such as autoregressive and moving average structures, Zheng et al. (2019) considered tests on linear structures, and Gou $\&$ Tang (2021) considered specification tests for covariance matrices with nuisance parameters in regression models.} These tests are not applicable to test graph structures. 
Moreover, compared with the above tests which usually 
involve the estimation of a finite number of nuisance 
parameters, one significant challenge associated with testing the graph structure in this paper is the need to estimate a high dimensional nuisance parameter.

The main novelty of this paper lies in a new goodness-of-fit test that explores the difference between a graph structure specified under the null and the true underlying graph structure, based on an appropriate maximum norm distance. We overcome the challenge of estimating the high dimensional nuisance parameter by employing a simple and direct plug-in method, thus bypassing the need of choosing tuning parameters in many regularization methods in the literature for estimating a  graph. Despite its simplicity, our test has a limitation in that it is not consistent whenever the graph under the null encompasses but is not equal to the true graph. To tackle this, we carefully develop a novel consistency-empowered test statistic by amplifying the noise, in the sense that small stochastic noises as a result of estimating zero entries in the graph will be enlarged. This modified test statistic is shown to be consistent for testing all types of graphs.

\setcounter{equation}{0} 

\section{Basic Setting and Our Proposed Test Statistic}

Let $\bX_1,\ldots,\bX_n$ be independent and identically distributed realizations of a $p$-dimensional random vector $\bX$ with mean $\boldsymbol{\mu}$ and covariance matrix $\bsig^*=(\sigma_{ij}^*)$. The corresponding precision matrix is denoted as $\bOmg^*=(\omega_{ij}^*)=\bsig^{*-1}$. It is known that $\bOmg^*$ naturally induces a graph denoted as $\mathcal{G}^*=(\bV, \bbE^*)$, where $\bV=\{1,\ldots, p\}$ is the set of nodes and $\bbE^* =\{(i,j): \omega_{ij}^*\not= 0 \}
\subset \bV \times \bV$ is the set of edges consisting of node pairs whose corresponding entries in $\bOmg^*$ are not zero. The absence of a pair of nodes in $\bbE^*$ indicates that the corresponding variables are conditionally independent given all the others (Lauritzen, 1996). 

While graph $\bbE^*$ is rarely known, in practice it can be estimated via the penalized likelihood methods discussed in Introduction or assumed a priori. For the latter, when the dimension $p$ is high, a convenient assumption popular in the literature is that $\bOmg^*$ admits some simple structure such as a banding or a block diagonal structure. We will denote the corresponding graph under the assumption as $\bbE_0$ and the main aim of this paper is to ascertain whether this assumption is valid. That is, we 
 consider the following hypothesis
$$H_0: \bbE^*=\bbE_0 \quad\mbox{vs.}\quad H_1: \bbE^*\neq\bbE_0,$$
where in our high-dimensional setup, $\bbE_0$ usually has a cardinality much smaller than $p^2$.  {\color{black} Our hypothesis corresponds to a  hypothesis for testing the precision matrices $\Omega^*$ with its  non-zero elements  completely unspecified. The number of the unknown parameters under the null is allowed to grow with $p$, which is drastically different from existing tests in the literature for testing a covariance matrix $\Sigma^*$ with its inverse under the null often specified up to a finite number of unknown parameters (e.g., Zhong et al. (2017); Zheng et al. (2019)).}

Let $\bOmg_0=(w_{ij,0})$ be a $p\times p$ precision matrix of $\bX$ compatible with $\bbE_0$ under the null, in the sense that if $(i, j) \not\in \bbE_0$, then $w_{ij, 0}=0$. The exact definition of $\bOmg_0$ is not important for the discussion below. Our main idea is that if $\bbE_0$ is correctly specified, $\bOmg_0$ will be equal to $\bOmg^*$; that is,
\[ \bsig^* \bOmg_0- \bI_p =\boldsymbol{0}_p,
\]
where $\bI_p$ is the $p$-dimensional identity matrix and $\boldsymbol{0}_p$ is a $(p\times p)$-dimensional matrix with entries all being zero. That is, if $\bbE_0=\bbE^*$, we can write the above equation elementwise as
\begin{equation}\label{eq:1}
\max_{1\leq i, j\leq p}\l \bE_j^\T\bsig^*\bw_{i,0} - \bE_j^\T\bE_i \l =0,
\end{equation}
where $\bOmg_0=(\bw_{1,0},\ldots, \bw_{p,0})$ by denoting $\bw_{i,0}$ as the $i$-th column of $\bOmg_0$ and $\bI_p=(\bE_1,\ldots, \bE_p)$ with $\bE_i$ being the $i$-th basis vector. On the other hand, if $\bbE_0$ is not correctly specified in the sense that $\bbE_0\not=\bbE^*$, the maximum element of  $ \bsig^* \bOmg_0- \bI_p$ may be different from zero.

Thus, to assess whether $H_0$ is true is equivalent to check \eqref{eq:1}. If $\bOmg_0$ and so $\bw_{i,0}$ is known in advance, 
an estimator of $(\bE_j^\T\bsig^*\bw_{i,0} - \bE_j^\T\bE_i)^2$ may be obtained by replacing $\bsig^*$ by the sample covariance matrix 
$\bSn =\sum_{i = 1}^n (\bX_i-\bar\bX) (\bX_i-\bar{\bX})^\T/(n-1)$ with $\bar{\bX}=\sum_{i=1}^n \bX_i/n$. Then, we may use the following statistic $D_n$ to distinguish $H_0$ and $H_1$, 
$$D_n= \max_{1\leq i, j\leq p} D_{ij}^2, \quad D_{ij}^2:= ( \bE_j^\T \bSn \bw_{i,0}-\bE_j^\T \bE_i)^2 /\theta_{ij, 0},$$
where $\theta_{ij, 0} = \mbox{var}(\bE_j^\T \bSn \bw_{i,0} -\bE_j^\T\bE_i)$, whose leading order term is given in the following Lemma.

\begin{lemma*}\label{covlm1}
For all $1 \leq i, j \leq p$, we have (a) $\mbox{var}({\bE_j^\T\bSn \bw_{i,0}-\bE_j^\T\bE_i})=\omega_{ii}^*\sigma_{jj}^*/n$, for $1 \leq i \neq j \leq p$
and (b) $\mbox{var}({\bE_i^\T\bSn \bw_{i,0}-\bE_i^\T\bE_i})=(\omega_{ii}^*\sigma_{ii}^* + 1)/n,$ for $1 \leq i \leq p$.
\end{lemma*}


However, $D_{n}$ is not directly applicable because several quantities involved are unknown. Noting that 
under the null hypothesis, $\bOmg_0$ is a sparse matrix, we denote the number of nonzero entries in the $j$th column of $\bOmg_0$ as $s_j$, where $\max_{1\le j\le p} s_j=o(\surd{n})$ is a typical assumption made in estimating high-dimensional precision matrices ( Cai et al. (2011), Liu $\&$ Wang (2017) ). Let $\bw_{i1, 0}$ and $\bw_{i0, 0}$ represent sub-vectors of $\bw_{i, 0}$ with only nonzero and only zero elements respectively, and $(\bB_{i, 0})$ be a $p\times s_i$ matrix with elements being either 0 or 1 such that $\bB_{i, 0} \bw_{i1, 0} = \bw_{i, 0}$. 
The precision matrix $\bOmg_0$ can be estimated in the following column-by-column fashion. 
Denote $\bsig_0=\bOmg_0^{-1}$. By definition, $\bsig_0\bw_{i,0}=\bsig_0\bB_{i, 0}\bw_{i1,0}=\bE_i$
and then $\bB_{i, 0}^\T\bsig_0\bB_{i, 0}\bw_{i1,0}=\bB_{i, 0}^\T\bE_i$. Thus, $\bw_{i1,0}=(\bB_{i, 0}^\T\bsig_0\bB_{i, 0})^{-1}\bB_{i, 0}^\T\bE_i$. Under $H_0$, 
$\bB_{i, 0}^\T\bX_1,\ldots,\bB_{i, 0}^\T\bX_n$ are $s_i$-dimensional independent and identically distributed random vectors with covariance $\bB_{i, 0}^\T\bsig_0\bB_{i, 0}$. 
Because $s_i$ are of smaller order of $\surd{n}$, $\bB_{i, 0}^\T\bsig_0\bB_{i, 0}$ can be consistently estimated by the sample covariance of $\bB_{i, 0}^\T\bX_1,\cdots,\bB_{i, 0}^\T\bX_n$  given by 
$\bB_{i, 0}^\T \bSn \bB_{i, 0}$ under $H_0$. Then, $\hat{\bw}_{i1,0}=(\bB_{i, 0}^\T \bSn \bB_{i, 0} )^{-1} \bB_{i, 0}^\T \bE_i$ 
and $\hat{\bw}_{i,0} = \bB_{i, 0} \hat{\bw}_{i1,0}$ is a consistent estimator of $\bw_{i,0}$ under $H_0$. By assembling $\hat{\bw}_{i,0}$ as $\hat{\bOmg}_0$, we have a consistent estimator of ${\bOmg}_0$. 
The technical detail of the preceding argument can be found in Le and Zhong (2021).

Based on Lemma \ref{covlm1}, we can then estimate $\theta_{ij, 0}$ as 
 $\hat{\theta}_{ij,0}  = (\hat{\omega}_{ii,0}s_{jj}+\delta_{ij})/n$ where $\hat{\omega}_{ii,0}$ is the 
$(i,i)$th element of  $\hat{\bOmg}_0$, $s_{jj}$ is the $(j,j)$th element of matrix $\bSn$, and $\delta_{ij}=1$ if $i\not=j$ and $\delta_{ij}=0$ if $i=j$. 
Replacing the unknown parameters by their estimators,  we construct a 
 test statistic $\hat{D}_n$ using the plug-in estimators
$\hat{\bw}_{i,0}$,
$$\hat{D}_n =\max_{1\leq i, j\leq p} {\hat{D}_{ij}}^2,$$ 
with $\hat{D}_{ij}^2 =(\bE_j^\T \bSn \hat{\bw}_{i,0}  - \bE_j^\T \bE_i)^2/\hat{\theta}_{ij,0}$. 
The test statistic $\hat{D}_n$ is free of tuning and extremely easy to calculate for practical use. These advantages should be compared to those penalized likelihood methods for which the choice of tuning parameters for estimating ${\bOmg}_0$ is critical for the performance of the resulting estimator.

\begin{figure}[h]
     \centering
     \begin{subfigure}{0.2\textwidth}
         \centering
         \includegraphics[width=\textwidth]{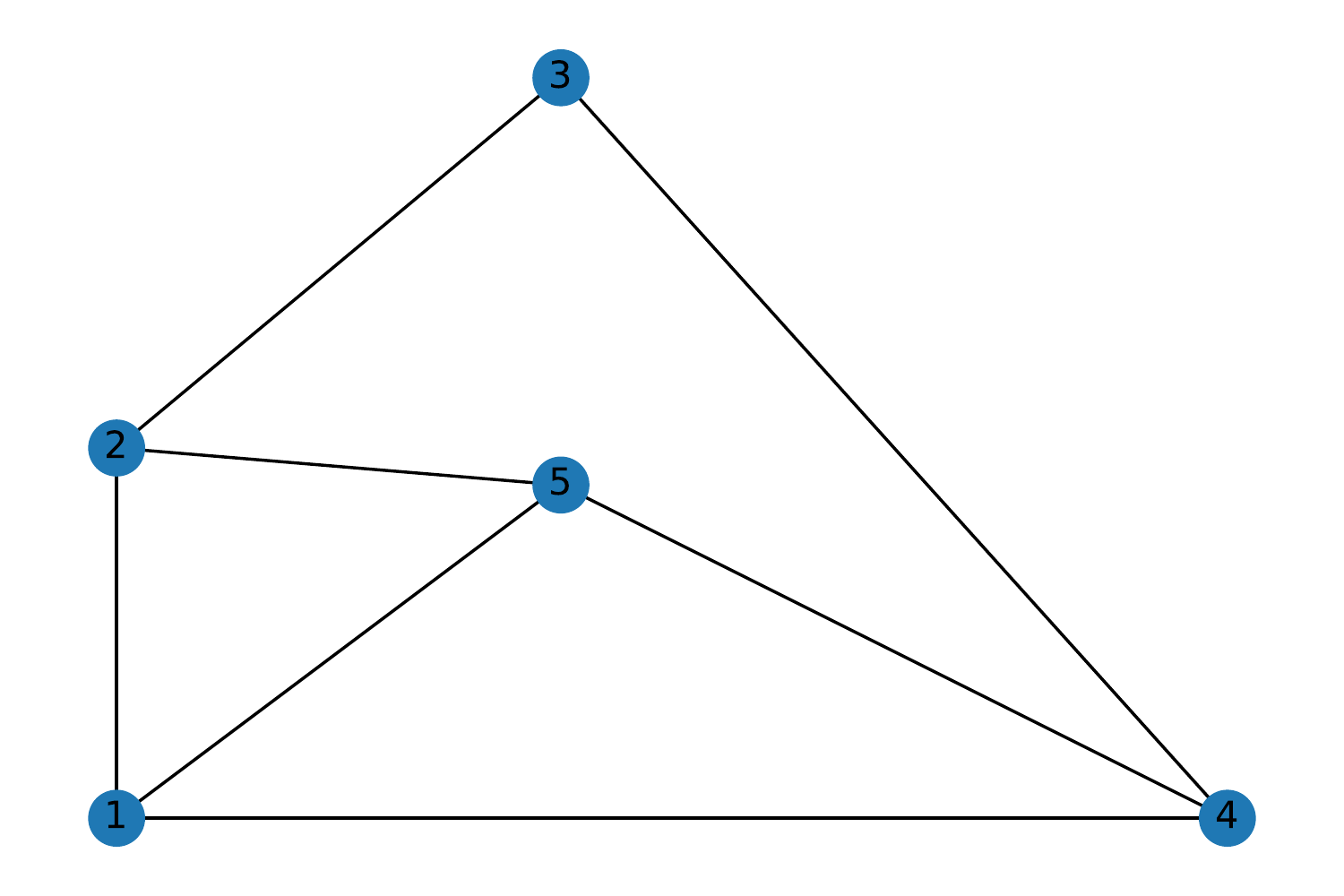}
         \caption{}
     \end{subfigure}
     \begin{subfigure}{0.2\textwidth}
         \centering
         \includegraphics[width=\textwidth]{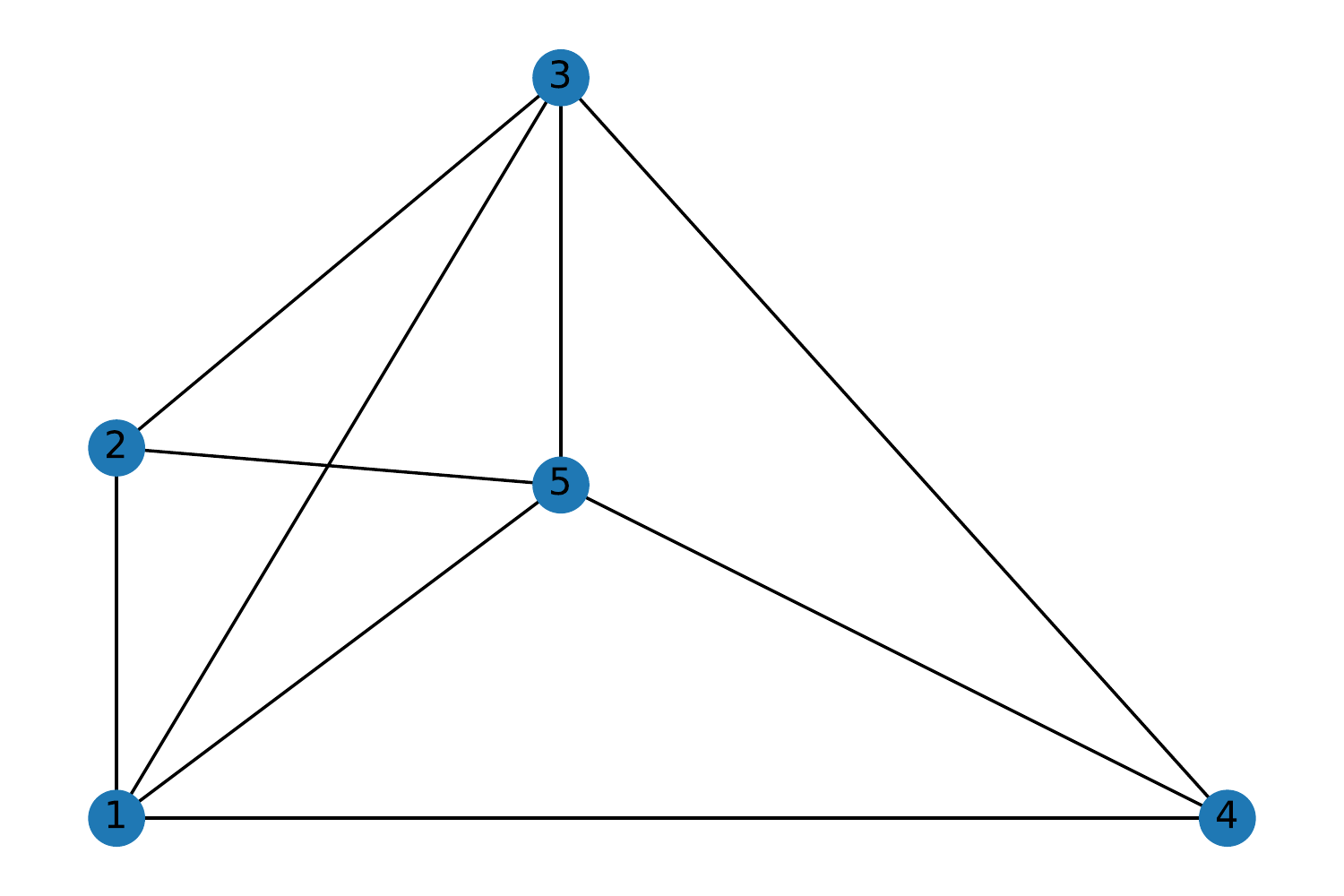}
          \caption{}
     \end{subfigure}
     \begin{subfigure}{0.2\textwidth}
         \centering
         \includegraphics[width=\textwidth]{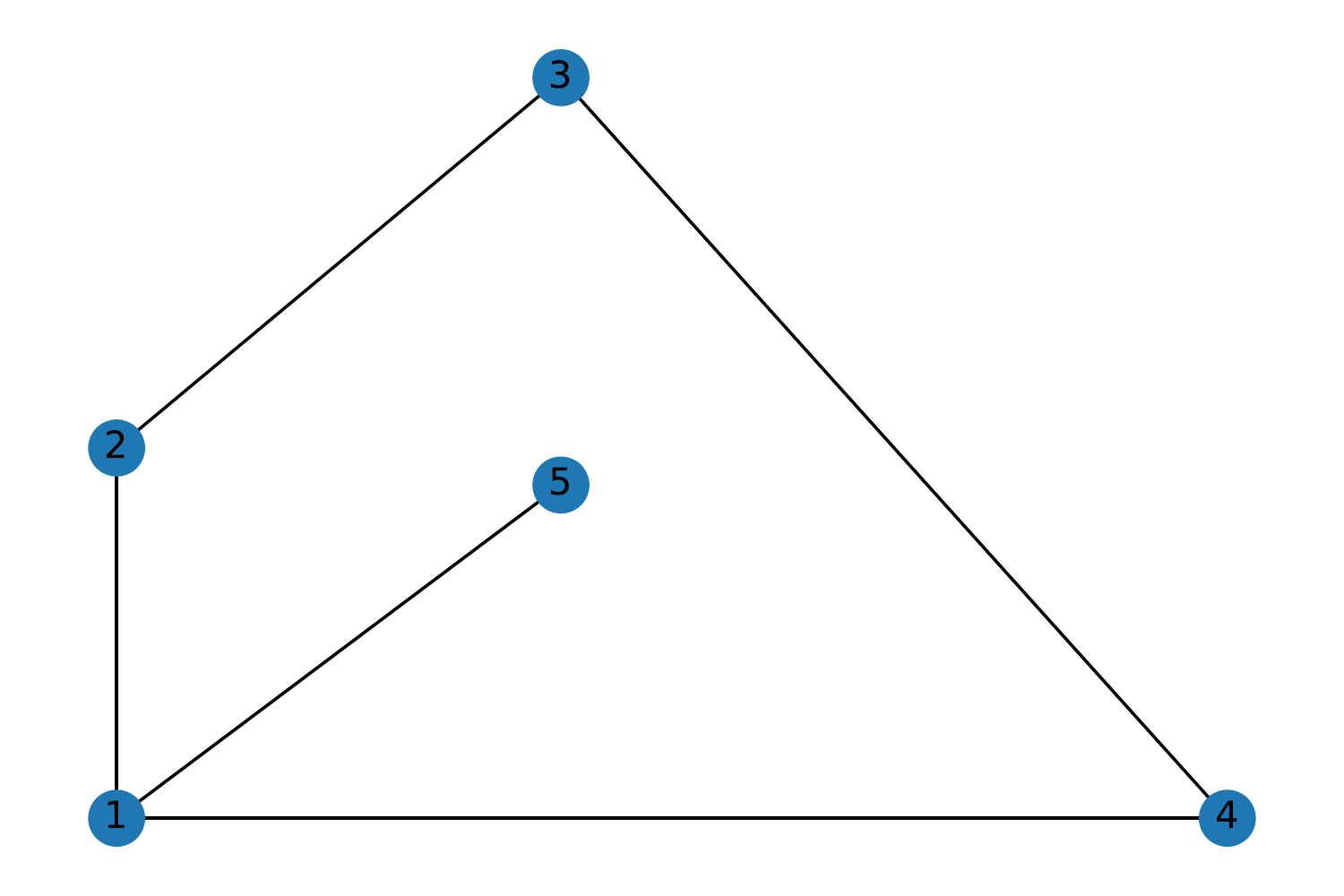}
          \caption{}
     \end{subfigure}
     \begin{subfigure}{0.2\textwidth}
         \centering
         \includegraphics[width=\textwidth]{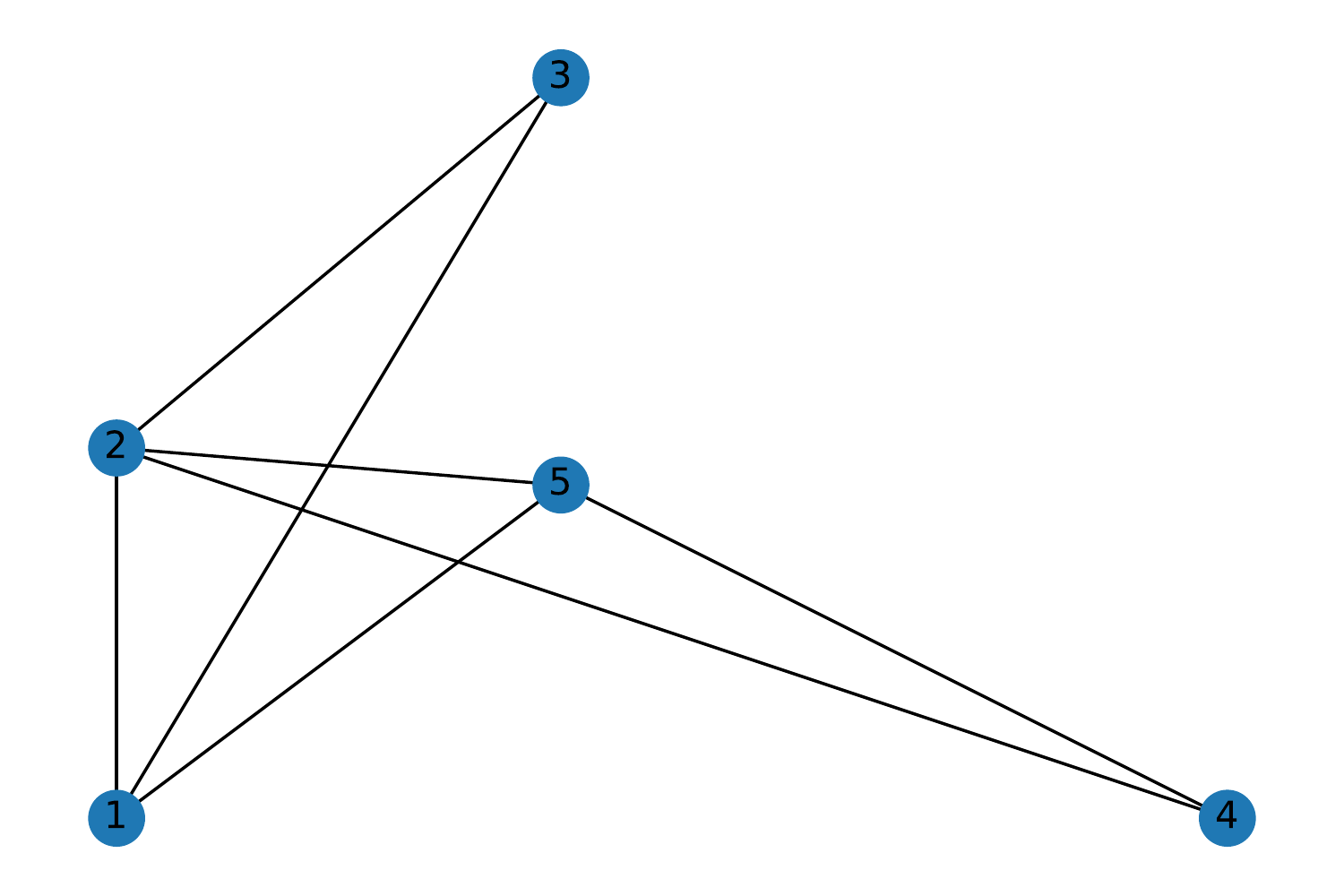}
          \caption{}
     \end{subfigure}
        \caption{Different dependence structures: (a) the true  graph $\bbE^*$; (b) $\bbE_0$ that satisfies $\bbE_0 \supsetneq \bbE^*$; (c)  $\bbE_0$ that satisfies  $\bbE_0 \subsetneq \bbE^*$;  (d) $\bbE_0$ that is not nested within or outside $\bbE^*$.   }
         \label{networkprespecified}    
\end{figure}

Despite the above advantages, $\hat{D}_n$ does not have much power in rejecting $\bbE_0$ if $\bbE^* \subsetneq \bbE_0 $; that is, when $\bbE^*$ is a subset of $\bbE_0$ but they are not equal. 
 An example is given in Figure \ref{networkprespecified} where $\hat{D}_n$ will have no power in rejecting the $\bbE_0$ in (b). This is simply because under the null hypothesis that $\bbE=\bbE_0$, any reasonable estimator of $\bOmg_0$ of $\bOmg^*$ denoted as $\hat\bOmg_0$, including the one discussed above, will asymptotically converge to $\bOmg^*$, making $\bsig^* \hat\bOmg_0- \bI_p$ very small stochastically.

%

\subsection{\color{black} A novel consistency-empowered test statistic}

When $\bbE_0 \supsetneq \bbE^*$ as illustrated in  Figure \ref{networkprespecified}, we know that its compatible estimator $\hat\bOmg_0$ will be close to $\bOmg^*$ loosely speaking. Thus, if $(i, j)\in \bbE_0$ but $(i, j)\not\in \bbE^*$, $\hat\omega_{ij,0}$ will be close to zero. For the test to have power, we need to offset the effect of those small estimates. Our idea is to augment those small estimates with a constant that is just large enough for us to reject the null. Thus, in certain sense, we are amplifying those small noises as a means to empower the consistency of a new test statistic. 
Of course, how close is close to zero for a small noise should be gauged against its standard error, which motivates the development of the following {\color{black} consistency-empowered} test statistic. 

Let $\hat{\omega}_{i1,0}^{(j)}$ be the $j$th component of $\hat{\bw}_{i1,0}$ with the associated standard error $\sigma_{i1,0}^{(j)}$, where
$\hat{\bw}_{i1,0}$ is defined in the previous section.  
Let $\hat{\sigma}_{i1,0}^{(j)}$ be a consistent estimator of $\sigma_{i1,0}^{(j)}$ which will be defined shortly.  Define 
$\tilde{\bw}_{i1,0}=(\tilde{\omega}_{i1,0}^{(1)},\ldots, \tilde{\omega}_{i1,0}^{(s_i)})^\T$ where
$$
\tilde{\bw}_{i1,0}^{(j)}=\hat{\bw}_{i1,0}^{(j)}+\Delta_{i1}^{(j)},
$$
with 
\[\Delta_{i1}^{(j)}=C_nI\{|\hat{\omega}_{i1,0}^{(j)}|/\hat{\sigma}_{i1,0}^{(j)}\leq\delta_n\}.\]  
Here $C_n\neq 0$ and $\delta_n$ are tuning parameters which will be discussed in the next section. Clearly, what this procedure does is to add a constant to those elements of $\hat{\bw}_{i1,0}$ that are stochastically small. Or put differently, it simply amplifies the noise, as opposed to the usual notion of filtering out noises for better estimation accuracy. 

Recall the definition of $B_{i, 0}$ in the last section. 
Let $\tilde{\bw}_{i,0}=B_{i,0}\tilde{\bw}_{i1,0}$ and $\Delta_{i}=B_{i,0}\Delta_{i1}$ where $\Delta_{i1}=(\Delta_{i1}^{(1)},\ldots, \Delta_{i1}^{(s_i)})^\T$.
Our proposed consistency-empowered test statistic is then
$$\tilde{D}_n = \max_{1\leq i, j\leq p}{(\bE_j^\T \bSn \tilde{\bw}_{i,0} - \bE_j^\T\bE_i)^2}/{\hat{\theta}_{ij,0}}=\max_{1\leq i, j\leq p} {\tilde{D}_{ij}}^2,$$ 
where $\tilde{D}_{ij}^2 =(\bE_j^\T \bSn \tilde{\bw}_{i,0}  - \bE_j^\T \bE_i)^2/\hat{\theta}_{ij,0}$. {\color{black} Interestingly, the form of the consistency-empowered estimator $\tilde{\bw}_{i1,0}^{(j)}$ appears to be similar to the 
power-enhanced test statistic proposed by Fan et al. (2015) on surface. In essence however, they are fundamentally different because
our modified estimator is to ensure the consistency of the proposed test for all types of null hypotheses,  
while Fan et al. (2015) focuses on improving power for some type of alternatives. Having said this, we point out that the test statistic $\hat{D}_n$ without consistency empowerment is powerless for testing those null hypotheses in which the true graph is nested within the graph under the null, regardless of the sample size.} 

We now discuss a consistent estimator of $\sigma_{i1,0}^{(j)}$ required 
in $\Delta_{i1}^{(j)}$. 
Any non-zero element $\hat{\omega}_{i1,0}^{(j)}$ of $\hat{\bw}_{i1,0}$ corresponds to  $\hat{\omega}_{ik,0}$ ($k = 1,\ldots, p$), $(i,k)$-th component of $\hat{\bOmg}_0$, such that $\hat{\omega}_{i1,0}^{(j)}$ = $\hat{\omega}_{ik,0}$.  Le $\&$ Zhong (2021) established that $\hat{\omega}_{i1,0}^{(j)}$ is asymptotically normal, in the sense that
\begin{equation}
\label{asynormOmg}
\surd{n} (\hat{\omega}_{i1,0}^{(j)} -\omega_{i1,0}^{(j)}) = \surd{n} (\hat{\omega}_{ik,0} -\omega_{ik,0}) \to N(0,h_{ik})
\end{equation}
in distribution, 
where  
$h_{ik}= \omega_{ii}^* \omega_{kk}^* +\omega_{ik}^*$.
Thus,  a consistent estimator of $\sigma_{i1,0}^{(j)}$ is simply $\hat{\sigma}_{i1,0}^{(j)} = \surd{(\hat{\omega}_{ii,0} \hat{\omega}_{kk,0} + 
\hat{\omega}_{ik,0}^2)}/\surd{n}$. 


\section{Asymptotic Distributions}

In this section, we study the asymptotic distributions of $\hat{D}_n$ and the consistency-empowered test statistic $\tilde{D}_n$, when $p\to\infty$ as $n\to\infty$. We assume the following regularity conditions. 

\begin{itemize}
\item[] (C1) There exist constants $C_1, C_2 > 0$ such that $\vert \vert \bsig^* \vert \vert_1 \leq C_1$ and $\vert \vert \bOmg^* \vert \vert_1 \leq C_2$,
where $\vert \vert \bM \vert \vert_1= \max_{1\leq j\leq n}\sum_{i=1}^m \l M_{ij}\l$ for any $m \times n$ matrix $\bM=(M_{ij})$; 
\item[] (C2) $s_0\surd{(\log p/n)} = o(1)$, where $s_0=\max_{1\le j\le p} s_j$.
\end{itemize}
These two conditions are commonly employed in the literature (Zhou et al. (2011), Liu $\&$ Lou (2015), cf). 
Many commonly assumed precision matrix structures such as polynomial decay, 
exponential decay, banded and factor models satisfy Condition (C1). See the
Appendix for details.

A node is called \textit{isolated} if it does not connect with any other nodes. That is,  node $i$ is isolated if and only if $\omega_{ij}^*= 0$ for  all $ j \neq i$. We have the following  results on the asymptotic distribution of $\hat{D}_n$.

\begin{theorem}
\label{generalTheo}
Under conditions (C1)-(C2),  if the graph $\bbE^*$ has $k$ isolated nodes where 
$\lim_{p \rightarrow \infty} {k}/{p} 
= \beta, 0 \leq \beta<1$,  then under the null $H_0$, 
$$ \mbox{pr}\{ \hat{D}_n - 4\log p + \log(\log p) \leq t \} \to  \exp\{-\exp(-{t}/{2})/\surd{(2\gamma\pi)}\}$$ 
where $\gamma = (1 - {\beta^2}/{2})^{-2}$.
\end{theorem}

Interestingly, the asymptotic distribution of $\hat{D}_n$ depends on $k$, the number of isolated nodes in $\bbE^*$. 
From Theorem \ref{generalTheo}, if the number of isolated nodes $k$ is of a smaller order of the number of variables $p$ as $k = o(p)$, then
 $\hat{D}_n$ converges to the following Gumbel distribution
\begin{equation}
\label{TestProblem}
 \mbox{pr}\{ \hat{D}_n - 4\log p + \log(\log p) \leq t \} \to \exp\{-\exp(-{t}/{2})/\surd{(2\pi)}\}.
\end{equation}

Example 1 and Example 2 below further illustrate Theorem \ref{generalTheo}. 

 \begin{example}
Assume that $\bbE^*$ has a Toeplitz structure 
$\bbE = \{(i,j), \l i - j \l \leq s_0 \}$.
The number of the isolated node in $\bbE^*$ is 0 and hence $\gamma=1$. Then under the null $H_0:\bbE_0=\bbE$, we have 
$\mbox{pr}\{ \hat{D}_n - 4\log p + \log(\log p) \leq t\} \rightarrow \exp\{-\exp({-t}/{2})/{\surd{(2\pi)}}\}.$
The limiting distribution of $\hat{D}_n$ is $\mbox{Gumbel}(-\log 2\pi, 2)$.
%
\end{example}

\begin{example}
Assume that $\bbE^*$ follows a factor model structure with
$\bOmg^*= \bI_p + \bu_1 \bu_1^T$, where $\bu_1 = (1,1,1,0,\ldots,0) \in \mathbb{R}^{p}$. 
 The limiting distribution of $\hat{D}_n$ satisfies 
$\mbox{pr}\{\hat{D}_n - 4\log p + \log(\log p) \leq t \}\to \exp\{-\exp({-t}/{2})/{\surd{(8\pi)}}\}.$ In this case, the limiting distribution of $\hat{D}_n$ is $\mbox{Gumbel}(- \log  8\pi, 2)$.
\end{example}

As an application of Theorem \ref{generalTheo}, if one is interested in assessing the local graph structure associated with the $i$-th node by testing
\begin{equation}
\label{coltest}
H_0: \bbE_i=\bbE_{0i} \quad\mbox{vs.}\quad H_1: \bbE_{i}\neq \bbE_{0i}
\end{equation}
where $\bbE_{0i}:=\{(i, j): (i, j) \in \bbE^*, \forall j \}$ is the graph structure specified for the $i$-th node,  we can  modify our test statistic as
$$\hat{D}_{ni} = \max_{1\leq j\leq p}{( \bE_j^\T \bold{S}_n \hat{\bw}_{i,0}-\bE_j^\T\bE_i)^2}/{\hat{\theta}_{ij,0}}.$$
The asymptotic distribution of the test statistic $\hat{D}_{ni}$ is summarized in the following corollary.
\begin{corollary}\ 
For the testing problem in (\ref{coltest}), if conditions (C1)-(C2) are satisfied, under the null hypothesis $\hat{D}_{ni}$ converges to the following Gumbel distribution
$$ \mbox{pr}\{\hat{D}_{ni} - 2\log p + \log(\log p) \leq t\}\to \exp\{-\exp(-{t}/{2})/\surd{\pi}\}. $$
\end{corollary}

We are ready to show the asymptotic distribution of the consistency-empowered test statistic $\tilde{D}_n$ under $H_0$ and discuss the choices of tuning parameters $C_n $ and $\delta_n$ involved in its definition. Recall that 
the main reason for introducing $C_n$ and $\delta_n$ is to detect network structures whose edge sets are strictly supersets of that of the true graph.
Due to this, we need to choose $C_n$ and $\delta_n$ such that $\tilde{D}_n$ and $\hat{D}_n$ have the same asymptotic distribution under $H_0$, while the
test based on $\tilde{D}_n$ can reject network structures $\bbE_0\supset \bbE^*$ with probability one.

Under $H_0$ when $\bbE=\bbE_0$, $\omega_{i1,0}^{(j)}$ ($j=1,\ldots, s_i$) are all non-zeros. However, if 
$\bbE_0\supset \bbE^*$ but $\bbE^*\neq\bbE_0$,  $\omega_{i1,0}^{(j)}$ ($j=1,\ldots, s_i$) are supposed to be non-zeros because of the specification of $\bOmg_0$ but some of them will be estimated as zeros (in asymptotic sense).  Based on the asymptotic normality in (\ref{asynormOmg}), we have
$\hat{\omega}_{i1,0}^{(j)} = \omega_{i1,0}^{(j)} + O_p(1/\surd{n})$. This results hold when the true value of $\omega_{i1,0}^{(j)}$ is zero or non-zero.
If $\omega_{i1,0}^{(j)}\neq 0$, then $\hat{\omega}_{i1,0}^{(j)}/{\sigma}_{i1,0}^{(j)}= O_p(\surd{n})$ where ${\sigma}_{i1,0}^{(j)}$ is 
defined in equation (\ref{asynormOmg}). If $\omega_{i1,0}^{(j)}=0$, then $\hat{\omega}_{i1,0}^{(j)}/{\sigma}_{i1,0}^{(j)}= O_p(1)$.
Based on these observations, we may choose $C_n=C\surd{\log(p)}$ for some $C>0$ and $\delta_n = \surd{\log(n)}$ so that $\tilde{D}_n$ and $\hat{D}_n$ have the same asymptotic distribution under $H_0$,
and $\tilde{D}_n$ rejects any network structure satisfying $\bbE_0\supsetneq \bbE^*$ with probability one. Theorem \ref{generalTheoThreshold} below formally provides the asymptomatic distribution of $\tilde{D}_n$.

\begin{theorem}
\label{generalTheoThreshold}
Under conditions (C1)-(C2), if the true structure has $k$ isolated node where 
$\lim_{p\to\infty} {k}/{p}=\beta$ for some  $0 \leq \beta <1$, $\delta_n = \surd{\log(n)}$ and $C_n=C\surd{\log(p)}$ for some constant $C > 0$,  
then under the null $H_0$, 
$$ \mbox{pr}\{\tilde{D}_n - 4\log p + \log(\log p) \leq t \} \to  \exp\{-\exp(-{t}/{2})/\surd{(2\gamma\pi)}\},$$ 
where $\gamma = (1 - {\beta^2}/{2})^{-2}.$ Moreover, 
if $\bbE_0\supsetneq \bbE^*$, by choosing  $C>\max_{i,j}4(\omega_{ii}^*\sigma_{ii}^* + 1)/(\sigma_{ii}^*\sigma_{jj}^*+2\sigma_{ij}^{*2})$,  
 the consistency-empowered test statistic $\tilde{D}_n$ rejects the null hypothesis $H_0$ with probability tending to one.
\end{theorem} 

\section{Simulation}
\subsection{Numerical performance of the test statistic $\hat{D}_n$}
We perform numerical study to evaluate the finite sample performance of the proposed test statistic $\hat{D}_n$ in terms of its size and power properties. We generate $n$ i.i.d. multivariate normally distributed $p$-dimensional random vectors with mean vector 0 and covariance matrix $\bsig^*$ with its corresponding graph $\bbE^*$ admitting a banded structure such that 
$\bbE^*=\{(i,j): |i-j|<s_0\}$.
Let $\bOmg^*={\bsig^*}^{-1}=(\omega_{ij}^*)_{p \times p}$ be the precision matrix. Because different precision matrices can correspond to the same underlying graph, 
we specify two precision matrices to examine the performance of the proposed test statistic. For the first precision matrix, we set it as 
banded with its non-zero components decaying at an exponential rate away from its diagonals. More specifically, we set $\omega_{ij}^*= 0\!\cdot\! 6^{- \l i -j \l}$ for $ \l i -j \l < s_0$ 
 and $\omega_{ij}^*= 0$ otherwise. For the other precision matrix, we again set it as banded with its non-zero components decaying at the polynomial rate, that is,  $\omega_{ij}^*=(1 + \l i-j \l)^{-2}$ for $ \l i -j \l < s_0$ and $\omega_{ij}^*= 0$ otherwise. We consider two
different sparsity levels as $s_0 = 4$ or $6$.

To evaluate the empirical size and power of the proposed test, we consider various specification of $\bbE_0$, the structure specified in the null hypothesis. 
For evaluating the empirical size, we consider $\bbE_0=\bbE^{*}$.
To evaluate the the power of the proposed test, we consider the following four different specifications of $\bbE_0$. 
\begin{itemize}
\item[1)] (Isolated structure) Set $\bbE_{0}=\bbE_{0,1}=\{(i,j): i=j\}$. All the nodes are isolated. 
\item[2)] (Nested structure) Set $\bbE_{0}=\bbE_{0,2}=\{(i,j):|i-j|<3\}$. This structure is nested in the true network structure $\bbE^*$.
\item[3)] (1-diff: structure with edges to node 1 different) Set $\bbE_{0}=\bbE^{*}_1 \cup\bbE_{0,3}$, 
where $\bbE^{*}_1 = \bbE^{*}$ on the set of edges $\{ (i,j), i, j \neq 1$\}, and $\bbE_{0,3}=\{(1,3),(1,7),(1,8),(1,9)\}$ are edges connected with node 1.
\item[4)]  (2-diff: structure with edges to 2 nodes different) Set $\bbE_{0}=\bbE^{*}_2\cup\bbE_{0,4}$, where 
$\bbE^{*}_2 = \bbE^{*}$ on the set of edges $\{ (i,j), i, j \neq 1, 2$\}, and $\bbE_{0,4}=\{(1,3),(1,7),(1,8),(1,9),(2,4),(2,9), (2,12)\}$ are edges connected to nodes 1 and 2.

\end{itemize} 


To understand the effect of sample size and data dimension, we choose two different sample sizes $n =300$ and $n =1000$. 
For each sample size, data dimension is changed by setting $p/n$ at three different values $0\!\cdot\!5, 1$, and 2.
Because the true precision matrix $\bOmg^*$ specified above satisfies Conditions (C1)-(C2), we applied the results in Theorem \ref{generalTheo}.
More specifically, we rejected the hypothesis if the test statistic values $\hat{D}_n$ are greater than the $4 \log p - \log ( \log p ) + \mbox{Gumbel}_{.95}(- \log 2 \pi,2) $, 
where $\mbox{Gumbel}_{.95}(- \log 2 \pi,2)$ is the 95 $\%$ quantile value of the Gumbel distribution with location 
parameter $- \log 2 \pi$ and scale parameter 2. Simulation results are reported based on 500 simulation replications. 


Table \ref{Power of the test Toeplitz} reports the empirical sizes and power of the proposed test statistic $\hat{D}_n$ for testing different structures $\bbE_{0}$ specified in the above 1)-4). 
It can be seen that our proposed test controls type I error rate well at the nominal level $0\!\cdot\!05$ under various settings. The proposed test statistic is consistent as the power of the tests are one in many scenarios. 
Based on the pattern of empirical power, we see that power of the proposed test  $\hat{D}_n$ increases as $n$ increases or $p$ decreases. 
Table \ref{Power of the test Toeplitz} also shows that sparsity level has some impact on the power of the test statistic, the increasing of $s_0$ leads to a decreasing power.

\begin{table}[tbhp!]
\begin{center}
\captionof{table}{Type I error and power of the proposed test statistic $\hat{D}_n$ under different alternatives when the precision matrix 
has banded structure and decays at an exponential rate. \label{Power of the test Toeplitz}}
\begin{tabular}{ccrccccc}
     &    &  &  \multicolumn{1}{c}{Empirical} & \multicolumn{4}{c}{Power of the Test $\hat{D}_n$}\\
$s_0$ & $n$ & ${p}/{n}$  &  Size & Isolated & Nested  & 1-diff & 2-diff \\ 
4 & 300 &    0$\cdot $5   &  0$\cdot$034   &    1$\cdot $000   &  1$\cdot $000  &  1$\cdot $000   & 1$\cdot $000 \\
 &        &       1   &  0$\cdot $042   &    1$\cdot $000   &  1$\cdot $000  &  1$\cdot$000   & 1$\cdot $000 \\
 &        &       2   &  0$\cdot $030   &    1$\cdot $000   &  1$\cdot $000  &  1$\cdot $000   & 1$\cdot $000 \\
 & 1000 &  0$\cdot $5 &  0$\cdot $038     &  1$\cdot $000   &  1$\cdot $000  &  1$\cdot $000   & 1$\cdot $000 \\
 &          &    1  &  0$\cdot$032     &  1$\cdot $000   &  1$\cdot $000  &  1$\cdot $000   & 1$\cdot $000 \\
 &          &    2  &  0$\cdot$044     &  1$\cdot $000   &  1$\cdot $000  &  1$\cdot $000   & 1$\cdot $000 \\ 
6  & 300 &  0$\cdot $5 &   0$\cdot$026      &   1$\cdot $000  & 0$\cdot $824    &  1$\cdot $000   & 1$\cdot $000 \\
 &        &     1 &   0$\cdot$032      &   1$\cdot $000  & 0$\cdot $766    &  1$\cdot $000   & 1$\cdot $000 \\
 &        &     2 &   0$\cdot$028      &   1$\cdot $000  & 0$\cdot $648    &  1$\cdot $000   & 1$\cdot $000 \\
  & 1000 &    0$\cdot $5 &  0$\cdot$032 &   1$\cdot $000   &  1$\cdot $000  &  1$\cdot $000   & 1$\cdot $000 \\
  &          &       1 &  0$\cdot$046 &   1$\cdot $000   &  1$\cdot $000  &  1$\cdot $000   & 1$\cdot $000 \\
  &          &       2 &  0$\cdot$036 &   1$\cdot $000   &  1$\cdot $000  &  1$\cdot $000   & 1$\cdot $000 \\
\end{tabular}
\end{center}
\end{table}


\begin{table}[tbhp!]
\begin{center}
\captionof{table}{Type I error rate and power of the proposed test statistic $\hat{D}_n$ under different alternatives where the precision matrix has banded structure and decays at a polynomial rate. \label{Power of the test Polynomial decay}}
\begin{tabular}{ccrccccc}
     &    &  &  \multicolumn{1}{c}{Empirical} & \multicolumn{4}{c}{Power of the Test $\hat{D}_n$}\\
$s_0$ & $n$ & ${p}/{n}$  &  Size & Isolated & Nested  & 1-diff & 2-diff \\ 
4 & 300 &    0$\cdot $5 & 0$\cdot$022    &   1$\cdot $000 & 0.034 & 0$\cdot $348 & 0$\cdot $468   \\
 &        &      1 &  0$\cdot$034    &   1$\cdot $000 & 0.038 & 0$\cdot $226 & 0$\cdot $336 \\
 &        &       2 &  0$\cdot$024   &   1$\cdot $000 & 0$\cdot$024 & 0$\cdot $170 & 0$\cdot $248\\
 & 1000 &    0$\cdot $5 &   0$\cdot$034  &   1$\cdot $000 & 0$\cdot $274 & 0$\cdot $998  & 1$\cdot $000 \\
 &          &       1 &   0$\cdot$046  &   1$\cdot $000 & 0$\cdot $210 & 0$\cdot $994    & 1$\cdot $000 \\
 &         &        2 &  0$\cdot$038   &   1$\cdot $000 & 0$\cdot $204 & 0$\cdot $988  & 1$\cdot $000 \\
6  & 300 &  0$\cdot $5 &    0$\cdot$032   &   1$\cdot $000 &0$\cdot$034& 0$\cdot $320 &0$\cdot $474\\ 
 &        &     1 &      0$\cdot$030 &   1$\cdot $000 & 0$\cdot$028& 0$\cdot $248& 0$\cdot $328\\ 
 &        &     2 &      0$\cdot$024 &   1$\cdot $000 & 0$\cdot$024 &0$\cdot $164& 0$\cdot $224\\ 
 & 1000 &    0$\cdot $5 &  0$\cdot$046  &  1$\cdot $000 & 0$\cdot $130 & 0$\cdot $992 &   1$\cdot $000 \\
 &        &         1 &   0$\cdot$032 &  1$\cdot $000 & 0$\cdot $106 & 0$\cdot $994 &   1$\cdot $000 \\
 &        &          2 & 0$\cdot$024  &  1$\cdot $000 & 0$\cdot$084 & 0$\cdot $994 &   1$\cdot $000 \\
\end{tabular}
\end{center}
\end{table}

Table \ref{Power of the test Polynomial decay} summarizes the empirical size and power of the proposed test $\hat{D}_n$ under the polynomial rate decay structure. 
We see that its patterns are similar to that in Table \ref{Power of the test Toeplitz} where the empirical power increases as sample size increases, and decreases as $p$ or $s_0$ increases. 
We observe that in this case the power is not as high as those in Table \ref{Power of the test Toeplitz} where the precision matrix decays at an exponential rate in Table \ref{Power of the test Toeplitz}. 
This is also something expected because the signals of the precision matrix in Table \ref{Power of the test Polynomial decay} is weaker than the signals in the previous example. 
For example, when $s_0 = 4$, the non-zeros in the first column of polynomial decayed precision matrix $\bOmg^*$ is $(1, 0\!\cdot\!6, 0\!\cdot\! 36, 0\!\cdot\!216)^\T$ 
while the first column non-zeros are $(1, 0\!\cdot\!25, 0\!\cdot\!11, 0\!\cdot\!06)^\T$ in the exponentially decayed $\bOmg^*$. 

\subsection{Numerical comparison of $\hat{D}_n$ and $\tilde{D}_n$}
In this simulation studies, we evaluate the finite sample performance of $\hat{D}_n$ and $\tilde{D}_n$ in terms of empirical sizes and powers in detecting the nested network and the included network structures. 
Similar to Simulation Settings I, we generate $n$ IID multivariate normally distributed $p$-dimensional random vectors with mean vector 0 and covariance matrix $\bsig^*$. 
The corresponding precision matrix is $\bOmg^*={\bsig^*}^{-1}  = (\omega_{ij}^*)_{p \times p}$ where $\omega_{ij}^*= 0\!\cdot\! 6^{- \l i -j \l}$ for $ \l i -j \l < s_0$  and $\omega_{ij}^*= 0$ otherwise.
For evaluating the empirical sizes, we consider $\bbE_0=\bbE^*$. 
We consider the following two specified structures hypotheses in the simulation
\begin{itemize}
\item[5)] (Nested structure) Set $\bbE_0=\bbE_{0,5}=\{(i,j): |i-j|<s_0-1 \}$. The structure $\bbE_0$ is nested in the true structure $\bbE^*$.
\item[6)] (Included structure) Set $\bbE_0=\bbE_{0,6}=\{(i,j): |i-j|<s_0+1 \}$. The structure $\bbE_0$ includes in the true structure $\bbE^*$.
\end{itemize} 


Table \ref{Thresholdtestpower} summarizes the empirical sizes and powers of the tests based on $\hat{D}_n$ and $\tilde{D}_n$. Table \ref{Thresholdtestpower} demonstrates both tests have the similar power in rejecting the nested 
structure and control the type 1 error rate. The modified test statistics version $\tilde{D}_n$ has the ability to reject the pre-specified network structures that include the true network structure, while the test statistic $\hat{D}_n$ 
loses power for included networks. We chose $\delta_n=\surd{\log(n)}$ and $C_n=0\!\cdot\!05$ for the test statistics $\tilde{D}_n$ in our simulation studies. 


\begin{table}[tbhp!]
\begin{center}
\captionof{table}{Type 1 error and empirical power of the test statistics $\hat{D}_n$ and $\tilde{D}_n$ for both nested
and included structures \label{Thresholdtestpower}}
\begin{tabular}{ccccccccc}
      &  & &  \multicolumn{3}{c}{$\hat{D}_n$} & \multicolumn{3}{c}{$\tilde{D}_n$}\\
       &    &  &  \multicolumn{1}{c}{Empirical} & \multicolumn{2}{c}{Power of $\hat{D}_n$} &  \multicolumn{1}{c}{Empirical} & \multicolumn{2}{c}{Power of $\tilde{D}_n$}\\
 $s_0$ &$n$ & $p/n$  &  Size & Nested & Included  & Size & Nested & Included \\ 
4&500 & 0$\cdot $50 & 0$\cdot$034 & 1$\cdot $000 & 0$\cdot$032 & 0$\cdot$034 & 1$\cdot $000 & 0$\cdot $986 \\ 
  &       & 1$\cdot $00 & 0$\cdot$038 & 1$\cdot $000 & 0$\cdot$036 & 0$\cdot$038 & 1$\cdot $000 & 0$\cdot $996 \\ 
  &       & 2.00 & 0$\cdot$038 & 1$\cdot $000 & 0.038 & 0.038 & 1$\cdot $000 & 1$\cdot $000 \\ 
  &1000 & 0$\cdot $25 & 0$\cdot$030 & 1$\cdot $000 & 0$\cdot$036 & 0$\cdot$030 & 1$\cdot $000 & 1$\cdot $000 \\ 
  &         & 0$\cdot $50 & 0$\cdot$040 & 1$\cdot $000 & 0$\cdot$040 & 0$\cdot$040 & 1$\cdot $000 & 1$\cdot $000 \\ 
  &         & 1$\cdot $00 & 0$\cdot$032 & 1$\cdot $000 & 0$\cdot$034 & 0$\cdot$032 & 1$\cdot $000 & 1$\cdot $000 \\
  6&500 & 0$\cdot $50 & 0$\cdot$026 & 0$\cdot $116 & 0$\cdot$030 & 0$\cdot$030 & 0$\cdot $146 & 0$\cdot $696 \\ 
    &       & 1$\cdot $00 & 0$\cdot$042 & 0$\cdot$092 & 0$\cdot$044 & 0$\cdot$048 & 0$\cdot $116 & 0$\cdot $660 \\ 
    &       & 2$\cdot$00 & 0$\cdot$036 & 0$\cdot$066 & 0$\cdot$040 & 0$\cdot$032 & 0$\cdot$090 & 0$\cdot $578 \\ 
   &1000 & 0$\cdot $25 & 0$\cdot$036 & 0$\cdot $810 & 0$\cdot$034 & 0$\cdot$042 & 0$\cdot $818 & 1$\cdot $000 \\ 
   &         & 0$\cdot $50 & 0.036 & 0$\cdot $740 & 0$\cdot$034 & 0$\cdot$034 & 0$\cdot $688 & 0$\cdot $998 \\ 
   &         & 1$\cdot $00 & 0$\cdot$040 & 0$\cdot $614 & 0$\cdot$044 & 0$\cdot$044 & 0$\cdot $600 & 1$\cdot $000 \\ 
\end{tabular}
\end{center}
\end{table}

\section{Real Data Analysis}
We illustrate the use of the proposed test statistics for identifying the structure of a graphical model by applying them to a correlated data analysis. 
Towards this, we examined a COVID-19 dataset provided by The New York Times  (NYT (2022)) that is publicly available on https://github.com/nytimes/covid-19-data. The data set includes daily confirmed COVID-19 cases observed over 51 states of the U.S. from January 1, 2021 to December 31, 2021. We aggregated the data on a weekly basis such that  the data contain 
52 weekly confirmed cases in thousands from 51 states, which is denoted as 
a matrix of size 51$\times$ 52. 
Our interest was to understand how the numbers of COVID cases depend on geographical locations. Towards this, we coded three dummy variables according to whether a state is in the North East, West, Mid West, or the South. The following linear regression was postulated
\begin{equation}
\label{GEE}
   E(y_{ij}|x_i)  = \beta_0 + \beta_1 x_{i1} + \beta_2 x_{i2} + \beta_3 x_{i3}, (i = 1,\ldots,51; j = 1,\ldots,52),
\end{equation}
where $x_{i1}$ is an indicator variable whether the state is in the North East, $x_{i2}$ is for the Mid West, and $x_{i3}$ is for the West. Denote 
$\boldsymbol{Y_i} = (y_{i1},\ldots,y_{i52})^\T$, $\boldsymbol{\beta} = (\beta_0, \beta_1, \beta_2, \beta_3)^\T$, $\boldsymbol{X_i} =\boldsymbol{1} \otimes (1,x_{i1},x_{i2},x_{i3})$, where $\boldsymbol{1} = (1,\ldots,1)^\T$ is a $52 \times 1$ matrix, $\otimes$ is the Kronecker product, and $\boldsymbol{X_i}$ is a $52 \times 4$ matrix. Since the components of $\boldsymbol{Y_i}$ are correlated, we applied the method of generalized estimation equations (Liang $\&$ Zeger, 1986) for estimating $\boldsymbol{\beta}$ by incorporating the correlation structure of $\boldsymbol{Y_i}$. That is, we estimate $\boldsymbol{\beta}$ by solving
\begin{equation}\label{GEE1}
    \sum_{i=1}^{51}\boldsymbol{X_i}^\T
    \boldsymbol{V}^{-1}(\boldsymbol{Y_i} - \boldsymbol{X_i} \boldsymbol{\beta} ) = 0,
\end{equation}
where $\boldsymbol{V}$ is the so-called working covariance matrix. It is known that correct specification of $\boldsymbol{V}$ improves the estimation efficiency of the resulting estimator.

To choose an appropriate graph corresponding to $\bOmg=\boldsymbol{V}^{-1}$, we test if the one of the following specified graphical structures fits the data well. 
\begin{itemize}
\item[a)] (Isolated structure) Set $\bbE_0=\bbE_{0,1}=\{(i,j), i = j\}$. 
\item[b)] (Banded structure with bandwidth 3, denoted as Band(3)) Set $\bbE_0=\bbE_{0,2}=\{(i,j), |i-j| < 3\}$.

\item[c)]  (Structure learned from TIGER method, denoted as TIGER) Set $\bbE_0=\bbE_{0,3}$, where $\bbE_{0,3}$ is the network learned from TIGER method as in the default setting by the flare R package by Li et al. (2020). 
\item[d)]  (Structure learned from GLASSO method, denoted as GLASSO) Set $\bbE_0=\bbE_{0,4}$, where $\bbE_{0,4}$ is the network learned from the GLASSO R package with tuning value $\rho = 10$ by Friedman et al. (2019). 

\end{itemize}




We applied our proposed methods to test the above hypothetical structures. 
Since the number of 
isolated nodes of the true structure is unknown, we chose $\gamma=1$ for the limiting distribution in Theorem \ref{generalTheo} so that our test 
is conservative because we 
only reject the null hypothesis if our test statistic value is large enough. 
The test statistic values and its corresponding p-values (in parentheses) for testing Isolated, Band(3), TIGER and GLASSO are, respectively, $ 49\!\cdot\!57  (<0\!\cdot\!0001)$, $17\!\cdot\!61 (0\!\cdot\!08)$, $32\!\cdot\!58(<0\!\cdot\!0001)$ and $16\!\cdot\!34  (0\!\cdot\!14)$. 
Therefore we reject the null hypothesis that the true  structure is the Isolated or the TIGER structure with 95$\%$ confidence. However, we cannot reject the null hypothesis that the true structure is the Band(3) network or the GLASSO network at the 95$\%$ confidence level.



\begin{table}[tbhp!]
\begin{center}
\captionof{table}{Estimated coefficients parameters under four different pre-specified structures, standard errors of the estimated parameters in the parentheses, and * denotes p-value less than 0$\cdot$05. \label{coefficients}}
\begin{tabular}{crlll}
    Coefficients &  \multicolumn{1}{c}{Isolated} & \multicolumn{1}{c}{Band(3)} &\multicolumn{1}{c}{TIGER} & \multicolumn{1}{c}{GLASSO}\\
${\beta}_1$ & -0$\cdot $78 (1$\cdot$59) & 0$\cdot$64 (0$\cdot$73) & 1$\cdot $58 (1$\cdot$07) & 0$\cdot$56 (0$\cdot$47)  \\

${\beta}_2$ & -0$\cdot $68 (1$\cdot$30) & 1$\cdot$01 (0$\cdot$80) & 1$\cdot $91 (1$\cdot$00) & 1$\cdot$99 (0$\cdot$47)*  \\ 
${\beta}_3$ & -0$\cdot $85 (1$\cdot$47) & 1$\cdot$91 (0$\cdot$66)* & 2$\cdot $80 (1$\cdot$04)* & 1$\cdot$34 (0$\cdot$62)*  \\

\end{tabular}
\end{center}
\end{table}

We then used these four pre-specified structures to obtain the estimating coefficients for the model (\ref{GEE}). Table \ref{coefficients} reports the estimated results for model (\ref{GEE}), including the estimated coefficients, their standard errors, and their statistical significance (p-value less than 0$\cdot$05). When using the Isolated structure, all the coefficients $\beta_1, \beta_2$ and $\beta_3$ are not significant which indicates
that there is no significant difference in COVID cases among the four regions in the U.S. However, under the pre-specified GLASSO, there are significant differences in COVID cases between the Mid West and the South, and between the West and the South. In addition, the estimated standard errors under the GLASSO pre-specified network
is also smallest, followed by the Band(3), TIGER, and Isolated structure. These results are  consistent with our proposed test statistics since they suggest that the GLASSO structure fits
the data best, followed by the Band(3), TIGER and Isolated structure.
This suggests that the proposed test statistics can be used as a powerful tool to identify a good pre-specified structure for further analysis. 

Finally, we applied a bootstrap method to further 
evaluate the standard errors of 
coefficient estimators, and compared
the efficiency gain in terms of the standard errors when using different pre-specified network structures. More
specifically, we subsampled 40 states from 51 states without replacement for 100 times. At each time we used subsampled data in the GEE equation (\ref{GEE1}) to estimate the coefficients. Note that here to increase the stability of the procedure, we reuse the estimated precision matrix $\boldsymbol{V}^{-1}$ based on the data from all 51 states. At the $i$-th replication, we denote the corresponding standard errors of each coefficient by $(\mbox{Sd}_{1,i},\mbox{Sd}_{2,i}, \mbox{Sd}_{3,i} ) $. To evaluate the variability of standard errors from the subsampling process, we then calculate the corresponding means and standard deviations of $(\mbox{Sd}_{1,i},\mbox{Sd}_{2,i}, \mbox{Sd}_{3,i} ) $, for $i = 1,\cdots,100$ as follows:
\begin{equation*}
    \mbox{AVE}_j = \sum_{i=1}^{100}\mbox{Sd}_{j,i} / 100, \ 
    \mbox{SD}_j = \surd{ \big\{ \sum_{i=1}^{100}(\mbox{Sd}_{j,i} - \mbox{AVE}_j)^2/ 100}\big\}\; \mbox{for} \; j = 1,2,3.
\end{equation*}

\begin{figure}
     \centering
     \begin{subfigure}{0.45\textwidth}
         \centering
         \includegraphics[width=\textwidth]{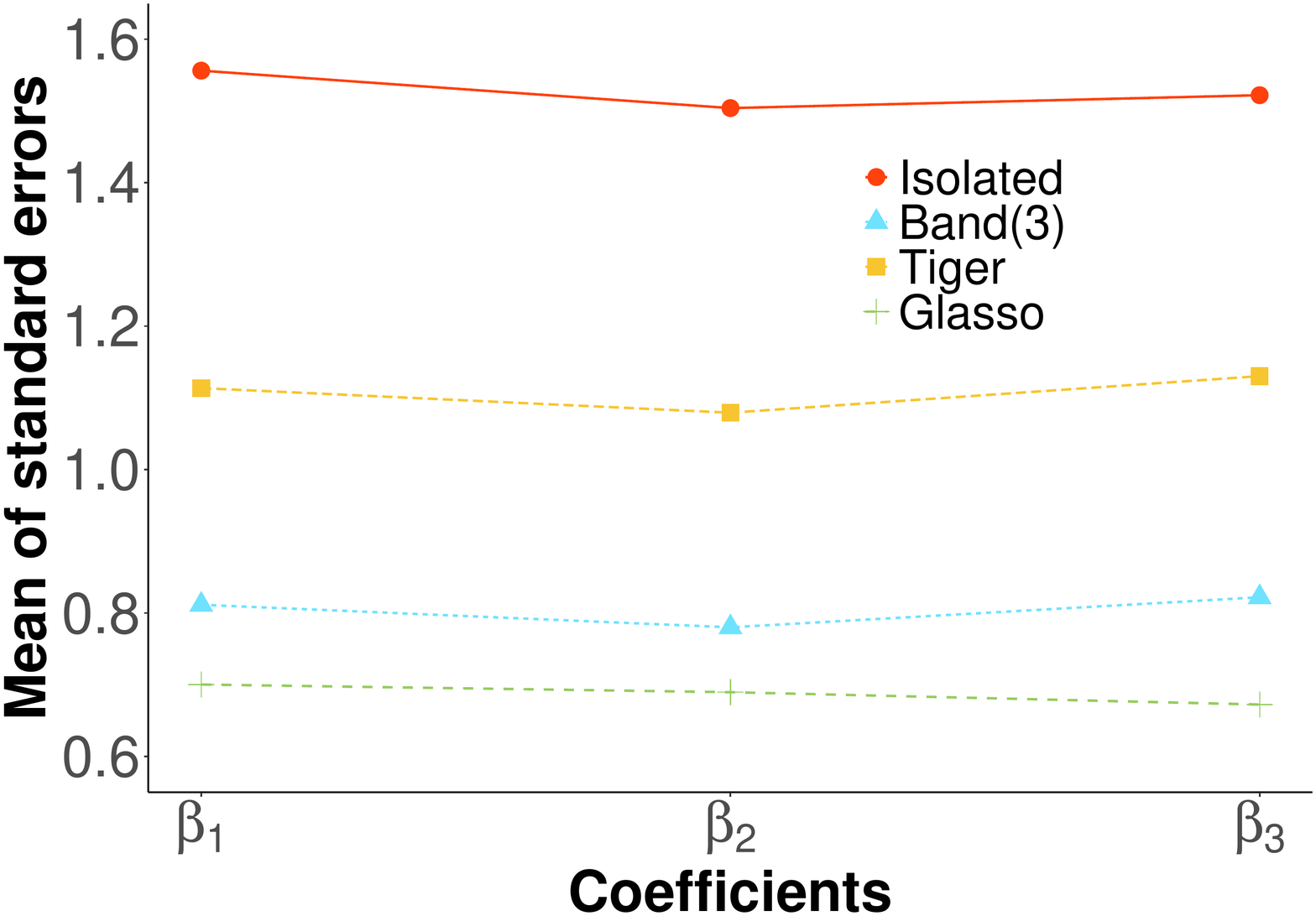}
         \caption{}
     \end{subfigure}
     \begin{subfigure}{0.45\textwidth}
         \centering
         \includegraphics[width=\textwidth]{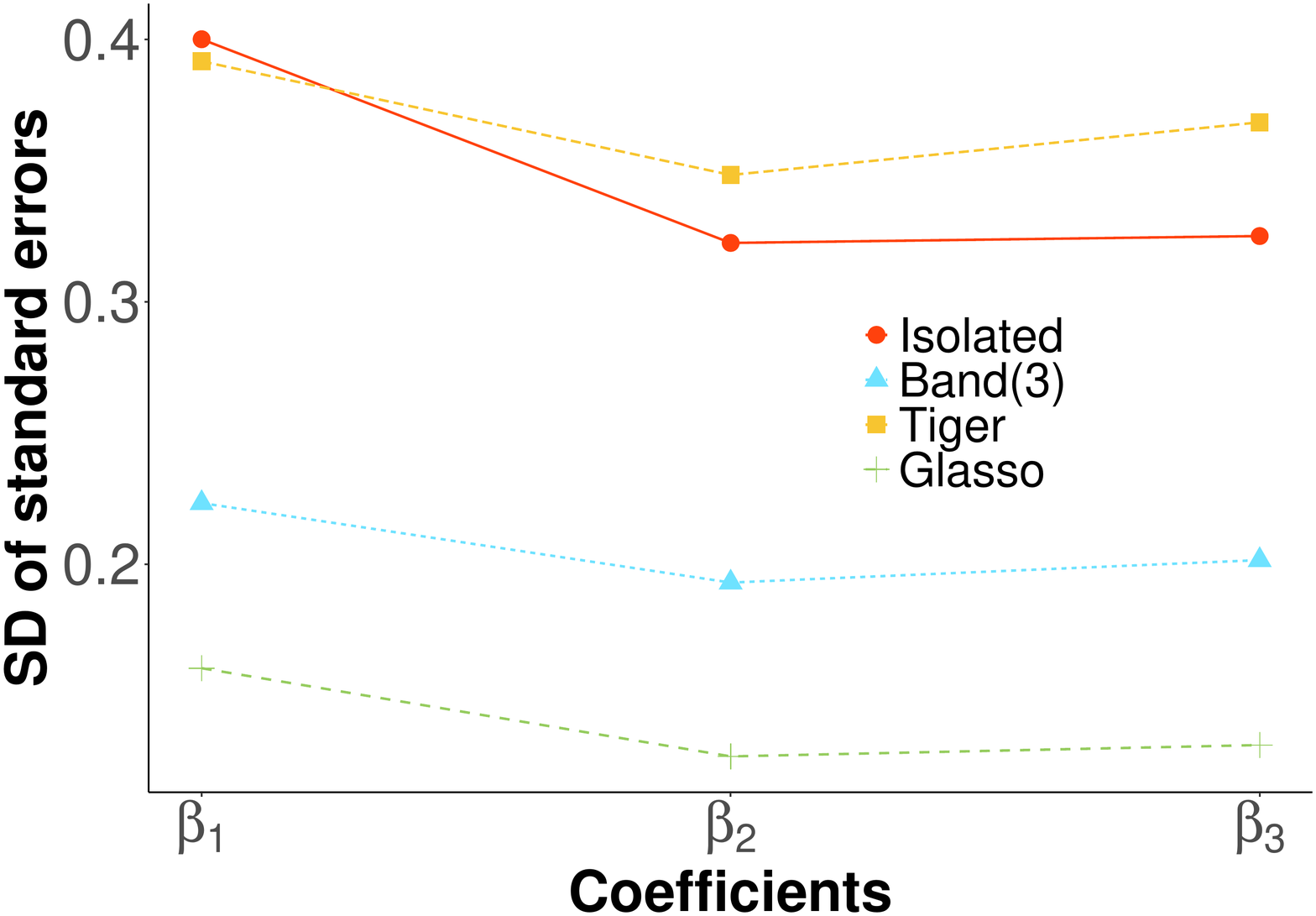}
          \caption{}
     \end{subfigure}
        \caption{Average (a) and standard error (b) of absolute prediction errors obtained
from four graphic structures: Isolated, Band(3), TIGER and GLASSO. }
        \label{Standarderror}
\end{figure}

 Figure \ref{Standarderror} shows the mean and standard deviation of the coefficients obtained from the above subsampling procedure. Both panels 2(a) and 2(b) of the figure demonstrate that the standard error obtained from the pre-specified GLASSO structure is the smallest, followed by Band(3), TIGER, and Isolated network. The result again agrees with our test statistics obtained and confirms that choosing a good pre-specified graphical structure is essential for efficiency gain when using the GEE method. Therefore, the proposed test statistics can serve as a valuable tool to help select a reliable pre-specified graphic structure. 

\vskip 14pt
\noindent {\large\bf Acknowledgements}

This research was partially supported by an NSF grant. The authors thank Dr. Peter Song of University of Michigan for sharing the code for the real data analysis.

\section*{Supplementary material}
\label{SM}

The supplementary material includes detailed proofs of the main results and examples of precision structures satisfying technical condition (C1). 
\appendix

\newpage
\section{Supplementary Material}

Because $\bSn$ is invariant to $\bmu$, without loss of generality, we assume that $\bmu=0$ in the rest of the proof. We replace $\bSn$ by $\bVn=\sum_{i = 1}^n \bX_i\bX_i^\T/n$ because the terms related to $\bar{\bX}$ in $\bSn$ 
are small order of $\bVn$.

\subsection{Examples of precision structures satisfying Condition (C1)}
The technical proof of Theorem \ref{generalTheo} is given in the Section A$\cdot$4 of the supplementary material. The main idea of the proof is to approximate the test statistics $\hat{D}_n$ by a modified version of $D_n$. 
Denote $A = \{(i, j), 1 \leq i, j \leq p\}$ be the set of all pairs of indices that $\hat{D}_n$ will be maximized over and write $\hat{D}_{n} = \max_{(i, j) \in A} \hat{D}_{ij}^2$.  
Let $A_0 = \{ (i, j), \omega_{ij}^* \neq 0 \}$ be the set of indices that excluding the sparse set of non-zeros in $\bOmg^*$. 
Let $A_1 = \cup_{i=1}^p \{(i, k): \lim_{p \to\infty} s_0\sigma_{ik} \neq 0, \forall (i, k) \notin A_0 \}$ be the set of indices that variables $(i, k)$ having covariance larger than $1/s_0$. 
Define $B_0 = A_0 \cup A_1$ as the union of $A_0$ and $A_1$. 
For convenience, denote , $\hat{D}_{n}^* = \max _{(i, j) \in A} \hat{D}_{ij}^{*2}$, $\hat{D}_{n1}^* = \max _{(i, j) \in A/A_0} \hat{D}_{ij}^{*2},$ $\hat{D}_{n2}^* = \max _{(i, j) \in A/B_0} \hat{D}_{ij}^{*2}$, where $ {\hat{D}_{ij}}^{*2} =( \bE_j^\T \bold{V}_n \hat{\bw}_{i,0} - \bE_j^\T \bE_i)^2/\theta_{ij}$. 
We will show that the distribution of $\hat{D}_n$ can be approximated by the distribution of $D_{n2}= \max _{(i, j) \in A/B_0} D_{ij}^{2}$.
Notice that for technical details in the proof we need $\text{card}(A/B_0) = p^2\{1+o(1)\}$, which followed by $\vert \vert \bsig^* \vert \vert_1 \leq C_1$, for some $C_1 > 0$.

We now provide some examples of classes of precision matrices that satisfying Condition (C1) and their corresponding forms of $A/B_0$. 
 
 \begin{example} \label{PolynomialPreEx} (Polynomial decay) Let $\bOmg^*=(\omega_{ij}^*)_{p \times p}$ be a banded polynomial precision matrix defined by
 $\omega_{ij}^*= {1}/{(1 + \l i-j\l )^\lambda}$, for $\l i - j \l < s_0$, $s_0 = o(\surd{n}), \lambda \geq 2$, and $\omega_{ij}^*=0$ otherwise. 
Lemma \ref{invHall} in Section A$\cdot$3 shows that $\l \bsig_{i,j}\l \leq C*(1+\l i-j \l)^{-\lambda}$.Therefore, $\vert \vert \bsig^* \vert \vert_1  < C_1$, for some $C_1$.   
Furthermore, we also have, $\sigma_{jk} = O({1}/{s_0^{\lambda}})=o({1}/{s_0})$ for $(j, k)$ such that $\l j-k\l \geq s_0$. So $A_1 = \cup_{i=1}^p \Big\{ (i, k): \max\{\l k - (i+ s0- 1) \l, \l k - (i- s0+ 1)\l\} \leq h, k \notin [i-s_0+1, i+s_0-1] \Big\},$
where $h = s_0$. Meanwhile $B_0 =  \cup_{i=1}^p \Big\{ (i, k), \max \{\l k - (i+ s0- 1) \l,  \l k - (i- s0+ 1)\l\} \leq h \Big\}$. Therefore,  
$A/B_0 = \cup_{i=1}^p \Big\{ (i, k),  \min\{\l k - (i+ s0- 1)\l, \l k - (i- s0+ 1)\l\} \geq h \Big\}$. As a result, $\mbox{card}(A/B_0) = p^2\{1+o(1)\}.$
\end{example}
\begin{example} \label{ExponentialPreEx} (Exponential decay) Let $\bOmg^*= (\omega_{ij}^*)_{p \times p}$ be a precision matrix decaying at an exponential rate 
so that $\omega_{ij}^*= \theta^{\l i - j \l}$ 
for $\l i - j \l < s_0, s_0 = o(\surd{n}), 0<\theta < 1$,
and $\omega_{ij}^*=0$ otherwise. Lemma \ref{invExp} in Section A$\cdot$3 shows that $\sigma_{jk} = O\{\exp (-\beta \l j-k \l)\},$ for some $0 < \beta < - \log\theta$. Therefore, $\vert \vert \bsig^* \vert \vert_1 < C_1$, for some $C_1 > 0$. 
So $A/B_0 = \cup_{i=1}^p \{ (i,k), \min (\l k - (i+ s0- 1)\l, \l k - (i- s0+ 1)\l) \geq h \}$, 
where $h = s_0^\gamma$, for some small $\gamma >0.$ As a result, $\mbox{card}(A/B_0) = p^2\{1+o(1)\}.$
 \end{example}
 
 \begin{example} \label{BandedPreEx} (Banded) Assume that precision matrix  $\bOmg^*$ has a banded structure such that $\omega_{ij}^*= 0 $, for $\l i-j \l \geq s_0$ where $s_0 = o(\surd{n})$.
Then 
$$A/B_0 = \cup_{i=1}^p \{ (i,k), \min (\l k - (i+ s0- 1)\l,  \l k - (i- s0+ 1)\l) \geq h \}$$
 where $h = s_0^{1+\gamma} $, for some small $\gamma>0$. 
Lemma \ref{invDemp} in Section A$\cdot$3 implies that  $\vert \sigma_{ij}\vert \leq C\lambda_1^{\vert i - j\vert}$,  for $0 < \lambda_1 = {(\surd{\mbox{cond}(\bOmg^*)} -1})/{(\surd{\mbox{cond}(\bOmg^*)} + 1)} < 1,$ 
 where $\mbox{cond}(\bOmg^*) = \vert \vert \bOmg^*\vert \vert \vert \vert {\bOmg^*}^{-1} \vert \vert.$ Therefore $\vert \vert \bsig^* \vert \vert_1  < C_1$, for some $C_1 > 0$, and $\sigma_{jk} <\lambda_1^{{2\l j - k \l}/{s_0} }= \lambda_1 ^{2s_0^\gamma} = o({1}/{s_0})$ on $A/B_0$. We also have that $\mbox{card}(A/B_0) = p^2\{1+o(1)\}.$
 \end{example} 
 \begin{example} \label{FactorPreEx} (Factor model)
Assume that $\bOmg^*$ is generated from a factor model. Specifically, $\bOmg^*= \bI_p + \sum_{i=1}^{k} \alpha_i \bu_i \bu_i^\T$ where $\bI_p$ is the identity matrix and 
 for each $i =1,\ldots,k$ ($k \in \mathbb{Z}^{+}$), $\alpha_i \in \mathbb{R}$, $\bu_i$ is a $p$-dimensional vector in $\mathbb{R}^{p}$ such that $\vert \vert \bOmg^*\vert \vert_1 = O(1)$. 
Lemma \ref{invFac} in Section A$\cdot$3 shows that $ A/B_0 = A/ A_0$, since $\sigma_{jk} = 0$ for  $(j,k) \in A/B_0$. As a result, $\vert \vert \bsig^* \vert \vert_1 <C_1$ for some $C_1>0$ and $\mbox{card}(A/B_0) = p^2\{1+o(1)\}.$
\end{example}

\subsection{Proof of Lemmas}

\noindent \textit{Proof of Lemma \ref{covlm1}:} (1) We have
\begin{align*}
\mbox{var}(\bE_j^\T \bold{V}_n \bw_i^*-\bE_j^\T \bE_i) 
&=  \mbox{var}(\bE_j^\T \sum_{i = 1}^n \bX_i \bX_i ^\T \bw_i^*)/n^2 = \mbox{var}(\bE_j^\T \bX_1 \bX_1^\T \bw_i^*)/n \\
&= \mbox{E}(\bE_j^\T\bX_1\bX_1^\T \bw_i^* {\bw_i^*}^\T\bX_1\bX_1^\T \bE_j)/n - (\bE_j^\T \bsig^* \bw_i^*)^2/n.
\end{align*}
We write $\bX_1$ as $\bgam^\T \bZ$, where $\bZ$ is a $p$-dimensional standard normally distributed random vector and $\bsig^*=\bgam^\T\bgam$.
Then we have
\begin{align*}
\mbox{E}(\bE_j^\T\bX_1\bX_1^\T \bw_i^* {\bw_i^*}^\T\bX_1\bX_1^\T \bE_j) 
&= \mbox{E}(\bE_j^\T \bgam^\T \bZ \bZ^\T \bgam \bw_i^* {\bw_i^*}^\T \bgam^\T \bZ \bZ^\T \bgam \bE_j) \\
&= \mbox{E}( \bZ^\T \bgam \bw_i^* {\bw_i^*}^\T \bgam^T \bZ \bZ^\T \bgam \bE_j \bE_j^\T \bgam^\T \bZ) \\
&= \mbox{tr}(\bgam \bw_i^* {\bw_i^*}^\T \bgam^\T)\mbox{tr}(\bgam \bE_j \bE_j^\T \bgam^\T) + 2\mbox{tr}(\bgam \bw_i^* {\bw_i^*}^\T \bgam^\T \bgam \bE_j \bE_j^\T \bgam^\T) \\
& = {\bw_i^*}^\T \bsig^*\bw_i^* \bE_j^\T \bsig^* \bE_j + 2({\bw_i^*}^\T \bsig^* \bE_j \bE_j^\T \bsig^* \bw_i^*)\\
& = \omega_{ii}^* \sigma_{jj}^* + 2({\bw_i^*}^\T \bsig^* \bE_j \bE_j^\T \bsig^* \bw_i^*).
\end{align*}
Since  $\bE_j^\T \bsig^* \bw_i^* = 0, {\bw_i^*}^\T \bsig^* \bE_j \bE_j^\T \bsig^* \bw_i^* = 0$, for $1 \leq i \neq j \leq p$.
 This yields $\mbox{var}({\bE_j^\T \bold{V}_n \bw_i^*-\bE_j^\T \bE_i})$ $ = \omega_{ii}^* \sigma_{jj}^*/n$.
(2) If $1 \leq i =j \leq p$, we have $\bE_j^T \bsig^* \bw_i^* = 1$ and  ${\bw_i^*}^\T \bsig^* \bE_j \bE_j^\T \bsig^* \bw_i^* = 1$. 
So $\mbox{var}({\bE_i^\T\bold{V}_n \bw_i^*-\bE_i^\T\bE_i})=(\omega_{ii}^*\sigma_{ii}^* + 1)/n$ \hfill$\square$
\smallskip

\subsection{Technical lemmas and their proofs
\label{A3}}

We include the following Lemmas A1- A3 and Lemmas \ref{MaxA0}-\ref{TechLem} that are needed for the proof of the main theorems.

\begin{lemma}[Bonferroni Inequality]\ 
\label{bonferri}
Let $B = \cup_{t=1}^p B_t$
 we have
$$\sum_{t=1}^{2k}(-1)^{t-1} E_t \leq pr(B) 
\leq \sum_{t =1}^{2k-1} (-1)^{t-1} E_t$$
where 
$E_t = \sum_{1 \leq i_1 \leq\ldots\leq i_t \leq p} 
pr(B_{i_1} \cap\cdots\cap B_{i_t})$ and  $k < [p/2]$.
\end{lemma}

\begin{lemma}[Berman (1962)] \ 
If X and Y are bi-variate normally distributed with expectations 0, unit variance and correlation $\rho$, then
$$ \lim \limits_{c \rightarrow \infty} 
\frac{pr(X > c, Y > c)}{\{2 \pi (1-\rho)^{1/2}c^2\}^{-1} 
\exp\{{-c^2}/{(1+\rho)}\}(1+\rho)^{1/2}} = 1,$$
uniformly for all $\rho$ such that $\l \rho \l < \delta$, for any $0 < \delta < 1$.

\end{lemma}

\begin{lemma}[Zaitsev (1987)]
\label{Zaitsepapp}
Let $\tau >0$, $\bxi_1,\ldots,\bxi_n \in \mathbb{R}^k$ are independent random variables such that $\mathcal{L}(\bxi_i) \in \mathcal{B}_1(k,\tau)$, for $i=1,\ldots,n$, where $\mathcal{B}_1(k,\tau) = \{\mathcal{L}(\bxi) \in \mathcal{F}_k: E\bxi = 0 ,  \vert E(\bxi,\bt)^2(\bxi,\bu)^{m-2} \vert \leq  m! \tau^{m-2} \vert \vert \bu \vert \vert ^{m-2}E(\bxi,\bt)^2/2,\ \mbox{for every integer} \ m \geq 3 \ \mbox{and for all} \ \bt, \bu$ $\}$, $\mathcal{L}(\bxi)$ is the distribution of random variable $\bxi$, $\mathcal{F}_k$ is the class of random distribution on  $\mathbb{R}^k$, $(\bxi,\bt)$ is the inner product of $\bxi$ and $\bt$. Denote $S = \bxi_1 + \bxi_2 +\cdots+ \bxi_n, F =\mathcal{L}(S)$. Let $\Phi $ is Gaussian distribution with mean vector 0, and the same covariance matrix with $F$. Define $\pi (F,\Phi;\lambda) = \mbox{sup}_{H \in \mathcal{B}_k}\mbox{max} \{F(H) - \Phi(H^\lambda), \Phi(H) - F(H^\lambda) \}$, where $\mathcal{B}_k$ is the $\sigma$-field of Borel subsets of $\mathbb{R}^k$, $H^\lambda = \{y \in \mathbb{R}^k: \mbox{inf}_{x \in H} \vert \vert  y - x \vert \vert \leq \lambda \}$. Then 
$$\pi (F,\Phi;\lambda) \leq c_1k^{5/2}\exp(-\frac{\lambda}
{ \tau c_2 k^{5/2}}),$$
for all $\lambda > 0.$
\end{lemma}

The following Lemmas \ref{invHall} - \ref{invFac} are used in Examples \ref{PolynomialPreEx}-\ref{FactorPreEx} for some special classes of precision matrices.

\begin{lemma}[Hall $\&$ Lin (2010)]
\label{invHall}
For $\lambda \geq 1, c_0 > 0, M >0$. For any sequence of matrices $\bsig_n$ such that
$$\bsig_n \in \Theta_n^\star(\lambda, c_0, M) = \{ \bsig_n: \l \bsig_n(j,k) \l \leq M*(1+ \l j-k \l)^{-\lambda}, \vert \vert  \bsig_n\vert \vert  \geq c_0  \}.$$ 
There exists a constant $C = C(\lambda, c_0, M)$ such that for any $n$ and any $1 \leq j, k \leq n$, 
$$\l \bsig_n^{-1}(j,k)\l \leq C*(1+\l j-k \l)^{-\lambda}.$$
\end{lemma}

\begin{lemma}[Gr\"{o}chenig $\&$ Leitner (2006)]
\label{invExp}
Let $\bA = (a_{ij})_{p \times p}, \bA^{-1} = (b_{ij})_{p \times p}$, $\lambda_{\max}(\bA)$ 
and $\lambda_{\max}(\bA^{-1})$ are bounded. If $a_{ij} = O\big\{\exp(- \alpha \l i - j \l )\big\}$, 
 then $b_{ij} = O\big\{\exp (- \beta \l i - j \l ) \big \}$ for some $\beta$ such that $0 < \beta < \alpha$.
\end{lemma}

 \begin{lemma}[Demko et al. (1984)]\ 
 \label{invDemp}
 Let $\bA = (a_{ij})_{p \times p}$ and $\bA^{-1} = (b_{ij})_{p \times p}$. Assume that $\lambda_{max}(\bA)$ and $\lambda_{max}(\bA^{-1})$ are bounded. 
 If $\bA$ is positive definite and $m$-banded, then we have 
 $\vert b_{ij}\vert \leq C\lambda^{\vert i - j\vert}$
  where $\lambda = \big[\{\surd{\mbox{cond}(\bA)} -1\}/\{\surd{\mbox{cond}(\bA)} + 1\}\big]^{2/m},$ 
  $\mbox{cond}(\bA)= \vert \vert \bA \vert \vert \ \vert \vert \bA^{-1}\vert \vert,$
   $C = \vert \vert \bA^{-1}\vert \vert  \max[1$ and $  \{1+ \surd{\mbox{cond}(\bA)}\}^2$$/\{2\mbox{cond}(\bA)\}].$

 \end{lemma}
 

\begin{lemma}
\label{invFac}
Let $\bI_p$ be an identity matrix and $\bA = \bI_p + \sum_{i=1}^{k} \alpha_i \bu_i \bu_i^\T$
for any vector  $\bu_i \in \mathbb{R}^{p \times 1}, \alpha_i \in \mathbb{R}, i =1,\ldots,k$. 
 Then outside the support of $\bA$, and $\bA^{-1}$ have the same zeros pattern.
 \end{lemma}
\noindent \textit{Proof:}
Let us denote
 $\bU_{p \times k} = (\alpha_1 \bu_1,\ldots,\alpha_k \bu_k) , 
 \bV_{k \times p} = (\bu_1,\ldots,\bu_k)^\T$,
  then
    $\sum_{i=1}^{k} \alpha_i \bu_i \bu_i^\T = \bU \bV$. 
    So $\bA = \bI_p + \bU \bV.$
Applying Woodbury formula from page 211 in Hager (1989) we have:
$$\bA^{-1} = (\bI_p + \bU \bV)^{-1}
 = \bI_p + \bU(\bI_k - \bV \bU)^{-1} \bV.$$
Denote 
$\bM = (\bI_k - \bV \bU)^{-1}, \bH = \bU \bM \bV$, 
then $\bA^{-1} = \bI_p + \bH.$
It can be checked that the zero patterns of $\bH$ and $\bU \bV$ are the same.
For easy to understand, let us consider a special case
$ \bA = \bI_p + \bu_1 \bu_1^\T + \bu_2 \bu_2^\T $
 where
  $\bu_1 = \bE_1 + \bE_2 \in \mathbb{R}^{p \times 1}$ 
  and
   $\bu_2 = \bE_3 + \bE_4 \in \mathbb{R}^{p \times 1}$. Then 
   $$\bV = \begin{pmatrix}
      1 & 1 & 0 & 0 & 0& \ldots & 0 \\
      0 & 0 & 1 & 1 & 0& \ldots &0 
    \end{pmatrix}_{2 \times p} \mbox{and} \quad \bU = \bV^\T$$

For $(i,j)$th position of $\bH$ where $i \notin \{1, 2, 3, 4\}$ 
or $j \notin \{1, 2, 3, 4\} $, we have
$\bH(i,j) =  \bU(i,) \bM \bV(,j) = 0$.
Since the zero patterns on $\bH$ and $\bU \bV$ are the same, using this fact together with $\bA = \bI_p + \bU \bV$ and $\bA^{-1} = \bI_p + \bH$ completes the proof of this lemma. \hfill $\square$


\begin{lemma}\ 
\label{MaxA0} 
$\max\limits_{(i, j) \in A_0}\hat{D}_{ij}^{*}
= o_p(1).$
\end{lemma}
\noindent \textit{Proof:}
Recall that
$\hat{D}_{ij}^{*} ={ \l \bE_j^\T \bold{V}_n \hat{\bw}_{i,0}
 - \bE_j^\T \bE_i \l}/{\surd{\theta_{ij}}}.$
Consider the numerator
\begin{align*}
 \bE_j^\T \bold{V}_n \hat{\bw}_{i,0} - \bE_j^\T \bE_i  
 &= \frac{n-1}{n} (\bE_j^\T \bS_n \hat{\bw}_{i,0} 
  - \bE_j^\T \bE_i) - \frac{1}{n} \bE_j^\T \bE_i + \bE_j^\T \bar{\bX} \bar{\bX}^\T \hat{\bw}_{i,0}\\
  &= \frac{n-1}{n}\{ \bE_j^\T \bS_n \bB_{i,0}(\bB_{i,0}^\T \bS_n \bB_{i,0})^{-1} \bB_{i,0} \bE_i - \bE_j^\T \bE_i \} \\
 &\quad- \frac{1}{n} \bE_j^\T \bE_i + \bE_j^\T \bar{\bX} \bar{\bX}^\T \hat{\bw}_{i,0}.
\end{align*}
Notice that the first term is indeed 0.
  For notation convenience, consider $i = 1$ and suppose that 
  $$\bw_{1,0}=(w_{11}, w_{12}, w_{13}, w_{14},0,\ldots,0 )^\T  = \bB_{1,0} \bw_{11,0} \in \mathbb{R}^{p \times 1}, $$
 where
   $\bw_{11,0} = (w_{11}, w_{12}, w_{13}, w_{14})^\T \in \mathbb{R}^{4 \times 1}$ is non zero components of $\bw_{1,0}$
   and 
$$ \bB_{1,0} = 
\begin{pmatrix}
   1 & 0 & 0 & 0 & 0 &\ldots&0 \\
    0 & 1 & 0 & 0 & 0 & \ldots &0 \\
   0 & 0 & 1 &0 & 0 & \ldots &0 \\
   0 & 0 & 0 & 1 & 0& \ldots&0 \\
\end{pmatrix}^\T = 
\begin{pmatrix}
   \bE_1  & \bE_2 & \bE_3  & \bE_4  \\
\end{pmatrix} \in R^{p \times 4}.
$$
Then
 $$ \bB_{1,0}^\T \bS_n \hat{\bw}_{1,0}  
= \bB_{1,0}^\T \bS_n \bB_{1,0}(\bB_{1,0}^\T \bS_n \bB_{1,0})^{-1} \bB_{1,0}\bE_1 = \bB_{1,0}\bE_1.$$
So
$$\bE_j^\T \bS_n \bB_{1,0}(\bB_{1,0}^\T \bS_n \bB_{1,0})^{-1} \bB_{1,0}\bE_1 - \bE_j^\T \bE_1 = 0, \ \mbox{for} \ j = 1,2,3,4.$$
So we have
$$
\hat{D}_{ij}^{*} = \frac{\bE_j^\T \bold{V}_n \hat{\bw}_{i,0} 
- \bE_j^\T \bE_i}{\surd{\theta_{ij}}} 
=   -\frac{ \bE_j^\T \bE_i/n}{\surd{\theta_{ij}}} 
+ \frac{\bE_j^\T \bar{\bX} \bar{\bX}^\T \hat{\bw}_{i,0}}{\surd{\theta_{ij}}}.
$$

So
\begin{equation}
\label{sepA0}
\max\limits_{(i, j) \in A_0} \hat{D}_{ij}^{*} 
 \leq \max\limits_{(i, j) \in A_0} \l \frac{ 
\bE_j^\T \bE_i/n}{\surd{\theta_{ij}}} \l
+ \max\limits_{(i, j) \in A_0} \l
\frac{\bE_j^\T \bar{\bX} \bar{\bX}^\T \hat{\bw}_{i,0}}{\surd{\theta_{ij}}} \l.
\end{equation}

From Lemma \ref{covlm1} in the maintext, we have the denominator $\surd{\theta_{ij}}$ is at the order of $1/\surd{n}$.
This gives us
\begin{equation}
\label{sepA01}
\max\limits_{(i,j) \in A_0}\frac{ \bE_j^\T \bE_i/n}{\surd{({1}/{n})}} = o(1).
\end{equation}

For the second term in (\ref{sepA0}), we note that
$\bE_j^\T \bar{\bX} \bar{\bX}^\T \hat{\bw}_{i,0}
 = \sum_{k=1}^{s_0} \bar{X}_j\bar{X}_{i_k} \hat{w}_{ii_k} $
where $1 \leq i_1, i_2,\ldots,i_{s_0} \leq p$ are non zero positions in $\bw_{i,0}$.
From page 2582 in Bickel $\&$ Levina (2008), we have 
$\max_{1 \leq i \leq p} \bar{X}_i=O_p\{\surd{({\log\ p}/{n})}\}$. 
This gives us
 $$\max_{(i,j) \in A_0} \l \bE_j^\T \bar{\bX} \bar{\bX}^\T \hat{\bw}_{i,0} \l
\leq \max_{1 \leq i \leq p} \bar{X}_i^2 \sum_{k=1}^{s_0} \l \hat{w}_{ii_k} \l
=O_p(\log p/{n}).$$
This gives us
\begin{equation}
\label{sepA02}
\max\limits_{(i,j) \in A_0} \l \bE_j^\T \bar{\bX} \bar{\bX}^\T \hat{\bw}_{i,0} \l/{\surd{\theta_{ij}}} 
= O_p(\log p/{\surd{n}}) = o_p(1).
\end{equation}

The facts $(\ref{sepA0}), (\ref{sepA01})$, and $(\ref{sepA02})$ together verify the lemma \hfill $\square$

\begin{lemma}
\label{turnlem}
$\surd{n} \max\limits_{(i,j) \in A/A_0} 
\l \bE_j^\T \bold{V}_n  \hat{\bw}_{i,0}  \l
 = \surd{n} \max\limits_{(i,j) \in A/A_0} \l   \bE_j^\T \bold{V}_n \bw_i^* 
 +  \bE_j^\T  \bsig^*  \hat{\bw}_{i,0} \l + o_p(\surd{\log\ p}) .$
 \end{lemma}
\textit{Proof:}
On the one hand we have
\begin{align}
\label{l12a}
&\surd{n} \max\limits_{(i,j) \in A/A_0} \l \bE_j^\T 
(\bold{V}_n - \bsig^*)\bB_{i,0}(\bS_i^{-1} - \bOmg_i^*) \bf{f}_i \l \notag \\
&\leq \surd{n} s_0 \max\limits_{1 \leq i,j \leq p} \l v_{ij} 
- \sigma_{ij}^*\l \max\limits_{1 \leq i,j \leq p} \l \hat{w}_{ij,0} 
- \omega_{ij}^*\l \notag \\
 &= O_p(s_0{\log p}/{\surd{n}}) = o_p(\surd{\log p}).
\end{align}
On the other hand
\begin{align}
\label{l12b}
&\surd{n} \max\limits_{(i,j) \in A/A_0} \l \bE_j^\T (\bold{V}_n - \bsig^*) \bB_{i,0}(\bS_i^{-1} - \bOmg_i^*) \bf{f}_i \l \notag\\
& = \surd{n} \max\limits_{(i,j) \in A/A_0} \l \bE_j^\T \bold{V}_n \hat{\bw}_{i,0}  -  \bE_j^\T \bold{V}_n  \bw_i^* 
-  \bE_j^\T  \bsig^* \hat{\bw}_{i,0}  \l .
\end{align}

Combining (\ref{l12a}) and (\ref{l12b}), we get
$$\surd{n} \max\limits_{(i,j) \in A/A_0} \l
 \bE_j^\T \bold{V}_n  \hat{\bw}_{i,0}  \l 
 = \surd{n} \max\limits_{(i,j) \in A/A_0} \l   
\bE_j^\T \bold{V}_n \bw_i^* + \bE_j^\T \bsig^* \bB_{i,0} \hat{\bw}_{i,0} \l 
 + o_p(\surd{\log\ p}).$$ 
 This completes the proof the Lemma.\hfill $\square$
\begin{lemma}\ 
\label{VarASw}
For any $ (i,j) \in A/B_0$, let $\ba^\T = \bE_j^\T 
- \bE_j^\T  \bsig^* \bB_{i,0} \bOmg_i^* \bB_{i,0}^\T$, then 
$$Var(\ba^\T \bold{V}_n \bw_{i,0})  
= ( \omega_{ii}^* \sigma_{jj}^* -  \omega_{ii}^* \bE_j^\T  \bsig^*  \bB_{i,0} \bOmg_i^* \bB_{i,0}^\T \bsig^* \bE_j)/n.  $$
\end{lemma}

\noindent \textit{Proof:}
We first note that
 \begin{eqnarray*}
 \E(\ba^\T \bold{V}_n \bw_{i,0}) &=& \E(\ba \bX_1 \bX_1^\T \bw_{i,0})
  =  \ba^\T \bsig^*\bw_{i,0} \\
   &=& (\bE_j^\T - \bE_j^\T  \bsig^*\bB_{i,0} (\bB_{i,0}^\T \bsig^*\bB_{i,0})^{-1} \bB_{i,0}^\T) \bsig^*\bw_{i,0}\\
& =& - \bE_j^\T  \bsig^*\bB_{i,0} (\bB_{i,0}^\T \bsig^*\bB_{i,0})^{-1} \bB_{i,0}^\T \bE_i  = - \bE_j^\T  \bsig^*\bw_{i,0} = 0.\\
 \mbox{var}(\ba^\T \bold{V}_n \bw_{i,0}) 
 &=& \E \{ (\ba^\T \bold{V}_n \bw_{i,0} )^2 \} = \frac{1}{n}\E(\ba^\T \bX_1 \bX_1^\T \bw_{i,0} \bw_{i,0}^\T \bX_1 \bX_1^\T \ba).\\
 \E(\ba^\T \bX_1 \bX_1^\T\bw_{i,0} \bw_{i,0}^\T \bX_1 \bX_1^\T \ba) 
 &=& \E(\ba^\T \bgam^\T \bZ \bZ^\T \bgam \bw_{i,0} \bw_{i,0}^\T \bgam^\T \bZ \bZ^\T \bgam \ba)\\ 
 &=& \E(\bZ^\T \bgam \bw_{i,0} \bw_{i,0}^\T \bgam^\T \bZ \bZ^\T \bgam \ba \ba^\T\bgam^\T \bZ )\\
 & =& \mbox{tr}(\bgam \bw_{i,0} \bw_{i,0}^\T \bgam^\T)\mbox{tr}(\bgam \ba \ba^\T \bgam^\T) +  2\mbox{tr}(\bgam \bw_{i,0} \bw_{i,0}^\T \bgam^\T \bgam \ba^\T \ba \bgam^\T)\\
 & =& \omega_{ii}^* \ba^\T \bsig^*\ba + 2\bE_i^\T \ba \ba^\T \bE_i.
 \end{eqnarray*}
Thus, we have
\begin{equation}
\label{varAsw1}
\mbox{var}(\ba^\T \bold{V}_n \bw_{i,0}) =  
(\omega_{ii}^* \ba^\T \bsig^*\ba + 2\bE_i^\T \ba \ba^\T \bE_i)/n.
\end{equation} 
Recall that $\bOmg^*=\bB_{i,0}^\T \Omega^* \bB_{i,0}$. 
We note the following
\begin{align}
\label{AsigA}
\ba^T \bsig^*\ba &= (\bE_j^\T - \bE_j^\T  \bsig^*\bB_{i,0} \bOmg_i^* \bB_{i,0}^\T) \bsig^*(\bE_j - \bB_{i,0} \bOmg_i^* \bB_{i,0}^\T \bsig^*\bE_j) \notag \\
&= \bE_j^\T \bsig^*\bE_j -  2 \bE_j^\T  \bsig^*
\bB_{i,0} \bOmg_i^* \bB_{i,0}^\T \bsig^*\bE_j \notag \\
&+  \bE_j^\T  \bsig^*\bB_{i,0} (\bB_{i,0}^\T \bsig^*\bB_{i,0})^{-1} \bB_{i,0}^\T \bsig^*\bB_{i,0} \bOmg_i^* \bB_{i,0}^\T \bsig^*\bE_j\notag \\
&= \bE_j^\T \bsig^*\bE_j -  \bE_j^\T  \bsig^*\bB_{i,0} \bOmg_i^* \bB_{i,0}^\T \bsig^*\bE_j \notag\\
&= \sigma_{jj}^* -  \bE_j^\T  \bsig^*\bB_{i,0} \bOmg_i^* \bB_{i,0}^\T \bsig^*\bE_j.\\
\label{eAe}
\bE_i^\T \ba \ba^\T \bE_i 
&= \bE_i^\T(\bE_j - \bB_{i,0} \bOmg_i^* \bB_{i,0}^\T \bsig^*\bE_j)(\bE_j^\T - \bE_j^\T  \bsig^*\bB_{i,0} \bOmg_i^* \bB_{i,0}^\T) \bE_i \notag \\
&= \bE_i^\T \bE_j \bE_j^\T\bE_i 
- 2 \bE_i^\T \bB_{i,0} \bOmg_i^* \bB_{i,0}^\T \bsig^*\bE_j\bE_j^\T \bE_i \notag \\
&+ \bE_i^\T \bB_{i,0} \bOmg_i^* \bB_{i,0}^T \bsig^*\bE_j\bE_j^\T  \bsig^*\bB_{i,0} \bOmg_i^* \bB_{i,0}^\T \bE_i \notag \\
&= \bE_i^\T
\bB_{i,0} \bOmg_i^* \bB_{i,0}^\T \bsig^*\bE_j\bE_j^\T  \bsig^*\bw_i^* \notag \\
&= \bE_i^\T \bB_{i,0} \bOmg_i^* \bB_{i,0}^\T \bsig^*\bE_j\bE_j^\T  \bE_i \notag\\
& = 0.
\end{align}
Plugging (\ref{AsigA}) and (\ref{eAe}) in (\ref{varAsw1}), we get
 $\mbox{var}(\ba^\T \bold{V}_n \bw_i^*) 
 = (\omega_{ii}^* \ba^\T \bsig^*\ba + 2\bE_i^\T \ba \ba^\T \bE_i)/n.$\hfill  $\square$

\begin{lemma}\ 
\label{MaxA1}
$pr(\max\limits_{(i, j) \in A_1} \hat{D}_{ij}^{*2} \geq t_p)  
= o(1),$
where $t_p = t + 4 \log p - \log (\log p ).$
\end{lemma}
\noindent \textit{Proof:}
Lemma \ref{turnlem} gives us
$$\surd{n} \max\limits_{(i,j) \in A_1} \l
\bE_j^\T \bold{V}_n  \hat{\bw}_{i,0}  \l 
= \surd{n} \max\limits_{(i,j) \in A_1} \l
 \bE_j^\T \bold{V}_n \bw_i^* + \bE_j^\T  \bsig^*\hat{\bw}_{i,0} \l 
 + o_p(\surd{\log\ p}) .$$

We have
\begin{align*}
pr(\max\limits_{(i, j) \in A_1} \hat{D}_{ij}^{*2} \geq t_p)
 &= pr(\max \limits _{(i, j) \in A_1} \l \hat{D}_{ij}^{*} \l \geq \surd{t_p}) \\
 &=    pr(\sqrt{n}\max \limits _{(i, j) \in A_1} \l \frac{ \bE_j^\T \bold{V}_n \hat{\bw}_{i,0}}{\surd{(\omega_{ii}^* \sigma_{jj}^* })} \l \geq \surd{t_p})\\
&= pr(\surd{n} \max\limits_{(i,j) \in A_1} \frac{\l   \bE_j^\T \bold{V}_n  \bw_i^* +  \bE_j^\T  \bsig^*\hat{\bw}_{i,0}  \l}{\surd{(\omega_{ii}^* \sigma_{jj}^* })} + o_p(\surd{\log p})  \geq \surd{t_p})\\
&= pr(\surd{n} \max\limits_{(i,j) \in A_1} \frac{\l \bE_j^\T \bold{V}_n  \bw_i^* +  \bE_j^\T  \bsig^*\hat{\bw}_{i,0}  \l}{\surd{(\omega_{ii}^* \sigma_{jj}^* })}  \geq \surd{t_p}).\\
\end{align*}
We have
\begin{align}
\label{res1A1}
 \surd{\{n/(\omega_{ii}^* \sigma_{jj}^*)\}}
  \bE_j^\T  \bsig^*\hat{\bw}_{i,0} 
  &=  \surd{\{n/(\omega_{ii}^* \sigma_{jj}^*)\}} (\bE_j^\T  \bsig^*\hat{\bw}_{i,0} - \bE_j^\T  \bsig^*{\bw}_{i,0}) \notag\\
  &= \quad \surd{\{n/(\omega_{ii}^*\sigma_{jj}^*)\}} \bE_j^\T  \bsig^*\bB_{i,0} \Big\{ \bS_i^{-1} -  (\bB_{i,0}^\T \bsig^*\bB_{i,0})^{-1}\Big\} \mathbf{f}_i  \notag \\
  &= \quad \surd{\{n/(\omega_{ii}^*\sigma_{jj}^*)\}} \bE_j^\T  \bsig^*\bB_{i,0} ( \bS_i^{-1} -  \bOmg_i^*) \mathbf{f}_i.  
\end{align}

Applying Lemma 5 in Le and Zhong (2021), we have
\begin{align}
&\bE_j^\T  \bsig^*\bB_{i,0} \Big\{\bS_i^{-1} -  \bOmg_i^*\Big\} \mathbf{f}_i \notag \\
&= - \bE_j^\T  \bsig^*\bB_{i,0} \bOmg^* (\bS_i - \bsig_i^*) \bOmg_i^* \mathbf{f}_i  \notag \\
&- \bE_j^\T  \bsig^*\bB_{i,0} \bOmg^* (\bS_i - \bsig_i^*) (\bS_i^{-1} -  \bOmg_i^*)\mathbf{f}_i.\notag
\end{align}
Let us denote
 $R = \max_{(i,j)\in A_1} \l 
 \bE_j^\T  \bsig^*\bB_{i,0} \bOmg_i^* (\bS_i - \bsig_i^*)
 (\bS_i^{-1} -  \bOmg_i^*)\mathbf{f}_i  \l. $
Since 
$\vert \vert \bsig^*\vert \vert_1$ and $\vert \vert \bOmg_i^* \vert \vert_1$ are bounded, so $\vert \vert \bE_j^\T  \bsig^*\bB_{i,0} \bOmg_i^* \vert \vert_1 = O(1).$ Then we have
\begin{equation}
R \leq s_0 \max\limits_{1 \leq i,j \leq p} \l s_{ij} - \sigma_{ij}^* \l 
\max\limits_{1 \leq i,j \leq p} 
\l \hat{\omega}_{ij,0} - \omega_{ij}^*\l 
= O_p({s_0\log\ p}/{n}). \notag
\end{equation}

So
\begin{align}
\label{res2A1}
&\bE_j^\T  \bsig^*\bB_{i,0} [\bS_i^{-1} -  \bOmg_i^*] \bf_i \notag \\
&= - \bE_j^\T  \bsig^*\bB_{i,0} \bOmg_i^* \bB_{i,0}^\T\bS_n\bw_{i,0}  
+ O_p({s_0 \log\ p}/{n})\notag\\
&= - \bm^\T \bS_n\bw_{i,0}  + O_p({s_0 \log\ p}/{n}),
\end{align}
where
 $\bm^\T =\bE_j^\T\bsig^*\bB_{i,0} \bOmg_i^* \bB_{i,0}^\T.$
  Notice that 
  $\vert \vert \bm \vert \vert_1 = O(1)$ and 
   $\vert \vert \bw_{i,0} \vert \vert_1 = 1$.
   
 We have

 $$ \bm^\T \bS_n\bw_{i,0}  
 =  \bm^\T \bold{V}_n \bw_{i,0} + \frac{1}{n-1}\bm^\T \bold{V}_n \bw_{i,0}
  - \frac{n}{n-1} \bm^\T \bar{\bX} \bar{\bX}^\T \bw_{i,0}. $$

 In addition we have, 
 \begin{align*}
 \frac{1}{n-1} \l \bm^\T \bold{V}_n \bw_{i,0} \l & \leq \frac{1}{n-1} \l \bm^\T (\bold{V}_n - \bsig^*) \bw_{i,0} \l + \frac{1}{n-1} \l \bm^\T \bsig^*\bw_{i,0} \l  \notag \\
 & = \frac{1}{n-1} \l \bE_j^\T\bsig^* \bB_{i,0} \bOmg_i^* \{ \bB_{i,0}^\T (\bold{V}_n - \bsig^*) \bB_{i,0} \} \bw_{i1,0} \l + O(1/n) \notag \\
& = O_p(s_0/n) 
 \end{align*}
 In other words, we have
 \begin{align}
 \label{res3A1}
 \frac{1}{n-1} \l \bm^\T \bold{V}_n \bw_{i,0} \l = O_p(s_0/n)
 \end{align}

 Further more, we have 
 \begin{align}
 \label{res4A1}
 \frac{n}{n-1} \bm^\T \bar{\bX} \bar{\bX}^\T \bw_{i,0} 
 \leq \frac{n}{n-1} \max \limits_{i = 1,\ldots,p}\bar{X}_i^2\ 
 \vert \vert \bm \vert \vert_1 \ \vert \vert \bw_{i,0} \vert \vert_1 
 = O_p({\log p}/{n}).
 \end{align}

Applying (\ref{res1A1}), (\ref{res2A1}), (\ref{res3A1}), and (\ref{res4A1}), we get
 \begin{align*}
 pr(\max\limits_{(i, j) \in A_1} \hat{D}_{ij}^{*2} \geq t_p) 
 &= pr(\max\limits_{(i,j) \in A_1} 
 \frac{ \surd{n}  \l   \bE_j^\T \bold{V}_n  \bw_{i,0}
  +  \bE_j^\T  \bsig^*\hat{\bw}_{i,0}  \l}{\surd{(\omega_{ii}^* \sigma_{jj}^*) }} \geq \surd{t_p}) \\
&= pr( \max\limits_{(i,j) \in A_1}
 \frac{ \surd{n}  \l   \bE_j^\T \bold{V}_n  \bw_{i,0} 
 - \bm^\T \bS_n\bw_{i,0} \l}{\surd{(\omega_{ii}^* \sigma_{jj}^*) }} \geq \surd{t_p})\\
&= pr\{\max\limits_{(i,j) \in A_1} 
\frac{ \surd{n}  \l  (\bE_j^\T - \bm^\T) \bold{V}_n  \bw_{i,0}  \l}{\surd{(\omega_{ii}^* \sigma_{jj}^* )}} \geq \surd{t_p}\}\\
&= pr( \max\limits_{(i,j) \in A_1}
 \frac{ \sqrt{n}  \l   \ba^\T \bold{V}_n \bw_{i,0}  \l}{\surd{(\omega_{ii}^* \sigma_{jj}^* )}} \geq \surd{t_p})\\
 \end{align*}
  where $\ba^\T = \bE_j^\T - \bE_j^\T  \bsig^*\bB_{i,0}\bOmg_i^* \bB_{i,0}^\T$. 
   
Lemma \ref{VarASw} gives us
 $$\mbox{var}(\ba^\T \bold{V}_n \bw_{i,0})  = 
 ( \omega_{ii}^* \sigma_{jj}^* 
 -  \omega_{ii}^* \bE_j^\T  \bsig^* 
\bB_i \bOmg_i^* \bB_i^\T \bsig^*\bE_j)/n  
\leq  (\omega_{ii}^* \sigma_{jj}^*)/n.$$ 
where we notice that 
$\omega_{ii}^* \bE_j^\T  \bsig^* \bB_{i,0}
 \bOmg_i^* \bB_{i,0}^\T \bsig^*\bE_j \geq 0$, since $\bOmg_i^*$ is positive definite.
 
By the CLT, we get 
 $\ba^\T \bold{V}_n \bw_{i,0} = ( \sum\limits_{k=1}^n \ba^\T \bX_k \bX_k^\T \bw_{i,0} )/n
 \sim N(0, \mbox{var}(\ba^\T \bold{V}_n \bw_{i,0}))$.
 
In addition $\mbox{card}(A_1)=  o(p^2) $, this gives us
\begin{align*}
pr(\max\limits_{(i,j) \in A_1}
 \frac{ \surd{n}  \l   \bE_j^\T \bold{V}_n  \bw_{i,0}
  +  \bE_j^\T  \bsig^*\bB_i \hat{\bw}_{i,0}  \l}{\surd{(\omega_{ii}^* \sigma_{jj}^*) }} \geq & \surd{t_p}) 
 = pr( \max\limits_{(i,j) \in A_1} 
 \frac{ \surd{n}  \l   \ba^\T \bold{V}_n \bw_{i,0}  \l}{\surd{(\omega_{ii}^* \sigma_{jj}^* )}} \geq \surd{t_p})\\
&\leq pr\{ \max\limits_{(i,j) \in A_1} 
\frac{\l \ba^\T \bold{V}_n  \bw_{i,0}  \l}
{\surd{\{\mbox{var}(\ba^\T \bold{V}_n \bw_{i,0}) \}}} \geq \surd{t_p}\}\\
&\leq \sum \limits_{(i, j) \in A_1}
 pr\{  \frac{   \l   \ba^\T \bold{V}_n  \bw_{i,0}  \l}
 {\surd{\{\mbox{var}(\ba^\T \bold{V}_n \bw_{i,0}) \}}} \geq \surd{t_p}\}\\ 
&\leq o(p^{2}) e^{-t_p/{2}} 
= o(p^{2}) e^{-2 \log \ p} 
= o(1)
\end{align*}
where the last inequality is due to Gaussian tail Inequality. The lemma is proved \hfill $\square$

\begin{lemma}\ 
\label{eSigB}
$\max\limits_{(i,j) \in A/B_0} \l \bE_j^\T  \bsig^*\bB_{i,0} (\bS_i^{-1} - \bOmg_i^*) \mathbf{f}_i \l =  o_p(\surd{({\log \ p}/{n})})$.
\end{lemma}
\noindent \textit{Proof:} When the underlying network structure is a factor model, it can be seen that 
$ \bE_j^\T  \bsig^*\bB_{i,0}(\bS_i^{-1} - \bOmg_i^*) \mathbf{f}_i = 0, \ \mbox{for all} \ (i, j) \in A/B_0.$ 
So the lemma is satisfied.

Now we consider the case for other network structures with their covariance matrix and precision matrix satisfying conditions (C1).
Let us denote 
$\bb^\T = \bE_j^\T  \bsig^*\bB_{i,0} = (\sigma_{ji_1}^*,\ldots,\sigma_{ji_{s_0}^*})$.
  where $i_1,\ldots,i_{s_0} $ are nonzero positions at column $\bw_{i,0}$ of the precision matrix $\bOmg$.
  Since $\vert \vert \bb \vert \vert_1 = O(1)$, applying Theorem 4 in Le and Zhong (2021), we have
\begin{align}
\label{ANV}
\surd{(n/a_{ij})}  \bb^\T (\bS_i^{-1} - \bOmg_i^*) \mathbf{f}_i   \sim AN(0,1)
\end{align}
where 
$a_{ij} = \mbox{var}(\bb^\T \bOmg_i^* \bX_{1i} \bX_{1i}^\T \bOmg_i^*\mathbf{f}_i),  \ \mbox{for all} \ (i,j) \in A/B_0.$

Denote
 $\bOmg_i^*  = (\gamma_{ij})_{s_0 \times s_0}.$
By Lemma 7 in Le and Zhong (2021), we get
 \begin{align}
\label{var.aij1}
 a_{ij} 
 &= \mbox{var}(\bb^\T \bOmg_i^* \bX_{1i} \bX_{1i}^\T \bOmg_i^* \mathbf{f}_i)  
 = \bb^\T \bOmg_i^* \bsig_i^* \bOmg_i^* \bb \mathbf{f}_i^\T \bOmg_i^* \bsig_i^* \bOmg_i^* \mathbf{f}_i 
 + (\bb^\T \bOmg_i^* \bsig_i \bOmg_i^* \mathbf{f}_i)^2 \notag \\
 &= \bb^\T  \bOmg_i^* \bb \mathbf{f}_i^\T  \bOmg_i^* \mathbf{f}_i + (\bb^\T \bOmg_i^* \mathbf{f}_i)^2 \notag \\ 
 &=\omega_{ii}^* \sum\limits_{k,l \in \{i_1,\ldots,i_{s_0}\}}\sigma_{jk}^*\sigma_{jl}^*\gamma_{kl} 
 + \sum\limits_{k,l \in \{i_1,\ldots,i_{s_0}\}} \sigma_{jk}^*\sigma_{jl}^*\gamma_{ik}\gamma_{il}
 \end{align}
On $ A/B_0$ we have
\be
\label{var.aij2}
\sigma_{jk}^*\sigma_{jl}^* = o(1/s_0^2), \ \mbox{for all} \ k, l \in\{i_1,\ldots, i_{s_0}\}, j \neq k, l.
\ee
The facts (\ref{var.aij1}) and (\ref{var.aij2}) give us 
$ a_{ij} = o(1),  \ \mbox{for all} \ (i, j) \in A/B_0$. 

 Let us denote
 $a = \max_{(i,j) \in A/B_0} \surd{a_{ij}},$ so $a = o(1).$ 
Applying (\ref{ANV}), we have
 \begin{align}
 &pr\Big\{ \max\limits_{(i,j) \in A/B_0}
  \l \bE_j^\T  \bsig^*\bB_{i,0}(\bS_i^{-1} - \bOmg_i^*) \mathbf{f}_i \l \geq t\Big\}\notag \\
  & \leq pr\Big\{ \max\limits_{(i,j) \in A/B_0} \l \surd{(n/a_{ij})}\bE_j  \bsig^*\bB_{i,0}(\bS_i^{-1} - \bOmg_i^*) \mathbf{f}_i \l \geq (\surd{n}t/a)\Big\} \notag\\
&\leq  p^2 \exp\{- {nt^2}/{(2a^2)}\}. \notag
\end{align}
  Choose $t = M a \surd{\{(\log p)/n\}}$ for $M > 0$ sufficient large, then we have
  \begin{align}
 pr\Big\{ \max\limits_{(i,j) \in A/B_0} 
 \l \bE_j  \bsig^*\bB_{i,0}(\bS_i^{-1} - \bOmg_i^*) \mathbf{f}_i \l \geq M a \surd{\frac{\log p}{n}}\Big\}  
 &\leq p^2 \exp(- \frac{nM^2 a^2 \log \ p}{2a^2n}) \notag \\
 &= p^2 \exp(\log \ p^{-M/2}) \notag \\
 &= p^{2- M/2} \rightarrow 0. \notag
 \end{align}
 Or 
 $$ \max\limits_{(i,j) \in A/B_0} 
 \l \bE_j^\T  \bsig^*\bB_{i,0}(\bS_i^{-1} - \bOmg_i^*) \mathbf{f}_i \l 
 = O_p\{a \surd{{(\log p}/{n})}\}
 = o_p\{\surd{({\log p}/{n})}\}.$$
 The lemma is verified. \hfill $\square$

 \begin{lemma}\label{TechLem}
\be
\label{key}
\sum\limits_{1 \leq k_1 <\ldots< k_d \leq q } 
pr\{\l \bN_d \l_{\min} \geq t_p^{1/2} 
\pm \epsilon_n(\log p)^{-1/2} \} 
= \frac{1}{d!}\{\frac{1}{\surd{(2\pi)}}\exp(-\frac{t}{2})\}^d\{1 + o(1)\},
\ee
\end{lemma}
where $\bN_d = (N_{k_1},\ldots,N_{k_d})^\T$ is a $d$-dimensional multivariate Gaussian random variable with mean vector $0$ and covariance matrix 
$\mbox{cov}(\bN_d) = \mbox{cov}(\bW_1)$. Here $\bW_1$ is the random variable defined as in equation (\ref{Wlvariable}) of the proof of Theorem \ref{generalTheo}.

\noindent \textit{Proof:}
Notice that for $X \sim N(0,1)$, we have
 $$pr(\l X \l \geq x ) = 2\{1 + o(1)\}\frac{\exp^{-x^2/2}}{x\surd{(2\pi)}}.$$
So when $d =1$, we get
\begin{align*}
pr\Big\{ \l \bN_1 \l_{\min} \geq t_p^{1/2} \pm \epsilon_n(\log p)^{-1/2} \Big\} 
&=  2\{1 + o(1)\}\frac{\exp^{-t_p/2}}{\surd{t_p}\surd{(2\pi)}}\\
 &= \{1 + o(1)\}\frac{2\exp(-t/2 - 2  \mbox{log} p)(\mbox{log} p)^{1/2}}{2 \surd{(\mbox{log} p)} \surd{(2 \pi)}}\\
  &= \{1 + o(1)\}\frac{p^{-2}\exp^{-t/2}}{\surd{(2\pi)}}.
\end{align*}
This leads
\be
\label{de1}
\sum\limits_{1 \leq k_1 \leq q } 
pr\{ \l \bN_1 \l_{\min} \geq t_p^{1/2} 
\pm \epsilon_n(\mbox{log} p)^{-1/2} \} 
= \frac{\exp^{-t/2}}{\surd{(2\pi)}}\{1 + o(1)\}.
\ee
 The lemma is verified for $d =1$.
 
Let us consider when $d \geq 2$, we need to show that
\be
\label{Techd2}
pr\Big\{\l \bN_d \l_{\min} \geq t_p^{1/2} \pm \epsilon_n(\log p)^{-1/2} \Big\} 
= \{1 +o(1)\} \Big\{ \frac{1}{\surd{(2\pi)}}\exp(-\frac{t}{2})^dp^{-2d}\Big\}.\ee
Let $\bR = (\rho_{ij})_{p \times p}$ be the correlation matrix and 
$\tilde{\bOmg}= (\tilde{\omega}_{ij})_{p \times p}$ is the standardized version of the precision matrix $\bOmg^*$ 
where
 $\tilde{\omega}_{ij} = {\omega_{ij}^*}/{\surd{(\omega_{ii}^* \omega_{jj}^* })}$.
For a fixed constant $\alpha_0 > 0$, for $j = 1,2,\ldots,p$, define
 $$s_j = s_j(\alpha_0) = \mbox{card}\{ i: \l \rho_{ij} \l \geq (\log p)^{-1 - \alpha_0} \}, h_j = h_j(\alpha_0)
  = \mbox{card}\{ i: \l \tilde{\omega}_{ij} \l \geq (\log p)^{-1 - \alpha_0} \}.$$
We need two following conditions for our proof
 \be  
\label{eorder}
\max\limits_{j=1,\ldots,p}s_j(\alpha_0)
 = o(p^\gamma), \max\limits_{j=1,\ldots,p}h_j(\alpha_0)
  = o(p^\gamma), \forall \gamma > 0.
\ee
\be
\label{corrbound}
\mbox{There exists some} \ r \in (0, 1), \rho_{ij} < r, \tilde{\omega}_{ij} < r,  \ \mbox{for all} \ 1\leq i \neq j \leq p.
\ee
Notice that the above conditions are mild. Condition (\ref{eorder}) is met if $\bR$, and $\bOmg^*$ has maximum 
eigenvector bounded from the above. And condition (\ref{corrbound}) met once the off diagonal elements of $\bR$ and $\tilde{\bOmg}$ are bounded by $r$. 
 We have
  $E Z_{lk_1}Z_{lk_2} 
  =  \bE_{j_{k_2}}^\T  \bE_{i_{k_1}} \bE_{j_{k_1}}^\T \bE_{i_{k_2}} 
  +  \sigma_{j_{k_1}j_{k_2}}^*  \omega_{i_{k_1}i_{k_2}}^*.$
  When either $i_{k_1} \neq j_{k_2}$ or $i_{k_2} \neq j_{k_1}$, then
   $E Z_{lk_1}Z_{lk_2} 
   = \sigma_{j_{k_1}j_{k_2}}^*  \omega_{i_{k_1}i_{k_2}}^* $. Notice that
on 
$A/B_0$, we have $\omega_{i_{k_1}j_{k_1}}^* = \omega_{i_{k_2}j_{k_2}}^*
= 0$, 
so when 
$i_{k_1} = j_{k_2}$ and $i_{k_2} = j_{k_1}$, we get 
$E Z_{lk_1}Z_{lk_2} = \sigma_{i_{k_1}j_{k_1}}^*   \omega_{i_{k_1}j_{k_1}}^*  + 1 = 1$.

For two different pairs $(i_a, j_a), (i_b, j_b)$, we can establish a graph defined by
$G_{i_aj_ai_bj_b} = (V_{i_aj_ai_bj_b}, E_{i_aj_ai_bj_b})$  
where $V_{i_aj_ai_bj_b} = \{ i_a, j_a, i_b, j_b \}$
 is the set of vertices and $E_{i_aj_ai_bj_b}$ is the set of edges. 
 We say there is an edge (connection) between 
 $i \neq j \in \{ i_a, j_a, i_b, j_b\}$ 
 if 
 $ \l \rho_{ij} \l \geq (\log p)^{-1 - \alpha_0} $
  or 
  $ \l \tilde{\omega}_{ij} \l \geq (\log p)^{-1 - \alpha_0}$.
  
We say $G_{abcd}$ is a $k$-vertices graph ($k$-G) if the number of different vertices is $k$, in our case $k \in \{2, 3, 4\}$. 
 For sake of convenient, we denote ``3G-1E" for a three vertices graph when either $\rho_{i_{a} i_{b}}$  
 or $\tilde{\omega}_{j_{a} j_{b}}$ form an edge. We denote ``4G- 2E" for a four vertices graph when both
  $\rho_{i_{a} i_{b}}$  
  and
   $\tilde{\omega}_{j_{a} j_{b}}$ form edges.
We say a graph 
$G = G_{i_{m_1}j_{m_1}i_{m_2}j_{m_2}}$ satisfy condition ($\star$) if\\
$(\star): $
 Either 
 $\tilde{\omega}_{i_{m_1}i_{m_2}} \leq (\log p)^{-1 - \alpha_0}$
  or
   $\rho_{j_{m_1}j_{m_2}} \leq (\log p)^{-1 - \alpha_0}$.

\noindent \textit{Remark:} Those graphs satisfying ($\star$) also satisfy
 \begin{align}
 \label{boundedcov}
    \mbox{cov}(\tilde{Z}_{lm_1}, \tilde{Z}_{lm_2} ) \rightarrow  \rho_{j_{m_1}j_{m_2}}  \tilde{\omega}_{i_{m_1}i_{m_2}} 
 = O\{(\log p)^{-1 - \alpha_0}\}. 
 \end{align}
As shown above for any two different pairs 
$(i_{k_1}, j_{k_1}), (i_{k_2}, j_{k_2})$ we have
 $$ \mbox{cov}(\tilde{Z}_{lk_1}, \tilde{Z}_{lk_2} ) \rightarrow \surd{\{ 1 / (\omega_{i_{k_1}i_{k_1}}^* 
 \omega_{i_{k_2}i_{k_2}}^*  \sigma_{j_{k_1}j_{k_1}}^* \sigma_{j_{k_2}j_{k_2}}^*) \} } EZ_{lk_1}Z_{lk_2}. $$
For any matrices  
$ \bA = (a_{ij})_{p\times p}, \bB  = (b_{ij})_{p \times p} = \bA^{-1}$, page 472 in Robinson $\&$ Wahten (1992) tells us
$$b_{ii} \geq a_{jj}/(a_{ii}a_{jj} - a_{ij}^2), \
\mbox{for any} \ 1 \leq i \neq j \leq p.$$ 
 This gives us
  $$\omega_{i_{k_1}i_{k_1}}^* \omega_{j_{k_1}j_{k_1}}^* \sigma_{i_{k_1}i_{k_1}}^* \sigma_{j_{k_1}j_{k_1}}^* 
  \geq \{( \omega_{i_{k_1}i_{k_1}}^* \omega_{j_{k_1}j_{k_1}}^*)/(\omega_{i_{k_1}i_{k_1}}^* \omega_{j_{k_1}j_{k_1}}^* - \omega_{i_{k_1}j_{k_1}}^*)\}^2 > 1/r,$$ 
  for some $r \in (0, 1)$.
 So for a 2G- 1E of two pairs 
 $(i_{k_1}, j_{k_1}), (j_{k_1}, i_{k_1})$ 
 we have
 \be
 \label{lessr1}
  \mbox{cov}(\tilde{Z}_{lk_1}, \tilde{Z}_{lk_2} )
  \rightarrow \surd{\{ 1/ ( \omega_{i_{k_1}i_{k_1}}^* \omega_{j_{k_1}j_{k_1}}^* \sigma_{i_{k_1}i_{k_1}}^* \sigma_{j_{k_1}j_{k_1}}^*) \}} < r, 
  \ee
   for some $r \in (0,1)$. 
 
For "4G-2E" or "3G-1E" of two different pairs 
 $(i_{k_1}, j_{k_1}), (i_{k_2}, j_{k_2})$ we have
 
 \begin{align}
 \label{lessr2}
  \mbox{cov}(\tilde{Z}_{lk_1}, \tilde{Z}_{lk_2} )&\rightarrow \surd{\{ 1/ (\omega_{i_{k_1}i_{k_1}} \omega_{i_{k_2}i_{k_2}}^* \sigma_{j_{k_1}j_{k_1}}^* \sigma_{j_{k_2}j_{k_2}}^*) \}}\sigma_{j_{k_1}j_{k_2}}^*  \omega_{i_{k_1}i_{k_2}}^* \notag \\
  &= \rho_{j_{k_1}j_{k_2}} \tilde{\omega}_{i_{k_1}i_{k_2}} < r,
\end{align}

 for some $0 < r <1$.

Now we define the following sets
$I = \{1 \leq k_1 < k_2 <\ldots< k_d \leq q \}$, 
$d$ is a fixed positive integer.
$I_0 = \{1 \leq k_1 < k_2 <\ldots< k_d \leq q:$ 
for some $m_1 \neq m_2 \in {k_1,\ldots,k_d},  G = G_{i_{m_1}j_{m_1}i_{m_2}j_{m_2}} $
 does not satisfy $(\star)\}$.
$I_0^c = \{1 \leq k_1 < k_2 <\ldots< k_d \leq q:$ 
for any $m_1 \neq m_2 \in {k_1,\ldots,k_d},  G = G_{i_{m_1}j_{m_1}i_{m_2}j_{m_2}} $ 
satisfies $(\star)\}$.

Notice that $I = I_0 \cup I_0^c$. 
For any subset S of $\{ k_1,\ldots,k_d \}$, we say that S satisfies $ (\star \star)$ if
$(\star \star)$ for any 
$m_1 \neq m_2 \in S, G_{i_{m_1}j_{m_1}i_{m_2}j_{m_2}}$ satisfies $ (\star)$.
For $2 \leq l \leq d$, let
$I_{0l} = \{1 \leq k_1 < k_2 <\ldots< k_d \leq q:$
 $\mbox{card}(S) = l$, where S is largest subset of 
 ${k_1<\ldots< k_d},$ satisfies $(\star \star)\}$.

$I_{01} = \{1 \leq k_1 < k_2 <\ldots< k_d \leq q:$ for any 
$m_1 \neq m_2 \in {k_1,\ldots,k_d}, 
G = G_{i_{m_1}j_{m_1}i_{m_2}j_{m_2}} $ 
does not satisfy $(\star)\}$.
So $I_0^c = I_{0d}, I_0 = \cup_{l=1}^{d-1}I_{0l}$.
 
\noindent \textit{Claim:}
\be
\label{I0l}
\mbox{card}(I_{0l}) \leq C_d q^{l + 2\gamma(d -l)},
\ee 
where $C_d$ is a constant depends only on $d$.
In addition
\be
\label{I0C}
\mbox{card} (I_0^c) = \{1 +o(1)\}C_q^d.
\ee
\noindent \textit{Proof:}
First, we verify (\ref{I0l}), 
$\mbox{card}(I_{0l}) \leq C_dq^{l + 2\gamma(d-l)}$.
There are at most $C_q^l$ ways of choosing S with cardinality $l$ from $1,2...,q$. 
For a fixed element "a" in S, there is at most 
$p^\gamma p^\gamma = p^{2\gamma}$ 
choices for "b" which satisfies  $G_{i_aj_ai_bj_b}$ not satisfies $(\star )$. 
So there will be at most $Clp^{2\gamma}$ choices for values "b" not go with l elements of S for properties $(\star)$.
So we get 
$\mbox{card}(I_{0l}) \leq C_q^l (Clp^{2\gamma})^{d-l} \leq C_dq^{l + 2\gamma(d-l)}$ . 

The claim (\ref{I0l}) is verified. 

Second, we show ($\ref{I0C}$), 
$\mbox{card} (I_0^c) = \{1 +o(1)\}C_q^d.$
We have $\mbox{card}(I) = C_q^d$, since we are choosing d numbers from q number without order. 
$$\mbox{card}(I_0) \leq \sum\limits_{l=1}^{d-1}\mbox{card}(I_{0l}) 
\leq \sum\limits_{l=1}^{d-1} C_dq^{l + 2\gamma(d-l)} 
= o(q^d) = o(C_q^d).$$ 
This gives us
$$\mbox{card} (I_0^c) = C_q^d - o(C_q^d) = \{1 +o(1)\}C_q^d.  $$

This clarifies ($\ref{I0C}$).

\noindent We claim that \textit{the follows are true:}

\be
\label{Minor}
\sum\limits_{I_0 }pr\{\l \bN_d \l_{\min}
 \geq t_p^{1/2} 
 \pm \epsilon_n(\log p)^{-1/2} \} = o(1)
\ee
and
\be
\label{Major}
\sum\limits_{I_0^c }pr\{ \vert \bN_d \vert_{\min} 
\geq t_p^{1/2} 
\pm \epsilon_n(\log p)^{-1/2} \} 
= \frac{1}{d!}\{ \frac{1}{\surd{(2\pi)}}\exp(-\frac{t}{2})\}^d\{1 + o(1)\},
\ee
\noindent \textit{Proof:}
Before verify (\ref{Minor}), we need to divide our set $I_{0l}$ a bit further.
For $1 \leq a \neq b \leq q $, we define 
$d((i_a, j_a), (i_b, j_b)) = 1$, if $G_{i_aj_ai_bj_b} $ does not satisfies $(\star)$;
 $d((i_a, j_a), (i_b, j_b)) = 0$ otherwise. 
We further divide $I_{0l}$ as the following.
Let $(k_1, k_2, \ldots, k_d) \in I_{0l}$ and
 let $S_\star \subset (k_1,\ldots, k_d)$ be the largest cardinality subset satisfying $(\star \star)$ (if there are more than two subsets attain the largest cardinality, then we choose any of them).
Define
 $I_{0l1} = \{ (k_1, \ldots, k_d) \in I_{0l}: $ 
 there exists an $a \notin S_\star$, 
 such that for some $b_1 \neq b_2 \in S_\star$ with, $d ((i_a, j_a), (i_{b_1}, j_{b_1})) = 1,$
  and 
  $ d ((i_a, j_a), (i_{b_2}, j_{b_2})) = 1 \}$,
$I_{0l2} = I_{0l}/ I_{0l1}$.
We have $I_{011} = \emptyset, I_{012} = I_{01}$.
 Recall that d fixed and $l \leq d-1$.
We can show that
\be
\label{I0l11}
\mbox{card}(I_{0l1}) \leq Cq^{l-1 + 2\gamma(d- l +1)}.
\ee
\be
\label{I0l12}
\mbox{card}(I_{0l2}) \leq C_dq^{l + 2\gamma(d-l)}.
\ee
Write $S_\star = (b_1, b_2,\ldots,b_l)$,
for $(k_1, \ldots, k_d) \in I_{0l2}$. 
Since there exists an $a \notin S_\star$ such that 
$d ((i_a, j_a), (i_{b_1}, j_{b_1})) = 1$ 
and $ d ((i_a, j_a), (i_{b_2}, j_{b_2})) = 1 $
 for some $b_1 \neq b_2 \in S_\star$. 
 We consider $b_1$ is the first element in $S_\star$, there are at most q ways to choose $b_1$. There are at most $p^{2\gamma}$ to choose the second element in $S_\star$ not goes with "a" for $\star$. 
 For the other $l-2$ elements in $S_\star$ there are at most $C_q^{l-2}$ ways of choosing. For the rest $d- l $ elements outside $S_\star$, there are at most $p^{2\gamma(d-l)}$ ways of choosing. 
So on the whole we have  
 $$\mbox{card}(I_{0l1}) \leq q p^{2\gamma}
C_q^{l-2}p^{2\gamma(d-l)} \leq Cq^{l-1+ 2\gamma(d-l+1)},$$
which verifies (\ref{I0l11}).
We have

\be
\label{I0l2w1}
\mbox{card}(I_{0l}) = \mbox{card}(I_{0l1}) + \mbox{card}(I_{0l2}) 
\leq C_dq^{l + 2\gamma(d-l)}.
\ee
 On the other hand 
 \be
 \label{I0l2w2}
 \mbox{card}(I_{0l1})\leq Cq^{l-1+ 2\gamma(d-l+1)} = o(q^{l}).
 \ee
  Applying (\ref{I0l2w1}) and (\ref{I0l2w2}), we get (\ref{I0l12}).

We go back to check our claim (\ref{Minor})
$$\sum\limits_{I_0 }pr\{ \l \bN_d \l_{min} 
\geq t_p^{1/2} \pm \epsilon_n(\log p)^{-1/2} \} = o(1).$$ 
On $I_{0l}$ we have 
For any $k_1, \ldots, k_d \in I_{0l}$, 
write $S_{\star} = (b_1, b_2,\ldots,b_l)$, 
$\bU_l$ is the covariance matrix of $(N_{b_1},\ldots,N_{b_l})$, 
then 
$\vert \vert \bU_l - \bI_l \vert \vert = O\{ (\log  p)^{-1-\alpha_0}\} $ (by (\ref{boundedcov})). As a result, we also have $\l \bold{U}_l \l \rightarrow 1$ as $p \rightarrow \infty$. 
Let us denote 
$\l \by \l_{\max} = \max_{1 \leq i \leq l} \l y_i \l$,
  for $\by = (y_1,\ldots,y_l)^\T$ and $x_p = t_p^{1/2} \pm \epsilon_n(\log p)^{-1/2}$. 
We claim that
\begin{align}
\label{mingmaxg}
 &\frac{1}{(2\pi)^{l/2} \l 
 \bU_l\l^{1/2}}\int_{\l \by \l_{\min} \geq x_p, \l \by \l_{\max} \geq (\log p)^{1/2 
 + \alpha_0/4}}\exp(-\frac{1}{2}\by^\T \bU_l^{-1} \by) d\by \notag \\
 & = O\Big[\exp\{-\frac{1}{4} (\log p)^{1 +\alpha_0/2}\}\Big]
 \end{align}
 and
\begin{align}
 \label{c2}
 & \frac{1}{(2\pi)^{l/2} \l 
 \bU_l\l^{1/2}} \int_{\l \by \l_{\min} \geq x_p, \l \by \l_{\max} \leq (\log p)^{1/2  + \alpha_0/4}} \exp(-\frac{1}{2}\by^\T \bU_l^{-1} \by) d\by \notag \\
   &= \frac{1 + O(\log p)^{-\alpha_0/2}}{(2\pi)^{l/2} }\int_{\l \by \l_{\min} \geq x_p, \l \by \l_{\max} \leq (\log p)^{1/2 + \alpha_0/4}}\exp(-\frac{1}{2}\by^\T \by) d\by.
\end{align}

First, we check (\ref{mingmaxg}). 
 
 We have
\begin{align}
&\frac{1}{(2\pi)^{l/2} \l 
\bU_l \l^{1/2}}\int_{\l \by \l_{\min}
 \geq x_p, \l \by \l_{\max} \geq (\log p)^{1/2
  + \alpha_0/4}}  \exp(-\frac{1}{2}\by^\T \bU_l^{-1} \by)d\by \notag \\
  &=  pr \{\l \bN_d \l_{\min} \geq x_p, \l \bN_d \l_{\max} \geq (\log p)^{1/2 + \alpha_0/4}\} \notag \\
& \leq \sum\limits_{i=1}^l pr\{\l N_i \l \geq (\log p)^{1/2 + \alpha_0/4} \} \notag \\
 &= O\Big[  \exp\{-\frac{1}{2}(\log p)^{1 + \alpha_0/2}\}\Big] \notag \\
 &= O\Big[\exp\{-\frac{1}{4}(\log p)^{1 + \alpha_0/2}\}\Big] \notag,
\end{align}
which closes (\ref{mingmaxg}).

\noindent We now verify (\ref{c2}).
  We have
  \begin{align*}
     \vert \vert \bU_l^{-1} - \bI_l \vert \vert \leq \vert \vert \bU_l^{-1} \vert \vert \vert \vert \bU_l 
  - \bI_l \vert \vert = O\{(\log\ p)^{-1 - \alpha_0}\}.  
  \end{align*}
 
So, on set 
$\{ {\l \by \l_{\min} \geq x_p,
 \l \by \l_{\max} \leq (\log\ p)^{1/2 + \alpha_0/4}} \}$, using Taylor expansion we have:
\begin{align*}
    \exp\{-\frac{1}{2}\by^\T(\bU_l^{-1} - \bI_l) \by\}
= 1 + O\{-\frac{1}{2}\by^\T(\bU_l^{-1} - \bI_l)\by\}
= 1 +  O\{(\mbox{log} p)^{-\alpha_0/2}\}.
\end{align*}

Therefore,
\begingroup
\allowdisplaybreaks
 \begin{align*}
  &\frac{1}{(2\pi)^{l/2} \l 
  \bU_l \l^{1/2}}\int_{\l \by \l_{\min} \geq x_p, \l \by \l_{\max} \leq (\mbox{log} p)^{1/2 + \alpha_0/4}} 
  \exp(-\frac{1}{2}\by^\T \bU_l^{-1} \by) d\by \notag \\
 &= \frac{1}{(2\pi)^{l/2} \l 
 \bU_l \l^{1/2}}\int_{\l \by \l_{\min} \geq x_p, \l \by \l_{\max} \leq (\mbox{log} p)^{1/2 + \alpha_0/4}}
 \exp\{-\frac{1}{2}\by^\T(\bU_l^{-1}- \bI_l)\by\}
 exp(-\frac{1}{2}\by^\T \by) d\by \notag \\
&= \frac{1}{(2\pi)^{l/2} \l \bU_l \l^{1/2}}\int_{\l \by \l_{\min} \geq x_p, \l \by\l_{\max} \leq (\mbox{log} p)^{1/2 + \alpha_0/4}}
\exp(-\frac{1}{2}\by^\T\bU_l^{-1} \by) d\by \notag \\
&= \frac{1 + O((\mbox{log} p)^{-\alpha_0/2})}{(2\pi)^{l/2} }\int_{\l \by \l_{\min} \geq x_p, \l \by \l_{\max} 
\leq (\mbox{log} p)^{1/2 + \alpha_0/4}}
\exp(-\frac{1}{2}\by^\T \by) d\by . 
\end{align*}
\endgroup
So we proved (\ref{c2}).
The two claims are proved, we come back to show (\ref{Minor}).
\begingroup
\allowdisplaybreaks
\begin{align}
\label{Ndlower}
&pr\{\l \bN_d \l_{\min}  \geq t_p^{1/2}
 \pm \epsilon_n  (\mbox{log} p)^{-1/2}\} 
 \leq pr(\l N_{b_1} \l \geq x_p,\ldots, \l N_{b_l} \l \geq x_p)\notag \\
  &=  \frac{1}{(2\pi)^{l/2} \l 
  \bU_l \l^{1/2}}\int_{\l \by \l_{\min} \geq x_p} 
  \exp(-\frac{1}{2}\by^\T \bU_l^{-1} \by) d\by \notag \\
&= \frac{1}{(2\pi)^{l/2} \l \bU_l\l^{1/2}}
\int_{\l \by \l_{\min} \geq x_p, \l \by \l_{\max} \leq (\log p)^{1/2 + \alpha_0/4}} \exp(-\frac{1}{2}\by^\T \bU_l^{-1} \by) d\by\notag \\
&\quad+ \frac{1}{(2\pi)^{l/2} \l \bU_l \l^{1/2}}
\int_{\l \by \l_{\min} \geq x_p, \l \by \l_{\max} \geq (\log p)^{1/2 + \alpha_0/4}}
\exp(-\frac{1}{2}\by^\T \bU_l^{-1} \by) d\by \notag \\
&= \frac{1}{(2\pi)^{l/2} \l \bU_l\l^{1/2}}
\int_{\l \by \l_{\min} \geq x_p, \l \by \l_{\max} \leq (\log p)^{1/2 + \alpha_0/4}} \exp(-\frac{1}{2}\by^\T \bU_l^{-1} \by) d\by \notag \\
&\quad+ O\Big[\exp\{-\frac{1}{4} (\mbox{log} p)^{1 +\alpha_0/2}\}\Big] \notag \\
&= \frac{1 + O(\mbox{log} p)^{-\alpha_0/2}}{(2\pi)^{l/2} }
\int_{\l \by \l_{\min} \geq x_p, \l \by \l_{\max} \leq (\log p)^{1/2 + \alpha_0/4}}\exp(-\frac{1}{2}\by^\T \by) d\by \notag \\
&\quad+ O\Big[\exp\{-\frac{1}{4} (\mbox{log} p)^{1 +\alpha_0/2}\}\Big]\notag \\
&= \frac{1 + O(\mbox{log} p)^{-\alpha_0/2}}{(2\pi)^{l/2} }
\int_{\l \by \l_{\min} \geq x_p}\exp(-\frac{1}{2}\by^\T \by) d\by 
+ O\Big[\exp\{-\frac{1}{4} (\mbox{log} p)^{1 +\alpha_0/2}\}\Big].
\end{align}
\endgroup
We have
\begin{align}
\label{Ndp1}
\int_{\l \by \l_{\min} \geq x_p}\exp(-\frac{1}{2}\by^\T \by) d\by
&= \Big(\int_{\l u \l \geq x_p}\exp(-\frac{1}{2}u^2)du\Big)^l \notag \\
&= \Big(\frac{2\frac{1}{\surd{(2\pi)}}\exp[-\frac{1}{2}\{t_p^{1/2} \pm \epsilon_n(\mbox{log} p )^{-1/2}\}]}{\surd{\{t_p^{1/2} \pm \epsilon_n(\mbox{log} p )^{-1/2}}\}}\Big)^l\notag \\
& = \{1 + o(1)\}\{\frac{2}{\surd{(8\pi)}}\exp(-\frac{t}{2})\}^lp^{-2l}.
\end{align}
 In addition, \be
 \label{Ndp2}
 O\Big[\exp\{-\frac{1}{4} (\mbox{log} p)^{1 +\alpha_0/2}\}\Big] = o(p^{-2l}).
 \ee
 The facts (\ref{Ndlower}), (\ref{Ndp1}), and (\ref{Ndp2}) together give us
 \begin{align}
 \frac{1 + O(\mbox{log} p)^{-\alpha_0/2}}{(2\pi)^{l/2} }
 \int_{\l \by \l_{\min} \geq x_p}\exp(-\frac{1}{2}\by^\T \by)d\by 
 &+ O\Big[\exp\{-\frac{1}{4} (\mbox{log} p)^{1 +\alpha_0/2}\}\Big]\notag \\ 
 &= \{1 + o(1)\} \{\frac{2}{\surd{(8\pi)}}\exp(-\frac{t}{2})\}^lp^{-2l}\notag \\ 
 &= O(p^{-2l}).
 \end{align}

So we have
\begin{align} 
\label{sumI0l1}
\sum\limits_{I_{0l1} }pr\{ \l \bN_d \l_{\min} \geq t_p^{1/2} \pm \epsilon_n(\log\ p)^{-1/2} \} &\leq \mbox{card}(I_{0l1})\ O(p^{-2l}) \notag \\
& = O(p^{2l -2 + 4\gamma(d-l+1) -2l}) \notag \\&=  o(1).
\end{align}

Let 
 $\bar{a} = min \{a: a\in (k_1, k_2,\ldots,k_d), a \notin S_\star\}. $
 WLOG we assume 
 $d((i_{\bar{a}}, j_{\bar{a})}, (i_{b_1}, j_{b_1})) = 1$,
then
$I_{0l2} = \{ (k_1, \ldots, k_d) \in I_{0l2}: G_{i_{\bar{a}}j_{\bar{a}} i_{b_1} j_{b_1} } $ is $2E-1G$  
Or $"3G-1E"$ Or $"4G-2E" \}$.
On $I_{0l2}$, we have
\be
\label{sumI0l2a}
\sum\limits_{I_{0l2} }pr\{\l \bN_d \l_{\min} \geq t_p^{{1}/{2}} \pm \epsilon_n(\log p)^{-{1}/{2}} \} \leq 
\sum\limits_{I_{0l2} } pr\{\l N_{\bar{a}} \l \geq x_p, \l N_{b_1} \l \geq x_p,\ldots, \l N_{b_l}\l \geq x_p\}.
\ee
Now covariance matrix of $(N_{\bar{a}}, N_{b_1},\ldots,N_{b_l})$ is $\bold{V}_l$, and the covariance matrix satisfies 
$$\vert \vert \bold{V}_l - \mbox{diag}(\bold{D}, \bold{I_{l-1}})\vert \vert 
= O\{(\mbox{log} p)^{-1-\alpha_0}\} $$ 
where $\bold{D}$ is the covariance matrix of $(N_{\bar{a}}, N_{b_1})$.

 Applying (\ref{lessr1}), (\ref{lessr2}), and Lemma 2 in Berman (1962), we obtain 
 \be
  \label{sumI0l2b}
 pr(\l N_{\bar{a}} \l \geq x_p, \l N_{b_1} \l \geq x_p) \leq 
 C\exp(-\frac{4\mbox{log}\ p}{1+ r}) = Cp^{-\frac{4}{1+r}}.
 \ee

  Combining (\ref{sumI0l2a}) and (\ref{sumI0l2b}), we get
\begin{align}
\label{sumI0l2}
\sum\limits_{I_{0l2} } pr(\l N_{\bar{a}} \l \geq x_p, & \l N_{b_1} \l \geq x_p,\ldots, \l N_{b_l} \l \geq x_p) \notag \\
 &\leq C \sum\limits_{I_{0l2} } \Big[pr(\l N_{\bar{a}} \l \geq x_p, \l N_{b_1} \l \geq x_p)\times p^{-2l+2} + \exp\{-(\log p)^{1+\alpha_0/2}/4\}\Big] \notag \\ 
 & \leq  C \sum\limits_{I_{0l2} } \Big[p^{-2l - (2-2r)/(1+r)  } 
 + \exp\{-(\mbox{log} p)^{1+\alpha_0/2}/4\}\Big] \notag \\
& \leq C\ \ p^{-(2- 2r)/(1 + r) + 4\gamma (d-l)} \notag \\
&= o(1).
\end{align}  
 The facts (\ref{sumI0l1}) and (\ref{sumI0l2}) yield (\ref{Minor}).
 
Last but not least, we prove (\ref{Major}).
Repeat the above argument on $I_0^c$, and since $I_0^c = I_{0d}$, or $l = d$, we have
\begin{align}
pr\{&\l \bold{N}_d \l_{\min} \geq t_p^{1/2} 
\pm \epsilon_n(\mbox{log} p)^{-1/2}\} 
= P(\l N_{b_1} \l \geq x_p,\ldots, \l N_{b_l}\l \geq x_p) \notag \\
& =  \frac{1}{(2\pi)^{l/2} \l \bU_l\l^{1/2}}
\int_{\l \by \l_{\min} \geq x_p}\exp(-\frac{1}{2}\by^T \bU_l^{-1} \by)d\by \notag \\
 &= \frac{1}{(2\pi)^{l/2} \l \bU_l \l^{1/2}}
 \int_{\l \by \l_{\min} \geq x_p, \l \by \l_{\max} \leq (\log p)^{1/2 + \alpha_0/4}}
 \exp(-\frac{1}{2}\by^T\bU_l^{-1} \by) d\by \notag \\ 
 &+ \frac{1}{(2\pi)^{l/2} \l \bU_l \l^{1/2}}
 \int_{\l \by \l_{\min} \geq x_p, \l \by \l_{\max} \geq (\log p)^{1/2 + \alpha_0/4}}
 \exp(-\frac{1}{2}\by^\T \bU_l^{-1} \by)d\by \notag \\
 &= \frac{1}{(2\pi)^{l/2} \l \bU_l \l^{1/2}}
 \int_{\l \by \l_{\min} \geq x_p, \l \by \l_{\max} \leq (\log p)^{1/2 + \alpha_0/4}}
 \exp(-\frac{1}{2}\by^\T \bU_l^{-1} \by) d\by \notag \\
 &+ O\Big[\exp\{-\frac{1}{4} (\mbox{log} p)^{1 +\alpha_0/2}\}\Big]\notag\\
 &= \frac{1 + O(\mbox{log} p)^{-\alpha_0/2}}{(2\pi)^{l/2} }
 \int_{\l \by \l_{\min} \geq x_p, \l \by \l_{\max} \leq (\log p)^{1/2 + \alpha_0/4}}
 \exp(-\frac{1}{2}\by^\T \by) d\by \notag \\
 &+ O\Big[\exp\{-\frac{1}{4} (\mbox{log} p)^{1 +\alpha_0/2}\}\Big] \notag 
 \end{align}
 \begin{align}
 &= \frac{1 + O(\mbox{log} p)^{-\alpha_0/2}}{(2\pi)^{l/2} }
 \int_{\l \by \l_{\min} \geq x_p}\exp(-\frac{1}{2}\by^\T \by) d\by+ O\Big[\exp\{-\frac{1}{4} (\mbox{log} p)^{1 +\alpha_0/2}\}\Big] \notag \\
 &= \{1 + o(1)\}\{\frac{2}{\surd{(8\pi)}}\exp(-\frac{t}{2})\}^lp^{-2l}  \notag\\
 &= \{1 + o(1)\}\{\frac{2}{\surd{(8\pi)}}\exp(-\frac{t}{2})\}^d p^{-2d} . \notag
 \end{align}
 So
 \begin{align}
 \label{Changedbyq}
  \sum\limits_{I_0^c }pr\{\l \bN_d \l_{min} \geq t_p^{1/2} \pm \epsilon_n(\mbox{log} p)^{-1/2} \}
   &= \mbox{card}(I_0^c)\{1 + o(1)\}\{\frac{2}{\surd{(8\pi)}}\exp(-\frac{t}{2})\}^dp^{-2d} \notag \\
  & = \{1 + o(1)\}  C_q^d \{\frac{2}{\surd{(8\pi)}}\exp(-\frac{t}{2})\}^dp^{-2d} \notag \\  
  &= \{1 + o(1)\}  \frac{1}{d!} p^{2d} \{\frac{2}{\surd{(8\pi)}}\exp(-\frac{t}{2})\}^dp^{-2d} \notag \\
  &= \frac{1}{d!}\{\frac{1}{\surd{(2\pi)}}\exp(-\frac{t}{2})\}^d\{1 + o(1)\},
 \end{align}
which confirms (\ref{Major}).

Using (\ref{Minor}) and (\ref{Major}), we have

\be
\label{dg2}
\sum\limits_{1 \leq k_1 <\ldots< k_d \leq q }
pr\{\l \bN_d \l_{\min} \geq t_p^{1/2} 
\pm \epsilon_n(\mbox{log} p)^{-1/2} \}
 = \frac{1}{d!}\{\frac{1}{\surd{(2\pi)}}
 \exp(-\frac{t}{2})\}^d\{1 + o(1)\},
\ee
for any $d \geq 2$ and $t \in \mathbb{R}$. Lemma \ref{TechLem} now is verified due to (\ref{de1} ) and (\ref{dg2}) \hfill $\square$

\subsection{Proof of Main Results}
\label{proofofmain}

\noindent \textit{Proof of Theorem \ref{generalTheo}:} Let us first assume $\beta=0$ and then $\gamma=1$. The proof of the general case
is given at the end of this proof. Notice that, the main steps in proving the limiting distribution of our test statistics are based on a similar idea as the main steps of the proof for the main result in Cai et al. (2013).

We first approximate $\hat{D}_n$ by its counter part $\hat{D}_n^*$ defined by 
 $\hat{D}_n^*=\max_{1\leq i, j\leq p} \hat{D}^{*2}_{ij}$ and $\hat{D}^{*}_{ij}={( \bE_j^\T \bold{V}_n \hat{\bw}_{i,0} - \bE_j^\T \bE_i)}/\surd{\theta_{ij}}$.
Based on Lemma 9 in Le $\&$ Zhong (2021), we have 
$\max_{1\leq i, j \leq p} \l \hat{\omega}_{ij,0}- \omega_{ij}^*\l = O_p\{\surd{(\mbox{log} p/{n})}\}$. Moreover, by Lemma A.3 
in Bickel $\&$ Levina (2008), we have $\max_{1\leq i, j \leq p} \l v_{ij} - \sigma_{ij}^*\l=O_p\{\surd{(\log p/{n})}\}.$ 
Then we have 
  \begin{align*}
   |\hat{D}_n/ \hat{D}_n^*-1|  
 & \leq \max_{1\leq i, j\leq p}|\hat{\theta}_{ij,0}/\theta_{ij}-1|  \\
 &= \max_{1\leq i, j\leq p}(\hat{\omega}_{ii,0}v_{jj}-\omega_{ii}^* \sigma_{jj}^*)/\omega_{ii}^* \sigma_{jj}^*
=o_p\{\surd{(\log p/n)}\}.
  \end{align*}
Since $\log p/n\to 0$, we have $ \l \hat{D}_n - \hat{D}_n^*\l=o_p(\hat{D}^*_n)$. 
To prove (\ref{TestProblem}), it is sufficient to prove that
\begin{align*}
pr\{\hat{D}_n^* - 4\log(p) + \log (\log p) \leq t \} \to \exp\big\{-\exp(-{t}/{2})/\surd{(2\pi)}\big\}.
\end{align*}
Define $t_p = t + 4\log(p) - \log(\log p)$ and $\hat{D}_{n1}^* = \max _{(i, j) \in A/A_0} \hat{D}_{ij}^{*2}$ where $A_0 = \{ (i, j): \omega_{ij}^* \neq 0 \}$.
Applying Lemma \ref{MaxA0} in Section 3 of the supplemental material, it is enough to show that
\begin{equation}
\label{TestProblem5}
pr( \hat{D}_{n1}^*  \leq t_p ) \to \exp\big\{-\exp(-{t}/{2})/\surd{(2\pi)}\big\}.
\end{equation}

Define $\hat{D}_{n2}^* = \max _{(i, j) \in A/B_0} \hat{D}_{ij}^{*2}$ where $B_0=A_0\cup A_1$ and $A_1 = \cup_{i=1}^p \{(i, k): \lim_{p \to\infty} s_0\sigma_{ik} \neq 0,\ \mbox{for all} \ (i, k) \notin A_0 \}$.
Using Lemma \ref{MaxA1} in Section 3 of the supplemental material, we have
$$\l pr(\hat{D}^*_{n2} \geq t_p ) - pr( \hat{D}^*_{n1} \geq t_p)\l\leq pr(\max\limits_{(i,j) \in A_1}\hat{D}_{ij}^{*2} \geq t_p) = o(1).$$
It is then sufficient to show that
\begin{equation}
\label{TestProblem6}
pr( \hat{D}^*_{n2} \leq t_p )\to \exp\big\{-\exp(-{t}/{2})/\surd{(2\pi)}\big\}.
\end{equation}

Recall that $D_{ij}={( \bE_j^\T \bold{V}_n \bw_{i} - \bE_j^\T \bE_i)}/\surd{\theta_{ij}}$ and $D_{n2}=\max\limits_{(i, j) \in A/B_0} {D}_{ij}^2$. It then follows that
\begin{align}
 \l \hat{D}_{n2}^{*1/2} - D_{n2}^{1/2} \l 
 &= \l \max\limits_{(i, j) \in A/B_0} |\hat{D}_{ij}^*| - \max\limits_{(i, j) \in A/B_0} |{D}_{ij}| \l \leq  \max\limits_{(i, j) \in A/B_0} |\hat{D}_{ij}^*-{D}_{ij}|  \notag \\
 &\leq C\surd{n}\max\limits_{(i,j) \in A/B_0}
  \l \bE_j^\T \bold{V}_n(\hat{\bw}_{i,0} - \bw_i^*) \l \notag \\ 
 &= C\surd{n} \max\limits_{(i,j) \in A/B_0} \l \bE_j^\T \bold{V}_n \bB_{i,0}(\bS_i^{-1} - \bOmg_i^*) \mathbf{f}_i \l \notag \\
 &  \leq C\surd{n}\max\limits_{(i,j) \in A/B_0} \l \bE_j^\T (\bold{V}_n - \bsig^*) \bB_{i,0}(\bS_i^{-1} - \bOmg_i^*) \mathbf{f}_i \l \notag \\
 &\quad+ C\surd{n}\max\limits_{(i,j) \in A/B_0} \l \bE_j^\T  \bsig^* \bB_{i,0}(\bS_i^{-1} - \bOmg_i^*) \mathbf{f}_i \l \label{eSigBa}
 \end{align}
for some positive constant $C$ where $\bOmg_i^*=\bB_{i,0}^\T\bOmg^*\bB_{i,0}$ and $\bS_i = \bB_{i,0}^\T\bSn^*\bB_{i,0}$.
 
For the first term on the right-hand side of (\ref{eSigBa}), we have
\begin{align}
\label{eSigBb}
\surd{n} \max\limits_{(i,j) \in A/B_0} \l \bE_j^\T 
(\bold{V}_n - \bsig^*) & \bB_{i,0}(\bS_i^{-1} - \bOmg_i^*) \bf{f}_i \l \notag \\
&\leq \surd{n} s_0 \max\limits_{1 \leq i,j \leq p} \l v_{ij} 
- \sigma_{ij}^*\l \max\limits_{1 \leq i,j \leq p} \l \hat{w}_{ij,0} 
- \omega_{ij}^*\l \notag \\
 &= O_p(s_0\log p/\surd{n}) = o_p(\surd{\log p}).
\end{align}

Applying Lemma \ref{eSigB} in Section 3 of the supplemental material, the second term on the right-hand side of (\ref{eSigBa}) is
at the order of $o_p(\surd{\log p})$. Then we have $\l\hat{D}_{n2}^{*1/2} - D_{n2}^{1/2} \l = o_p(\surd{\log p})$.
Because of the inequality $\l \hat{D}_{n2}^* - D_{n2} \l \leq 2 \l D_{n2}^{1/2} \l \ \l \hat{D}_{n2}^{*1/2} - D_{n2}^{1/2} \l + \l \hat{D}_{n2}^{*1/2} - D_{n2}^{1/2} \l ^2$,
to verify (\ref{TestProblem6}) is sufficient to show
\be
\label{TestProblem7}
pr( {D}_{n2} \leq t_p ) \to \exp\big\{-\exp(-{t}/{2})/\surd{(2\pi)}\big\}.
\ee

Suppose there are $k$ isolated nodes in the true network, for any two nodes $i$ and $j$ belong to this isolated nodes set, we have
$(\bE_j^\T\bold{V}_n \bw_i^*)^2/(n\omega_{i i}^*\sigma_{j j}^*)=(\bE_i^\T \bold{V}_n \bw_j^*)^2/(n\omega_{jj}^*\sigma_{ii}^*).$
On the set of isolated nodes we are only maximizing over $[k^2/2]$ components. Thus, ${D}_{n2}$ involves the maximization of $p^2 - k^2/2$ components $D_{ij}^2$. 
For convenience, denote the set that ${D}_{n2}$ is maximizing over as $A/B_0^*$. For any $(i,j)\in A/B_0^*, D_{ij}^2 \neq D_{ji}^2$. 
It is clear that $A/B_0^* \subset A/B_0$.

Re-enumerate the index pairs $(i,j)$ in $A/B_0^*$ to $(i_k, j_k)$, where $k= 1,\ldots,q,$ for $q = \mbox{card}(A/B_0^*)$. 
Since $k = o(p)$ and $\mbox{card}(A/B_0^*)=p^2\{1+o(1)\}$, we have $q = p^2\{1 + o(1)\}$. Then, (\ref{TestProblem7}) is rewritten as
\be
\label{TestProblem8}
pr(\max \limits_{1 \leq k \leq q} {V_k}^2   \leq t_p)\to \exp\big\{-\exp(-{t}/{2})/\surd{(2\pi)}\big\}
\ee
where $V_k=\sum_{l = 1}^n \bE_{j_k}^\T \bX_l\bX_l^\T \bw_{i_k}^*/\surd{(n\omega_{i_k i_k}^*\sigma_{j_k j_k}^*)}$.

Define $Z_{lk}= \bE_{j_k}^\T\bX_l\bX_l^\T\bw_{i_k}^*$ and $\tau_n = 8C\eta^{-1} \mbox{log}(p+n)$
where $C$ is some positive constant and $\eta$ is a constant as in (\ref{Gausstail}). Define $\hat{Z}_{lk}= Z_{lk}I\{\l Z_{lk} \l \leq \tau_n\}- E[Z_{lk}I\{\l Z_{lk} \l \leq \tau_n\}]$
as the centralized truncated version of $Z_{lk}$, and $\hat{V}_k={\sum_{l = 1}^n\hat{Z}_{lk}}/\surd{(n\omega_{i_k i_k}^*\sigma_{j_k j_k})^*}.$ 
To show (\ref{TestProblem8}), it is sufficient to show
\be
\label{TestProblem9}
pr(\max_{1\leq k \leq q} {\hat{V}_k}^2  \leq t_p)\to \exp\big\{-\exp(-{t}/{2})/\surd{(2\pi)}\big\}.
\ee
The above claim is true if (\ref{TestProblem9}) implies (\ref{TestProblem8}). First, note that
\begin{align}
\label{truncate1}
\max \limits_{1 \leq k \leq q} \frac{1}{\surd{n}}
 \sum\limits_{l=1}^n E \vert Z_{lk} \l \ I\{\vert Z_{lk} \l \geq \tau_n  \}
&= \max \limits_{1 \leq k \leq q} \frac{1}{\surd{n} } \sum\limits_{l=1}^n E \l Z_{lk} \l \ I\{\l \eta Z_{lk} \l \geq 2C\ \mbox{log}(p+n)^4\} \notag \\
& \leq \max \limits_{1 \leq k \leq q}  \max \limits_{1 \leq l \leq n} \surd{n}\ E \l Z_{lk} \l \ I\{\l \eta Z_{lk} \l \geq 2C\ \mbox{log}(p+n)^4  \} \notag \\
& \leq \max \limits_{1 \leq k \leq q} \max \limits_{1 \leq l \leq n} \surd{n} (p + n)^{-4}\    E \vert Z_{lk} \l \ \exp \big ( \eta \vert Z_{lk} / (2C) \l \big ) .
\end{align}
The last inequality is due to
 $\exp \{ \l \eta Z_{lk}/(2C) \l \}  (p + n)^{-4} \geq 1 $ if $\l \eta Z_{lk}/(2C) \l \geq \log (p+n)^4$.

Assume that the first $s_0$ components $\bw_{i_k}^*$ are non-zeros to simplify notations, that is
 $\bw_{i_k}^*=(\omega_{i_k1}^*,\ldots, \omega_{i_ks_0}^*,0,\ldots,0)^\T$, then
\begin{align}
\l Z_{lk} \l &= \l \bE_{j_k}^\T \bX_l\bX_l^\T\bw_{i_k}^* \l = \l \omega_{i_k1}^* X_{lj_k}X_{l1} +\cdots+ \omega_{i_ks_0}^* X_{lj_k}X_{ls_0} \l \notag \\
& \leq \frac{1}{2}(X_{lj_k}^2+\max\limits_{a =1,\ldots,s_0}X_{la}^2) \sum_{a=1}^{s_0} \l \omega_{i_ka}^* \l \leq C\max_{h \in U_0} X_{lh}^2,
\end{align}
where $U_0 = \{ j_k, l_1,\ldots,l_{s_0}\}$. Because $\bX = (X_1, X_2,\ldots,X_p)^\T$ is multivariate Gaussian distributed, for some $\eta >0$ we have
\be
\label{Gausstail}
E\{\exp(\eta\ X_i^2)\} \leq 2, 
\ee 
for $i = 1, 2,\ldots,p.$ Applying (\ref{Gausstail}), we get
\begin{align}
\label{truncate2}
E [\vert Z_{lk} \l \exp \{ \eta \vert Z_{lk} /(2C) \l \}] &\leq C E \{\exp(\eta \l Z_{lk} / C \l)\}
= C E[\exp \{ \eta    \max_{h \in U_0}(X_{lh}^2) \}]\nonumber \\
 &= C E \max\limits_{h \in U_0}   \exp(\eta X_{lh}^2)= O(s_0).
\end{align}
Combining (\ref{truncate1}) and (\ref{truncate2}), we obtain
\begin{align}
\label{trunabove}
 \max \limits_{1 \leq k \leq q}\frac{1}{\surd{n}}\sum_{l=1}^n E \l Z_{lk} \l \ I\{\l Z_{lk} \l \geq \tau_n  \} =  O\{ s_0 \surd{n} nq(p + n)^{-4}\} = o\big\{(\mbox{log} p)^{-1}\big\}.
 \end{align}
Because 
$\max_{1 \leq k \leq q} \l\sum_{l=1}^n  E  Z_{lk}I\{\l Z_{lk} \l \geq \tau_n\}\l \leq  \max_{1 \leq k \leq q}  \sum_{l=1}^n E \l Z_{lk} \l I\{\l Z_{lk}\l \geq \tau_n\}$,
equation (\ref{trunabove}) gives us
 \be 
 \label{trunaboveorder}
 \max \limits_{1 \leq k \leq q} \frac{1}{\surd{n} } \l \sum\limits_{l=1}^n  E Z_{lk} \ I\{\l Z_{lk} \l \geq \tau_n\} \l = o\big\{ (\mbox{log} p)^{-1}\big\}.
 \ee
In addition, on the set $A/B_0^*$, we have $E  Z_{lk} = E (\bE_{j_k}^\T\bX_l\bX_l^\T\bw_{i_k}) = \bE_{j_k}^\T \bsig^* \bw_{i_k}^* =0$, therefore
\be \label{trunmean} \max \limits_{1 \leq k \leq q} \frac{1} {\surd{n} } \l\sum\limits_{l=1}^n E  Z_{lk} \l = 0. \ee
 
Using (\ref{trunaboveorder}) and (\ref{trunmean}), we get
\be
\label{trunorder}
\max \limits_{1 \leq k \leq q} \frac{1}{\surd{n} } \l 
\sum\limits_{l=1}^n E  Z_{lk} \ I\{\l Z_{lk} \l \leq \tau_n\}  \l = o\big\{ (\mbox{log} p)^{-1}\big\}. 
\ee
Hence 
\begin{align*}
&pr\{ \max\limits_{1 \leq k \leq q} \l V_k - \hat{V}_{k}\l\geq (\log p)^{-1}\} = pr\{\max\limits_{1 \leq k \leq q} \l 
\frac{1}{\surd{n}} \sum\limits_{l=1}^n (Z_{lk} -\hat{Z}_{lk})\l \geq (\mbox{log} p)^{-1} \}\\
& = pr\Big[\max\limits_{1 \leq k \leq q} \l \frac{1}{\surd{n}} \sum\limits_{l=1}^n (Z_{lk}I\{\l Z_{lk} \l \geq \tau_n\} 
+ E Z_{lk}\ I\{ \l Z_{lk} \l \leq \tau_n \})\l\geq (\mbox{log} p)^{-1} \Big]\\
& = pr\Big[\max\limits_{1 \leq k \leq q} \l\frac{1}{\surd{n}} \sum\limits_{l=1}^n Z_{lk}I\{\l Z_{lk} \l \geq \tau_n\} \l \geq (\mbox{log} p)^{-1} \Big].
\end{align*} 
It follows that
\begin{align} 
\label{Vkdiff} 
&pr\{\max\limits_{1 \leq k \leq q} \l V_k - \hat{V}_{k}\l\geq (\log p)^{-1}\} \leq pr(\max \limits_{1 \leq k \leq q} \max \limits_{1 \leq l \leq n} \l Z_{lk}\l \geq \tau_n) \notag \\
& \leq pr\Big[\max \limits_{1 \leq k \leq q} \max \limits_{1 \leq l \leq n}  C\{ X_{lj_k}^2 + \max(X_{l1}^2,\ldots, X_{ls_0}^2)\} \geq \tau_n\Big]\notag \\ 
&\leq  q\cdot pr\{ \max \limits_{1 \leq l \leq n}  X_{lj_k}^2 \geq \tau_n/(2C)\}
+ q\cdot pr\{ \max \limits_{1 \leq l \leq n}  \max(X_{l1}^2,\ldots, X_{ls_0}^2) \geq \tau_n/(2C)\}. 
\end{align}
For $j=1,\ldots, p$, we have
\begin{align*}
pr(X_j^2 \geq \tau_n/2C)&= pr\big\{ \exp(\eta X_j^2) \geq \exp(\eta \tau_n/2C)\big\}\\
 &\leq \E\Big\{\exp(\eta X_j^2)\Big\}\exp(-\eta\tau_n/2C) \leq 2(p +n)^{-4}.
\end{align*}
This gives us
\be
\label{Vkd1}
 q\cdot pr( \max \limits_{1 \leq l \leq n}   X_{lj_k}^2 \geq \tau_n/2C) 
 \leq nq \max\limits_{1\leq j \leq p }P(X_j^2 \geq {\tau_n}/{2C}) \leq {2nq}/{(n+p)^4} 
 = o(1)
\ee
and
\begin{align} 
\label{Vkd2} 
q\cdot pr\{ \max _{1 \leq l \leq n}  \max(X_{l1}^2,\ldots, X_{ls_0}^2) \geq \tau_n/(2C)\} 
&\leq nqs_0 \max_{1\leq j \leq p } pr(X_{j}^2 \geq \tau_n/2C) \notag \\
&\leq 2 nqs_0 /(n+p)^4  = o(1).
\end{align}

Combining (\ref{Vkdiff}), (\ref{Vkd1}), and (\ref{Vkd2}), we obtain $pr\{\max\limits_{1 \leq k \leq q}\l V_k-\hat{V}_{k}\l \geq (\mbox{log} p)^{-1} \}= o(1).$ This means
\be
\label{eq1}
\max\limits_{1\leq k \leq q} \l V_k - \hat{V}_k \l=O_P\big\{ (\mbox{log} p)^{-1}\big\}.
\ee
We have
\be
\label{Ineq1}
\vert \max\limits_{1\leq k \leq q}V_k^2 
- \max\limits_{1\leq k \leq q}\hat{V}_k^2 \l
 \leq  2 \max\limits_{1\leq k \leq q} \l \hat{V}_k \l 
 \max\limits_{1\leq k \leq q} \l V_k - \hat{V}_k \l 
 + \max\limits_{1\leq k \leq q} \l V_k - \hat{V}_k \l^2.
\ee
If (\ref{TestProblem9}) holds, then $\max\limits_{1\leq k \leq q}\hat{V}_k = O_p(\surd{\log p})$. 
In addition, (\ref{eq1}) and (\ref{Ineq1}) implies that  
$\l\max_{1\leq k \leq q}V_k^2-\max_{1\leq k \leq q}\hat{V}_k^2\l
=o_p(1).$
As a result, to prove (\ref{TestProblem8}), it is sufficient to prove (\ref{TestProblem9}).

Finally, we prove (\ref{TestProblem9}). Let us denote
\begin{equation}
\label{Wlvariable}
 \tilde{Z}_{lk} = { { \hat{Z}_{lk}}}/{\surd({ \omega_{i_k i_k}^* \sigma_{j_k j_k}^*}}), 
  \bW_l = (\tilde{Z}_{lk_1},\tilde{Z}_{lk_2},\ldots,\tilde{Z}_{lk_d}),  
\end{equation}
for $l =1,\ldots,n$, and denote $E_{kj} = \{ \hat{V}_{kj}^2 \geq t_p\}$ for any integer
 $1 \leq k_j \leq q$. 
Applying Bonferroni inequality in Lemma \ref{bonferri} in the supplemental material for $pr(\max_{1 \leq k \leq q} {\hat{V}_k}^2 \geq t_p)$, we have
\begin{align}
\label{Bon}
\sum_{d = 1}^{2m}{(-1)}^{d-1} 
\sum_{1 \leq k_1 <\cdots< k_d \leq q }
 pr(\bigcap^d_{j=1} E_{k_j}) 
&\leq pr(\max_{1 \leq k \leq q} {\hat{V}_k}^2   \geq t_p) \notag\\
&\leq  \sum_{d = 1}^{2m - 1}{(-1)}^{d-1}
 \sum_{1 \leq k_1 <\cdots< k_d \leq q }
 pr(\bigcap^d_{j=1}E_{k_j}),
\end{align}
for any fixed integer $m<[q/2]$.
 
Rewrite $pr(\bigcap^d_{j=1}E_{k_j})$ as
$pr(\bigcap^d_{j=1} E_{k_j}) 
= pr (\l n^{{-1}/{2}}\sum_{l=1}^n \bW_l \l_{\min} 
\geq t_p^{{1}/{2}}) $. 
We will apply Zaitsev approximation to approximate this probability. To this end, 
we first check the conditions for Zaitsev approximation in Lemma \ref{Zaitsepapp} in the supplemental material.
Define
\begin{align*}
\bxi_i &= n^{-{1}/{2}} \bw_i^* 
= n^{-{1}/{2}} (\tilde{Z}_{ik_1},\tilde{Z}_{ik_2},\ldots,\tilde{Z}_{ik_d}) \\
&= n^{-{1}/{2}}\{\hat{Z}_{ik_1}/(\omega_{i_{k_1} i_{k_1}}^* \sigma_{j_{k_1} j_{k_1}}^*)^{1/2},\ldots,
\hat{Z}_{ik_d}/(\omega_{i_{k_d} i_{k_d}}^* \sigma_{j_{k_d} j_{k_d}}^*)^{1/2}\}.
\end{align*}
We have $E\bxi_i = 0$, for $i =1,\ldots,n,$ and  $\bxi_1,\ldots,\bxi_n$ are independent. We also have
\begin{align*}
\vert (\bxi_i,\bu) \vert ^{m-2} &\leq \vert \vert \bxi_i \vert \vert ^{m-2} 
\vert \vert \bu \vert \vert ^{m-2} \leq \vert \vert \bu \vert \vert ^{m-2} 
(2\surd{(d/n)}  \tau_n) ^{m-2} 
= 2^{m-2} \tau^{m-2}\vert \vert \bu \vert \vert ^{m-2} \\
&\leq \frac{1}{2} m!\tau^{m-2}\vert \vert \bu \vert \vert^{m-2}
\end{align*}
where $m \geq 3$ and $\tau = \surd{(d/n)} \tau_n= 8C\eta^{-1}\surd{(d/n)}  \log (p + n)$. 
It follows that $\l \E (\bxi_i,\bt)^2(\bxi_i,\bu)^{m-2} \l\leq 1/2 m! \tau^{m-2} \vert \vert \bu \vert \vert^{m-2} \E (\bxi_i,\bt)^2$, 
for $i = 1,\ldots,n.$

Applying Lemma \ref{Zaitsepapp}, we have
\begin{align}
\label{zai1}
pr(\l \bN_d \l_{\min} \geq \surd{t_p}+\epsilon_n/\surd{\log p}) &-pr(\l\sum_{l=1}^n\bW_l \l_{\min} \geq \surd{(nt_p)}) \notag \\
 &\leq c_1d^{5/2}\exp\Big\{-\epsilon_n(d^5\log p)^{-1/2}/(\tau c_2)\Big\}
\end{align}
where $\bN_d = (N_{k_1}, N_{k_2},\cdots,N_{k_d})^\T$ is a $d$-dimensional multivariate normal
distributed random vector with mean vector $\E \bN_d =0$ and covariance matrix $\mbox{cov}(\bN_d) = \mbox{cov}(\bW_1)$.
Notice that $d$ is fixed and does not depend on $n, p$, and 
\begin{align}
\label{zai2}
c_1d^{5/2}\exp\{-\frac{\epsilon_n(\log\ p)^{-{1}/{2}}}{\tau c_2d^{5/2}}\} 
&= c_1d^{5/2}\exp\{-\frac{\epsilon_n(\mbox{log}\ p)^{-{1}/{2}}}{ 8C\eta^{-1}\surd{(d/n)} \mbox{log}(p + n) c_2d^{5/2}}\} \notag \\
& = O \Big[\exp\{-{\epsilon_n \surd{n}}/{ (\mbox{log} p)^{3/2} }\}\Big]
 = O(p^{-M}),
\end{align} 
for some $M >0$ and $\epsilon_n \rightarrow 0$ sufficient slow. The facts (\ref{zai1}) and (\ref{zai2}) give us
\be
\label{Bonplus}
pr(\l \bN_d \l_{\min} \geq \surd{t_p}+\epsilon_n/\surd{\mbox{log} p})-pr\{\l\sum_{l=1}^n\bW_l \l_{\min} \geq \surd{(nt_p)}\}=O(p^{-M}),
\ee
for some $M>0$. Similarly, we can prove
\be
\label{Bonminus}
pr\{\l\sum_{l=1}^n\bW_l \l_{\min} \geq \surd{(nt_p)}\}-pr(\l \bN_d \l_{\min} \geq \surd{t_p}-\epsilon_n/\surd{\mbox{log} p})=O(p^{-M}),
\ee
for some $M>0$.  
  
Applying (\ref{Bon}) and (\ref{Bonminus}), we get
\begin{align}
\label{Bonupper}
pr(\max_{1 \leq k \leq q} {\hat{V}_k}^2   \geq t_p)
 &\leq  \sum_{d = 1}^{2m - 1}{(-1)}^{d-1}\sum_{1 \leq k_1 <\cdots< k_d \leq q}pr(\bigcap^d_{j=1}E_{k_j}) \notag \\
 &=\sum_{d = 1}^{2m - 1}{(-1)}^{d-1}\sum_{1\leq k_1 <\cdots< k_d \leq q}pr(\l n^{{-1}/{2}}\sum\limits_{l=1}^n
  \bW_l \l_{\min} \geq t_p^{{1}/{2}}) \notag \\
 &=\sum_{d = 1}^{2m - 1}{(-1)}^{d-1}\sum_{1 \leq k_1 <\cdots< k_d \leq q }
 pr\{ \l\bN_d \l_{\min} \geq t_p^{{1}/{2}}-\epsilon_n(\mbox{log} p)^{-{1}/{2}} \}+o(1).
\end{align}
Similarly, applying (\ref{Bon}) and (\ref{Bonplus}), we get
\be
\label{Bonlower}
pr(\max\limits_{1 \leq k \leq q} {\hat{V}_k}^2   \geq t_p) 
\geq \sum\limits_{d = 1}^{2m - 1}{(-1)}^{d-1} 
\sum\limits_{1 \leq k_1 <\cdots< k_d \leq q }
pr\{ \l \bN_d \l_{\min} \geq t_p^{{1}/{2}} 
+ \epsilon_n(\mbox{log} p)^{-{1}/{2}} \} - o(1). 
\ee
Combining Lemma \ref{TechLem} in the supplemental material, (\ref{Bon}), (\ref{Bonupper}), and (\ref{Bonlower}), we obtain
\begin{align*}
\sum\limits_{d = 1}^{2m}{(-1)}^{d-1}
 \frac{1}{d!}\{\frac{1}{\sqrt{2\pi}}
 \exp(-\frac{t}{2})\}^d\{1 + o(1)\} 
& \leq pr(\max_{1 \leq k \leq q} {\hat{V}_k}^2   \geq t_p) \\
&\leq  \sum\limits_{d = 1}^{2m - 1}{(-1)}^{d-1} 
\frac{1}{d!}\{\frac{1}{\sqrt{2\pi}}
\exp(-\frac{t}{2})\}^d\{1 + o(1)\}.
\end{align*}
It follows that
 $$\lim\sup_{n\to\infty} 
 pr(\max_{1 \leq k \leq q} {\hat{V}_k}^2 \geq t_p)\leq  \sum_{d = 1}^{2m - 1}{(-1)}^{d-1}\frac{1}{d!}\Big\{\frac{1}{\surd{(2\pi)}}\exp(-{t}/{2})\Big\}^d.$$
Let $m \to \infty$ then
\be
\label{limsup}
\lim\sup_{n\to \infty}pr(\max_{1 \leq k \leq q} {\hat{V}_k}^2   \geq t_p) 
\leq 1- \exp\big\{-\exp(-{t}/{2})/\surd{(2\pi)}\big\}. 
\ee
Similarly, we get 
\be
\label{liminf}
\lim\inf_{n \to \infty}pr(\max_{1 \leq k \leq q} {\hat{V}_k}^2   \geq t_p) 
\geq 1- \exp\big\{-\exp(-{t}/{2})/\surd{(2\pi)}\big\}.  
\ee
The facts (\ref{limsup}) and (\ref{liminf}) give us
 $$\lim_{n\to \infty}pr(\max_{1 \leq k \leq q} {\hat{V}_k}^2   \geq t_p)
 =1-\exp\big\{-\exp(-{t}/{2})/\surd{(2\pi)}\big\}. $$ 
 
In other words, 
$$\lim_{n\to \infty}pr(\max_{1 \leq k \leq q} {\hat{V}_k}^2   \leq t_p)
 =\exp\big\{-\exp(-{t}/{2})/\surd{(2\pi)}\big\}.$$ 
This finishes the proof of equation (\ref{TestProblem9}) and then the result in equation (\ref{TestProblem}) holds.

The proof of the general case with $\beta\neq 0$ is similar to the above proof with $\beta=0$ but we are maximizing over $q = p^2 - {k^2}/{2} = p^2 (1 - {\beta^2}/{2})$ components which changes 
Lemma \ref{TechLem} in the supplemental material to
$$\sum\limits_{1 \leq k_1 <\cdots< k_d \leq q } 
pr\{ \l \bN_d \l_{\min} \geq t_p^{{1}/{2}} \pm \epsilon_n(\mbox{log} p)^{-{1}/{2}} \}
= \frac{1}{d!}\{\frac{1}{\surd{(2\gamma \pi)}}\exp(-{t}/{2})\}^d\{1 + o(1)\}.$$
To verify this, we can repeat the proof of Lemma \ref{TechLem} where equation (\ref{Changedbyq}) in the lemma is replaced by 
\begin{align*}
  \sum\limits_{I_0^c }pr\{\l \bN_d \l_{\min} 
  \geq t_p^{{1}/{2}} \pm \epsilon_n(\mbox{log} p)^{-{1}/{2}} \} 
  & = \{1 + o(1)\}  C_q^l \{ \frac{2}{\surd{(8\pi)}}
  \exp(-{t}/{2})\}^lp^{-2l} \\  
  &=  p^{2l} \{(1- \frac{\beta^2}{2})\frac{2}{\surd{(8\pi)}}\exp(-{t}/{2})\}^lp^{-2l}\{1 + o(1)\} \\
  &= \frac{1}{d!}\{\frac{1}{\surd{(2\gamma\pi)}}\exp(-{t}/{2})\}^d\{1 + o(1)\}. \qquad\qquad\qquad \square
 \end{align*}

\noindent \textit{Proof of Theorem \ref{generalTheoThreshold}:} We will show that, under $H_0$, the modified test statistic 
$\tilde{D}_n$ converges to the same distributions as $\hat{D}_n$ as in Theorem \ref{generalTheo}.
We first note that $\tilde{D}_{ij}^2 = \hat{D}_{ij}^2$ if $\Delta_i=0$ for all $i=1,\ldots, p$. Then,
\begin{align*}
pr(\tilde{D}_n  \leq t_p)&=pr(\tilde{D}_n  \leq t_p,\Delta_i=0)+pr(\tilde{D}_n  \leq t_p,\Delta_i\neq 0)\\
&=pr(\hat{D}_n  \leq t_p)+pr(\tilde{D}_n  \leq t_p,\Delta_i\neq 0).
\end{align*}
Since  $pr(\tilde{D}_n  \leq t_p,\Delta_i\neq 0)\leq pr(\Delta_i\neq 0)$, it is sufficient to show that $pr(\Delta_i\neq 0)=0$ under $H_0$.
For any $(i,j)\in \bbE$  but $(i,j)\notin\bbE_0$, under $H_0$, $\Delta_{ij}=0$ according to the definition of $B_{i,0}$. Thus, it is enough to show the
following:
\begin{align}
\label{toprove1}
pr(\max_{(i,j) \in \bbE_0}\Delta_{ij}=0)=1.
\end{align}
To this end, we note
\begin{align*}
pr(\max_{(i,j) \in \bbE_0}\Delta_{ij}=0)&=pr(\min_{i=1,\ldots, p, j=1,\ldots, s_i}|\hat{\omega}_{i1,0}^{(j)}|/\hat{\sigma}_{i1,0}^{(j)}>\delta_n)\\
&=1-\cup_{i=1,\ldots, p, j=1,\ldots, s_i}pr(|\hat{\omega}_{i1,0}^{(j)}|/\hat{\sigma}_{i1,0}^{(j)}\leq \delta_n)\\
&\geq 1-\sum_{i=1}^{p}\sum_{j=1}^{s_i} pr(|\hat{\omega}_{i1,0}^{(j)}|/\hat{\sigma}_{i1,0}^{(j)}\leq \delta_n).
\end{align*} 
Under $H_0$, $\omega_{i1,0}^{(j)}\neq 0$, and hence ${{w}_{i1,0}^{(j)}}/{{\sigma}_{i1,0}^{(j)}}=C_{ij}\sqrt{n}$ for some constants $C_{ij}$. Then, for $\delta_n\asymp \surd{\log(n)}$, we have
\begin{align*}
pr\big( \vert \hat{\omega}_{i1,0}^{(j)} \vert /\hat{\sigma}_{i1,0}^{(j)}\leq \delta_n \big)
&= pr\Big( -\delta_n  - \frac{\omega_{i1,0}^{(j)} }{ \sigma_{i1,0}^{(j)}} \leq \frac{  \hat{\omega}_{i1,0}^{(j)} - \omega_{i1,0}^{(j)} }{ \sigma_{i1,0}^{(j)}}\leq
 \delta_n  - \frac{\omega_{i1,0}^{(j)} }{ \sigma_{i1,0}^{(j)}} \Big)\\
&\leq \Phi(\delta_n-C_{ij}\surd{n})\asymp \exp(-C_{ij}^2n/2)/\surd{n},
\end{align*} 
where $\Phi(\cdot)$ is the CDF of the standard normal. Under Condition (C2) and $s_0\asymp o(\surd{n})$, we have
$$
\sum_{i=1}^{p}\sum_{j=1}^{s_i} pr(|\hat{\omega}_{i1,0}^{(j)}|/\hat{\sigma}_{i1,0}^{(j)}\leq \delta_n)\to 0.
$$
Therefore, under the null hypothesis $H_0$, (\ref{toprove1}) holds and the asymptotic distributions of $\hat{D}_n$ and $\tilde{D}_n$ are the same 
when $\delta_n\asymp \surd{\log(n)}$ and $C_n>0$. 

If $\bbE_0$ specified under the null hypothesis $H_0$ includes the true network structure $\bbE^*$, then there exist some $\omega_{i1,0}^{(j)}=0$, 
say $\omega_{i_01,0}^{(j_0)}=0$ and the corresponding $\hat{\omega}_{i_01,0}^{(j_0)}$ are consistent estimators of $\omega_{i_01,0}^{(j_0)}=0$ for some $i_0\in\{1,\ldots, p\}$
and $j_0\in\{1,\ldots, s_i\}$. This event happens with probability one because
\begin{align*}
pr(\Delta_{i1}\neq 0\;\mbox{for some $i=1,\ldots,p$})&=pr\Big(\cup_{i=1}^p\cup_{j=1}^{s_i}\{|\hat{\omega}_{i1,0}^{(j)}|/\hat{\sigma}_{i1,0}^{(j)}\leq \delta_n\}\Big)\\
&\geq pr\Big(\{|\hat{\omega}_{i_01,0}^{(j_0)}|/\hat{\sigma}_{i_01,0}^{(j_0)}\leq \delta_n\}\Big)
= pr\Big(-\delta_n  \leq \hat{\omega}_{i_01,0}^{(j_0)}/\hat{\sigma}_{i_01,0}^{(j_0)}\leq \delta_n \Big)\\
&=\Phi(\delta_n)-\Phi(-\delta_n)\to 1.
\end{align*}
This implies that $pr(\Delta_i=0\;\mbox{for all $i$})=0$. It follows that
\begin{align*}
pr(\tilde{D}_n  > t_p)&=pr(\tilde{D}_n  > t_p, \Delta_i=0\;\mbox{for all $i$})+pr(\tilde{D}_n  > t_p,\Delta_i\neq 0 \;\mbox{for some $i$})\\
&=pr(\tilde{D}_n  > t_p,\Delta_i\neq 0 \;\mbox{for some $i$}).
\end{align*}

When the event $\{\Delta_i\neq 0 \;\mbox{for some $i$}\}$ happens, there exists at least one $\Delta_{ij}=C_n\asymp \surd{\log(p)}\neq 0$. Without loss of generality, assume that there exists one
$\Delta_{ij^*}=C_n=C\surd{\log(p)}\neq 0$ and $\sigma_{jj^*}\neq 0$ for some $j$, then, in probability, we have
\begin{align}
\label{inproblimit}
(\bE_j^T \bold{V}_n\Delta_i)^2/\hat{\theta}_{ij,0}\geq C_n^2(\sum_{l=1}^nX_{lj}X_{lj^*})^2/n^2\hat{\theta}_{ij,0}\to C^2\log(p)(\sigma_{jj}^*\sigma_{j^*j^*}^*+2\sigma_{jj^*}^{*2})/(\omega_{ii}^*\sigma_{jj}^*+1) \;\;,
\end{align}
for some positive constant $C.$ 

Applying Theorem \ref{generalTheo}, for a small $\epsilon>0$, $pr\{\max_{i,j}\hat{D}_{ij}^2\leq (4+\epsilon)\log(p)\}\to 1$.
Using the definition of $\tilde{D}_{ij}^2$, we have the following decomposition of $\tilde{D}_{ij}^2$, 
$$\tilde{D}_{ij}^2 =\hat{D}_{ij}^2+\{(\bE_j^\T \bold{V}_n\Delta_i)^2 + 2(\bE_j^\T \bold{V}_n \hat{\bw}_{i,0}-\bE_j^\T \bE_i)\bE_j^\T \bold{V}_n\Delta_i  \}/\hat{\theta}_{ij,0}.$$   
If $C>4\max_{i,j}(\omega_{ii}^*\sigma_{jj}^*+1)/(\sigma_{ii}^*\sigma_{jj}^*+2\sigma_{ij}^{*2})$, then $\max_{i,j}\tilde{D}_{ij}^2\asymp \max_{i,j}(\bE_j^\T \bold{V}_n\Delta_i)^2/\hat{\theta}_{ij,0}$ with probability one and hence
\begin{align*}
pr(\tilde{D}_n  > t_p,\Delta_i\neq 0 \;\mbox{for some $i$})&=pr\big\{\max_{1\leq i,j\leq p}(\bE_j^\T \bold{V}_n\Delta_i)^2/\hat{\theta}_{ij,0}>t_p,\Delta_i\neq 0 \;\mbox{for some $i$}\big\}\\
&\geq pr\big\{C_n^2(\sum_{i=1}^nX_{ij}X_{ij^*})^2/n^2\hat{\theta}_{ij,0}>t_p\big\}\to 1.
\end{align*}
In summary, $pr(\tilde{D}_n  > t_p)\to 1$ if $\bbE_0$ includes $\bbE^*$. Theorem \ref{generalTheoThreshold} is proved.\hfill $\square$

\bibliographystyle{unsrt}  


\begin{thebibliography}{1}


\bibitem[{Cai et al. (2011)}]{cai2011}
\textsc{ Cai, T., Liu, W. \& Luo, X.} (2011).
\newblock A constrained L1 minimization approach to sparse precision matrix estimation.
\newblock \textit{Journal of the American Statistical Association} $\bold{106}$, 594-607.

\bibitem[{Chen et al.(2010)}]{chenetal2010}
\textsc{ Chen, S. X., Zhang, L. X. \& Zhong, P. S.} (2010).
\newblock Tests for high-dimensional covariance matrices.
\newblock \textit{Journal of the American Statistical Association} $\bold{105}$, 810-819. 

\bibitem[{Cheng et al.(2017)}]{chengetal2017}
\textsc{ Cheng, G., Zhang, Z. \& Zhang, B.} (2017).
\newblock Test for bandedness of high-dimensional precision matrices.
\newblock \textit{Journal of Nonparametric Statistics} $\bold{29}$, 884-902. 

\bibitem[{Drton \& Perlman(2004)}]{drton2004}
\textsc{ Drton, M. \& Perlman, M D.} (2004).
\newblock Model selection for Gaussian concentration graphs.
\newblock \textit{Biometrika} $\bold{91}$, 591-602.

\bibitem[{Edwards(2000)}]{edwards2000}
\textsc{ Edwards, D.} (2000).
\newblock \textit{Introduction to graphical modelling}.
\newblock Springer.



\bibitem[{Eftekhari et al.(2021)}]{eftekhari2021}
\textsc{ Eftekhari, A., Pasadakis, D., Bollh\"{o}fer, M., Scheidegger, S. \& Schenk, O.} (2021).
\newblock Block-enhanced precision matrix estimation for large-scale datasets.
\newblock \textit{Journal of Computational Science} $\bold{53}$, 2975-3026.


\bibitem[{Fan et al.(2015)}]{Fanetal2015}
\textsc{ Fan., J., Liao, Y. \& and Yao, J.} (2015).
\newblock Power enhancement in high-dimensional cross-sectional tests.
\newblock \textit{Econometrica} $\bold{83}$, 1497-1541.


\bibitem[{Friedman et al. (2007)}]{friedman2007}
\textsc{ Friedman, J., Hastie, T. \& Tibshirani, R.} (2007).
\newblock Sparse inverse covariance estimation with the graphical lasso.
\newblock \textit{Biostatistics} $\bold{9}$, 432-441.

\bibitem[{Friedman et al. (2019)}]{glasso2019}
\textsc{ Friedman, J., Hastie, T. \& Tibshirani, R.} (2019).
\newblock glasso: Graphical Lasso: Estimation of Gaussian graphical models.
\newblock \textit{R package version 1.11, https://cran.r-project.org/web/packages/glasso}.








\bibitem[{Guo \& Tang(2021)}]{GuoTang2021}
\textsc{Guo, X \& Tang, C.Y.} (2021) \newblock 
Specification tests for covariance structures in high-dimensional statistical models.
\newblock \textit{Biometrika}, {\bf 108}, 335–351.

\bibitem[{Jankov\'{a} \& Geer(2017)}]{jankova2017}
\textsc{ Jankov\'{a}, J. \& Geer, S. V. D.} (2017).
\newblock Honest confidence regions and optimality in high-dimensional precision matrix estimation.
\newblock \textit{Test} $\bold{26}$, 143-162.



\bibitem[{Lauritzen(1996)}]{lauritzen1996}
\textsc{ Lauritzen, S. L.} (1996).
\newblock \textit{Graphical models}.
\newblock Oxford: Clarendon Press.

\bibitem[{Le \& Zhong(2021)}]{le2021}
\textsc{ Le, T. M. \& Zhong, P. S.} (2021).
\newblock High-dimensional precision matrix estimation with a known graphical structure.
\newblock \textit{Stat}.


\bibitem[{Li \& Li(2008)}]{li2008}
\textsc{ Li, C.  \& Li, H.} (2008).
\newblock Network-constrained regularization and variable selection for analysis of genomic data.
\newblock \textit{Bioinformatics} $\bold{24}$, 1175-1182.


\bibitem[{Li et al.(2020)}]{flare2020}
\textsc{Li, X., Zhao, T., Wang, L., Yuan, X.  \& Liu, H.} (2020).
\newblock flare: Family of Lasso regression.
\newblock \textit{R package version 1.7.0, https://cran.r-project.org/web/packages/flare}.



\bibitem[{Liang \&  Zeger (1986)}]{Liang1986}
\textsc{ Liang, K. Y.  \&  Zeger, S. L.} (1986).
\newblock Longitudinal data analysis using generalized linear models.
\newblock \textit{Biometrika} $\bold{73}$, 13-22.


\bibitem[{Liu \& Wang(2017)}]{liu2017}
\textsc{ Liu, H.  \& Wang, L.} (2011).
\newblock TIGER: A tuning-insensitive approach for optimally estimating Gaussian graphical models.
\newblock \textit{Electronic Journal of Statistics} $\bold{11}$, 241-294.

\bibitem[{Liu \& Luo(2015)}]{Liu2015}
\textsc{ Liu, W.  \& Luo, X.} (2015).
\newblock Fast and adaptive sparse precision matrix estimation in high dimensions.
\newblock \textit{Journal of Multivariate Analysis} $\bold{135}$, 153-162.

\bibitem[{Liu(2013)}]{liu2013}
\textsc{ Liu, W.} (2013).
\newblock Gaussian graphical model estimation with false discovery rate control.
\newblock \textit{The Annals of Statistics} $\bold{41}$, 2948-2978.


\bibitem[{Ning \& Liu(2017)}]{ning2017}
\textsc{ Ning, Y.  \& Liu, H.} (2017).
\newblock A general theory of hypothesis tests and confidence regions for sparse high dimensional models.
\newblock \textit{The Annals of Statistics} $\bold{45}$, 158-195.

\bibitem[{NYT(2022)}]{datasource}
\textsc{ NYT} (2022).
\newblock The New York Times. (2021). Coronavirus (Covid-19) data in the United States. Retrieved on February 1, 2022, from  \textit{https://github.com/nytimes/covid-19-data}.


\bibitem[{Qiu  \& Chen(2012)}]{QiuChen2012}
\textsc{ Qiu, Y.  \& Chen, S. X.} (2012).
\newblock Test for bandedness of high-dimensional covariance matrices and bandwidth estimation.
\newblock \textit{The Annals of Statistics} $\bold{40}$, 1285-1314.

\bibitem[{Ren et al.(2015)}]{ren2015}
\textsc{ Ren, Z., Sun, T., Zhang, C. H.  \&  Zhou, H. H.} (2015).
\newblock Asymptotic normality and optimalities in estimation of large Gaussian graphical models.
\newblock \textit{The Annals of Statistics} $\bold{43}$, 991-1026.


\bibitem[{Wang et al.(2022)}]{Wangetal2022}
\textsc{  Wang, X., Xu, G.  \& Zheng, S.} (2022).
\newblock Adaptive tests for bandedness of
high-dimensional covariance matrices.
\newblock \textit{Statistica Sinica}.


\bibitem[{Xia et al. (2015)}]{Xia2015}
\textsc{Xia, Y., Cai, T.X. \& Cai, T.T.} (2015) 
Testing differential networks with applications to the detection of gene-gene interactions. 
\newblock \textit{Biometrika}
{\bf 102}, 247–266.

\bibitem[{Yuan  \& Lin(2007)}]{yuan2007}
\textsc{ Yuan, M.  \& Lin, Y.} (2007).
\newblock Model selection and estimation Gaussian graphical model.
\newblock \textit{Biometrika} $\bold{94}$, 19-35.

\bibitem[{Zheng et al.(2019)}]{Zhengetal2019}
\textsc{ Zheng, S., Chen, Z., Cui, H.  \&  Li, R.} (2019).
\newblock Hypothesis testing on linear structures of high-dimensional covariance matrix.
\newblock \textit{The Annals of Statistics} $\bold{47}$, 3300-3334.


\bibitem[{Zhong et al.(2017)}]{Zhong2017}
\textsc{ Zhong, P. S., Lan, W., Song, P. X. K.  \& Tsai, C. L.} (2017).
\newblock Tests for covariance structures with high-dimensional repeated measurements.
\newblock \textit{The Annals of Statistics} $\bold{45}$, 1185-1213.


\bibitem[{Zhou et al.(2011)}]{zhou2011}
\textsc{ Zhou, S., R\"{u}timann, P., Xu, M. \& B\"{u}hlmann, P.} (2011).
\newblock High-dimensional covariance estimation based on Gaussian graphical models.
\newblock \textit{Journal of Machine Learning Research} $\bold{12}$, 2975-3026.

\bibitem[{Zhou \& Song(2016)}]{zhou2016}
\textsc{ Zhou, Y. \& Song, P. X. K.} (2016).
\newblock Regression analysis of networked data.
\newblock \textit{Biometrika} $\bold{103}$, 287-301.

\bibitem[{Berman(1962)}]{Berman}
\textsc{ Berman, S. M.} (1962).
\newblock A law of large number for the maximum in a stationary Gaussian sequence.
\newblock \textit{The Annals of Mathematical Statistics} $\bold{33}$, 93-97.

  
   \bibitem[{Bickel \& Levina(2008)}]{Bickel2008}
\textsc{ Bickel, P. J. \& Levina, E.} (2008).
\newblock Regularized estimation of large covariance matrices.
\newblock \textit{The Annals of Statistics} $\bold{36}$, 199-227.

   \bibitem[{Cai et al.(2013)}]{Cai2013}
\textsc{ Cai, T., Liu, W. \& Xia, Y. } (2013).
\newblock Two sample covariance testing and support recovery in high-imensional and sparse settings.
\newblock \textit{Journal of the American Statistical Association} $\bold{108}$, 265-277.

\bibitem[{Demko et al.(1984)}]{Demko84}
\textsc{ Demko, S., Moss, W. S. \& Smith, P. W. } (1984).
\newblock Decay rates for inverses of band matrices.
\newblock \textit{Mathematics of Computation} $\bold{43}$, 491-499.

\bibitem[{Gr\"{o}chenig \& Leinert(2006)}]{Grochenig2006}
\textsc{ Gr\"{o}chenig, K. \& Leinert, M. } (2006).
\newblock Symmetry and inverse closedness of matrix algebras and functional calculus for infinite matrices.
\newblock \textit{Transactions of the American Mathematical Society} $\bold{358}$, 2695-2711.

  \bibitem[{Hager(1989)}]{Hager1989}
\textsc{ Hager, W. W. } (1989).
\newblock Updating the inverse of a matrix.
\newblock \textit{Society for Industrial and Applied Mathematics} $\bold{31}$, 221-239.

\bibitem[{Hall \& Lin(2010)}]{Hall2010}
\textsc{ Hall, P. \& Lin, J.} (2010).
\newblock Innovated higher criticism for detecting sparse signals in correlated noise.
\newblock \textit{The Annals of Statistics} $\bold{38}$, 1686-1732.



  \bibitem[{Robinson \& Wahten(1992)}]{Robinson1992}
\textsc{ Robinson,P. D. \& Wahten, A. J. } (1992).
\newblock Variational bounds on the entries of the inverse of a matrix.
\newblock \textit{IMA Journal of Numerical Analysis} $\bold{12}$, 463-486.


\bibitem[{Zaitsev(1987)}]{Zaitsev87}
\textsc{ Zaitsev, A. Y. } (1987).
\newblock On the Gaussian approximation of convolutions under multidimensional analogues of S.N. Berstein's inequality conditions.
\newblock \textit{Theory and Related Fields} $\bold{74}$, 535-566.




\end{thebibliography}

\end{document}